\documentclass[11pt]{article}
\usepackage{axodraw2}
\usepackage{epsfig}
\usepackage{amsfonts}
\usepackage{slashed}
\usepackage{amsmath,amssymb}
\usepackage{bbm,bm}
\usepackage{cite}
\allowdisplaybreaks[1]
\usepackage{xcolor}
\usepackage{tikz}
\usetikzlibrary{shapes,arrows,shadows,automata,positioning}
\usepackage{pgfplots}
\pgfplotsset{compat=1.18}
\newdimen\nodeDist
\usepackage{comment}
\usepackage[normalem]{ulem} % \sout

\usepackage[
  colorlinks=true,
  linkcolor={red!50!black},
  citecolor={blue!50!black},
  filecolor=black,
  urlcolor=black,
  breaklinks=true
  ]{hyperref}
\usepackage{orcidlink}

\newcommand{\nn}{\nonumber \\}
\renewcommand{\vec}[1]{{\bf #1}}
\newcommand{\rmi}[1]{{\mbox{\scriptsize #1}}}
\newcommand{\rmii}[1]{{\mbox{\tiny\rm{#1}}}}
\newcommand{\hatPip}{\hat{\Pi}^{\prime}}

\newcommand{\Tbar}{\overline{T}}

\newcommand{\LamD}{\bar\Lambda}
\newcommand{\LamDRef}{\LamD_{\text{ref}}}
\newcommand{\LamDInput}{\LamD_{0}}
\newcommand{\gammaE}{{\gamma_\rmii{E}}}
\newcommand{\Veff}{V_{\rmi{eff}}}

\newcommand{\dd}{{\rm d}}

\newcommand{\nG}{n_{\rmii{G}}}

\newcommand{\Tc}{T_{\rm c}}

\newcommand{\CA}{C_\rmii{A}}

\newcommand{\gs}{g_\rmi{s}}

\newcommand{\mDi}[1]{m_{\rmii{D#1}}}

\newcommand{\mH}{m_\rmii{$H$}}

\newcommand{\mZ}{m_\rmii{$Z$}}

\newcommand{\mpl}{m_\rmii{Pl}}
\newcommand{\Mpl}{M_\rmii{Pl}}

\newcommand{\Tsph}{T_\text{sph}}
\newcommand{\TB}{T_\rmii{$B$}}
\newcommand{\Teq}{T_\text{eq}}
\newcommand{\Tasym}{T_\text{asym}}
\newcommand{\Lb}{L_b}
\newcommand{\Lf}{L_f}

\newcommand{\tr}{\mbox{Tr}}

\newcommand{\msl}[1]{\,\slash\!\!\!{#1}\,}
\newcommand\MSbar{$\overline{\rm MS}$}

\newcommand{\geff}{g_{*}}
\newcommand{\heff}{h_{*}}

\newcommand{\thetaw}{\theta_\rmi{w}}
\newcommand{\thetaE}{\tilde\theta}

\newcommand{\nB}{n_\rmii{B}}

\renewcommand{\nB}{n_\rmii{$B$}}
\newcommand{\NB}{N_\rmii{$B$}}
\newcommand{\nBmL}{n_\rmii{$B$-$L$}}
\newcommand{\GammaB}{\Gamma_{\!\rmii{$B$}}}
\newcommand{\SB}{\mathcal{S}_{\rmii{$B$}}}
\newcommand{\Gammaws}{\Gamma_\text{ws}}

\newcommand{\nBeq}{\nB^\rmi{eq}}

\newcommand{\kaLO}{\kappa^\rmii{LO}}

\newcommand{\nBeqNLOstar}{\nB^{\text{eq},\rmii{NLO$^\star$}}}
\newcommand{\kaNLOstar}{\kappa^\rmii{NLO$^\star$}}

\newcommand{\Fsph}{\mathcal{F}_\text{sph}}
\newcommand{\Csph}{C_\text{sph}}
\newcommand{\Fspheq}{\Fsph^\text{eq}}
\newcommand{\Cspheq}{\Csph^\text{eq}}
\newcommand{\FspheqLO}{\Fsph^{\text{eq},\rmii{LO}}}
\newcommand{\CspheqLO}{\Csph^{\text{eq},\rmii{LO}}}
\newcommand{\CspheqNLOstar}{\Csph^{\rmi{eq},\rmii{NLO$^\star$}}}
\newcommand{\FspheqNLOstar}{\Fsph^{\rmi{eq},\rmii{NLO$^\star$}}}
\newcommand{\sigmaPauli}{\tau}

\newcommand{\sumint}[1]{{\hbox{$\sum$}\!\!\!\!\!\!\!\int\,}_{\!\!\!\!\raise-0.9ex\hbox{$\scriptstyle{#1}$}}}
\newcommand{\Tint}[1]{{\hbox{$\sum$}\!\!\!\!\!\!\!\int\,}_{\!\!\!\!\raise-0.9ex\hbox{$\scriptstyle{#1}$}}}
\newcommand{\Tinti}[1]{{{\Sigma}\!\!\!\!\raise0.3ex\hbox{$\int$}_\rmii{${#1}$}}}
\newcommand{\Tintip}[1]{{{\Sigma'}\!\!\!\!\!\raise0.3ex\hbox{$\int$}_\rmii{${#1}$}}}

\newcommand{\Nsamp}{N_\rmi{sample}}
\newcommand{\phibarMin}{\phibar_\rmi{min}}

\renewcommand{\rmi}[1]{{\mbox{\scriptsize #1}}}
\newcommand{\yt}{h_{t}}

\newcommand{\yeAlpha}{h_{e_\alpha}}
\newcommand{\yeBeta}{h_{e_\beta}}
\newcommand{\CF}{C_\rmii{F}}
\newcommand{\muB}{\mu_{\rmii{$B$}}}
\newcommand{\muBpL}{\mu_{\rmii{$B\!\!+\!\!L$}}}
\newcommand{\muBmL}{\mu_{\rmii{$B\!\!-\!\!L$}}}
\newcommand{\muY}{\mu_\Ysub}
\newcommand{\muA}{\mu_\Asub}

\newcommand{\muLAlpha}{\mu_{\rmii{$L_\alpha$}}}
\newcommand{\n}{n} 
\newcommand{\nal}{n_{\Delta_\alpha}} 
\newcommand{\olsi}[1]{\,\overline{\!{#1}}} % overline short italic
\newcommand{\phibar}{\olsi\varphi} 
\newcommand{\phibarT}{\olsi\varphi_{3}}
\newcommand{\Phibar}{\bar\Phi} 
\newcommand{\Psibar}{\bar\Psi}
\newcommand{\F}{F} 
\newcommand{\A}{A} 
\newcommand{\At}{A}
\newcommand{\B}{B} 
\newcommand{\Bt}{B}
\newcommand{\X}{X} 
\newcommand{\Xt}{X}
\newcommand{\Abg}{\bar\A} 
\newcommand{\Abgt}{\bar{\A}^3} 
\newcommand{\Bbg}{\olsi\B} 
\newcommand{\Xbg}{\olsi\X}
\newcommand{\Atbg}{\bar{\At}}
\newcommand{\Atbgt}{\Atbg^3}
\newcommand{\Btbg}{\olsi{\Bt}}
\newcommand{\Xbgt}{\olsi{\Xt}}

\newcommand{\pMom}{k}
\newcommand{\vol}{\mathcal{V}}
\newcommand{\Ysub}{\rmii{$Y$}}
\newcommand{\Asub}{\rmii{$T_3$}}

\def\TsphNum{133.5}

\def\betaSlope{0.858}
\def\betaIntercept{153.1}
\def\geffFo{103.4}
\def\kappaFo{1.164}
\def\fqcdFoNum{1.058}
\def\TBNum{138.2}

\def\CsphNum{0.3328}
\def\CsphNumErr{0.3328(5)}
\def\CsphErrNum{0.0005}
\def\FsphNum{0.0279}
\def\FsphNumErr{0.0279(19)}
\def\FsphErrNum{0.0019}

\def\CspheqTBNum{0.3328}
\def\FspheqTBNum{0.0279}
\def\CspheqTsphNum{0.3313}
\def\FspheqTsphNum{0.0320}
\def\nBobsNum{8.72}
\def\nBobsErrNum{0.05}

\def\scfc{0.7}  % picture scale factor
\def\phgt{21}   % all picture height 30 * \scfc
\def\pwc{21}    % c   picture width  30 * \scfc
\def\pwcb{31.5} % cb  picture width  45 * \scfc
\def\pwcc{42} % cb  picture width  60 * \scfc
\newcommand{\PIC}[4]{\;\parbox[c]{#2 pt}{\begin{picture}(#2,#3)(0,0)
\SetWidth{1.0}\SetScale{#4} #1 \end{picture}}\;}
\newcommand{\pic}[1]{\PIC{#1}{\pwc}{\phgt}{\scfc}}
\newcommand{\picb}[1]{\PIC{#1}{\pwcb}{\phgt}{\scfc}}
\newcommand{\picc}[1]{\PIC{#1}{\pwcc}{\phgt}{\scfc}}

\def\Lwidth{1}

\newcommand*{\Lge}[1]{\csname Lge\romannumeral#1\endcsname}
\def\Aglx(#1,#2)(#3,#4,#5){\PhotonArc(#1,#2)(#3,#4,#5){\Lwidth}
  {6.283 #3 mul 360 div #4 #5 sub #4 #5 sub mul sqrt mul Ldensity mul}}
\def\Lglx(#1,#2)(#3,#4){\Photon(#1,#2)(#3,#4){\Lwidth}
  {#1 #3 sub #1 #3 sub mul #2 #4 sub #2 #4 sub mul add sqrt Ldensity mul}}
\def\Lgli(#1,#2)(#3,#4){
  \Photon(#1,#2)(#3,#4){\Lwidth}
  {#1 #3 sub #1 #3 sub mul #2 #4 sub #2 #4 sub mul add sqrt Ldensity mul 2 div}}
\def\Lglii(#1,#2)(#3,#4){
  \ZigZag(#1,#2)(#3,#4){\Lwidth}
  {#1 #3 sub #1 #3 sub mul #2 #4 sub #2 #4 sub mul add sqrt Ldensity mul}}
\def\Lgliii(#1,#2)(#3,#4){
  \Gluon(#1,#2)(#3,#4){\Lwidth}
  {#1 #3 sub #1 #3 sub mul #2 #4 sub #2 #4 sub mul add sqrt Ldensity mul}}
\def\Aqu(#1,#2)(#3,#4,#5){\ArrowArc(#1,#2)(#3,#4,#5)}
\def\Aqq(#1,#2)(#3,#4,#5){\CArc(#1,#2)(#3,#4,#5)}
\def\Asai(#1,#2)(#3,#4,#5){\CArc(#1,#2)(#3,#4,#5)}
\def\Lsai(#1,#2)(#3,#4){\Line(#1,#2)(#3,#4)}
\def\Asrii(#1,#2)(#3,#4,#5){\DashArc(#1,#2)(#3,#4,#5){1.5}}
\def\Lsrii(#1,#2)(#3,#4){\DashLine(#1,#2)(#3,#4){1.5}}
\def\Lcsi(#1,#2)(#3,#4){\DashArrowLine(#3,#4)(#1,#2){3}}

\def\fextB(#1,#2){%
  \def\Lgex{#2}%
  \def\Lgez{\Lge3}%
  \def\Lgey{\Lge2}%
  #1(0,15)(7.5,15)
  \ifx\Lgex\Lgez
    #2(45,15)(37.5,15)
  \else
    \ifx\Lgex\Lgey
      #2(45,15)(37.5,15)
    \else
      #2(37.5,15)(45,15)
    \fi
  \fi}
\def\fextT(#1,#2){
  \def\Lgex{#2}%
  \def\Lgez{\Lge3}%
  \def\Lgey{\Lge2}%
  #1(0,0)(22.5,0)
  \ifx\Lgex\Lgez
    #2(45,0)(22.5,0)
  \else
    \ifx\Lgex\Lgey
      #2(45,0)(22.5,0)
    \else
      #2(22.5,0)(45,0)
    \fi
  \fi}

\def\TopoVR(#1){\pic{%
  #1(15,15)(15,0,180)%
  #1(15,15)(15,180,360)%
}}
\def\TopoVRo(#1){\;\pic{
  #1(15,15)(15,-90,270)%
  \GCirc(15,0){2}{0}
  \;}}
\def\ToptVS(#1,#2,#3){\pic{
  #1(15,15)(15,0,180)%
  #2(15,15)(15,180,360)%
  #3(30,15)(0,15)%
}}
\def\ToptVE(#1,#2){\picc{
  #1(15,15)(15,0,360)
  #2(45,15)(15,-180,180)}}
\def\TopoOT(#1,#2){\picb{
  #1(0,15)(7.5,15)
  #2(22.5,15)(15,-180,180)%
  }}
\def\ToptOM(#1,#2,#3,#4,#5){\picb{
  #1(0,15)(7.5,15)
  #2(22.5,15)(15,180,270)%
  #3(22.5,15)(15,-90,90)%
  #4(22.5,15)(15,90,180)%
  #5(22.5,0)(22.5,30)%
  }}
\def\TopoSBo(#1,#2,#3){
  \fextB(#1,#1)
  #2(22.5,15)(15,0,180)%
  #3(22.5,15)(15,180,360)}
\def\TopoSB(#1,#2,#3){\picb{
  \TopoSBo(#1,#2,#3)}}
\def\TopoST(#1,#2){\picb{
  \fextT(#1,#1)
  #2(22.5,15)(15,-90,270)}}
\def\ToptSEo(#1,#2,#3,#4,#5){
  \fextB(#1,#1)
  #2(30,15)(7.5,0,180)
  #3(15,15)(7.5,0,180)
  #4(15,15)(7.5,180,360)%
  #5(30,15)(7.5,180,360)}
\def\ToptSE(#1,#2,#3,#4,#5){\picb{
  \ToptSEo(#1,#2,#3,#4,#5)}}
\def\ToptSSo(#1,#2,#3,#4){
  \TopoSBo(#1,#2,#3)
  #4(37.5,15)(7.5,15)}
\def\ToptSS(#1,#2,#3,#4){\picb{
  \ToptSSo(#1,#2,#3,#4)}}
\def\ToptSBTo(#1,#2,#3,#4,#5){
  \fextB(#1,#1)
  #2(22.5,15)(15,0,90)
  #3(22.5,15)(15,90,180)
  #4(22.5,15)(15,180,360)%
  #5(22.5,37.5)(7.5,-90,270)}
\def\ToptSBT(#1,#2,#3,#4,#5){\picb{
  \ToptSBTo(#1,#2,#3,#4,#5)}}
\def\ToptSTBo(#1,#2,#3,#4,#5){
  \fextT(#1,#1)
  #2(22.5,15)(15,-90,60)
  #3(22.5,15)(15,120,270)%
  #4(22.5,15 60 sin 15 mul add)(7.5,0,180)
  #5(22.5,15 60 sin 15 mul add)(7.5,180,360)}
\def\ToptSTB(#1,#2,#3,#4,#5){\picb{
  \ToptSTBo(#1,#2,#3,#4,#5)}}
\def\ToptSTTo(#1,#2,#3,#4){
  \fextT(#1,#1)
  #2(22.5,15)(15,-90,90)
  #3(22.5,15)(15,90,270)
  #4(22.5,37.5)(7.5,-90,270)}
\def\ToptSTT(#1,#2,#3,#4){\picb{
  \ToptSTTo(#1,#2,#3,#4)}}
\def\TopoVBlr(fex(#1),#2,#3){\picb{
  \def\fex(##1,##2,##3,##4){
    ##1(22.5 30 cos 22.5 mul sub,15 30 sin 22.5 mul add)(10,15)%
    ##2(22.5 30 cos 22.5 mul sub,15 30 sin 22.5 mul sub)(10,15)%
    ##3(22.5 30 cos 22.5 mul add,15 30 sin 22.5 mul add)(35,15)%
    ##4(22.5 30 cos 22.5 mul add,15 30 sin 22.5 mul sub)(35,15)%
  }%
  \fex(#1)
  #2(22.5,15)(12.5,0,180)
  #3(22.5,15)(12.5,180,0)}}
\def\ToptVEo(#1,#2){\picc{%
  #1(15,15)(15,0,360)%
  #2(45,15)(15,-180,180)%
  \GCirc(15,0){2}{0}}}

\makeatletter \@addtoreset{equation}{section} \makeatother
\renewcommand{\theequation}{\arabic{section}.\arabic{equation}}
\makeatletter
\renewcommand\section{\@startsection{section}{1}{\z@}%
  {-5.5ex \@plus -1ex \@minus -.2ex}% bfr-skip
  {2.3ex \@plus.2ex}%
  {\normalfont\large\bfseries}}
\renewcommand\subsection{\@startsection{subsection}{2}{\z@}%
  {-3.25ex\@plus -1ex \@minus -.2ex}%
  {1.5ex \@plus .2ex}%
  {\normalfont\normalsize\bfseries}}
\renewcommand\thesection{\@arabic\c@section}
\renewcommand\thesubsection{\thesection.\@arabic\c@subsection}
\renewcommand{\@seccntformat}[1]{%
  \csname the#1\endcsname.\hspace{1.0em}}
\makeatother

\begin{document}

\flushbottom

\begin{titlepage}

\begin{flushright}
  September 2026 \\
  MS-TP-26-25,
  KEK-TH-2869
\end{flushright}

\bigskip

\begin{centering}

{\Large{\bf
  Baryon number freeze-out
  in the Standard Model, precisely
}}

\vspace{0.5cm}

\DeclareRobustCommand{\fnDiag}{\raisebox{0.50ex}{\scalebox{0.21}{$\ToptVS(\Aqu,\Aqu,\Lglx)$}}}
\renewcommand{\thefootnote}{\fnDiag}
Cristina Benso%
\orcidlink{0000-0003-1922-8534},%
$^{\rm a}$
Dietrich B\"odeker%
\orcidlink{0000-0002-1459-6023},%
$^{\rm b}$
Kohei Kamada%
\orcidlink{0000-0001-5095-5911},%
$^{\rm c,d,e}$
\\
Kyohei Mukaida%
\orcidlink{0000-0003-1951-9497},%
$^{\rm f,g}$
Laura Sagunski%
\orcidlink{0000-0002-3506-3306},%
$^{\rm h}$
Philipp Schicho%
\orcidlink{0000-0001-5869-7611},%
$^{\rm i,}$%
\footnote[1]{Corresponding author: Philipp Schicho,
  \href{mailto:philipp.schicho@unige.ch}{\tt philipp.schicho@unige.ch}}
Kai Schmitz%
\orcidlink{0000-0003-2807-6472}%
\,$^{\rm j,k}$%

\vspace{0.5cm}

$^\rmi{a}$%
{\em
  Institute for Theoretical Particle Physics, KIT, 76128 Karlsruhe, Germany
}

\vspace{0.2cm}

$^\rmi{b}$%
{\em
Fakultät für Physik, Universität Bielefeld, 33501 Bielefeld, Germany
}

\vspace{0.2cm}

$^\rmi{c}$%
{\em School of Fundamental Physics and Mathematical Sciences,\\
%Hangzhou Institute for Advanced Study, University of Chinese Academy of Sciences,
HIAS-UCAS, 310024 Hangzhou, China%
}

\vspace{0.2cm}
 
$^\rmi{d}$%
{\em International Centre for Theoretical Physics Asia-Pacific, Hangzhou/Beijing, China%
}

\vspace{0.2cm}

$^\rmi{e}$%
{\em RESCEU, The University of Tokyo, Bunkyo-ku, Tokyo 113-0033, Japan%
}

\vspace{0.2cm}

$^\rmi{f}$%
{\em
  KEK Theory Center, Tsukuba 305-0801, Japan%
}

\vspace{0.2cm}

$^\rmi{g}$%
{\em
  Graduate University for Advanced Studies (Sokendai), Tsukuba 305-0801, Japan
}

\vspace{0.2cm}

$^\rmi{h}$%
{\em
  Institute for Theoretical Physics, Goethe University, 60438 Frankfurt am Main, Germany
}

\vspace{0.2cm}

$^\rmi{i}$%
{\em
  Département de Physique Théorique, Université de Genève, 1211 Genève 4, Switzerland
}

\vspace{0.2cm}

$^\rmi{j}$%
{\em
  Institute for Theoretical Physics, University of M\"unster, 48149 M\"unster, Germany
}

\vspace{0.2cm}

$^\rmi{k}$%
{\em
  Kavli IPMU (WPI), UTIAS, The University of Tokyo, Kashiwa, Chiba 277-8583, Japan
}

\vspace*{0.6cm}

\mbox{\bf Abstract}

\end{centering}

\vspace*{0.3cm}

\noindent
Weak sphaleron transitions turn a lepton asymmetry of
the Standard Model plasma in the early Universe into
a baryon asymmetry,
conserving baryon-minus-lepton number $B-L$ and
its individual flavored charges.
A baryon asymmetry can thus also arise from
flavored lepton asymmetries with vanishing $B-L$.
Standard equilibrium calculations
in the symmetric and broken phases are performed at constant temperature and hence
neglect the fact that both
the Higgs expectation value and
the sphaleron rate vary as functions of temperature across
the electroweak crossover.
We derive a Boltzmann equation for
the baryon number evolution across the crossover and
calculate the freeze-out abundance
including higher-order corrections to both
the grand canonical partition function and 
the perturbative Higgs expectation value.
This yields two sphaleron conversion factors:
$\Csph = \CsphNumErr$ for $B-L$ and
$\Fsph = \FsphNumErr$ for
the flavored charges
weighted by the charged-lepton Yukawa couplings.

\vfill

\end{titlepage}

%%%%%%%%%%%%%%%%%%%%%%%%%%%%%%%%%%%%%%%%%%%%%%%%%%
{\hypersetup{hidelinks}
\tableofcontents
}
\renewcommand{\thefootnote}{\arabic{footnote}}
\setcounter{footnote}{0}

\clearpage
% %%%%%%%%%%%%%%%%%%%%%%%%%%%%%%%%%%%%%%%%%%%%%%%%%%
\section{Introduction}
\label{sec:intro}

The origin of the observed~\cite{Cooke:2013cba,Planck:2015fie}
baryon asymmetry of the Universe is
one of the central open questions in cosmology.
Any successful explanation of this asymmetry must
ultimately account for the processes that violate baryon number in
the Standard Model~(SM) plasma at high temperatures.
In the electroweak sector of the SM, such violation is mediated by
non-perturbative sphaleron transitions, which efficiently convert
baryon ($B$) and lepton ($L$) number during the electroweak epoch.
While they violate $B$ and $L$ separately,
these transitions exactly conserve the difference, baryon-minus-lepton number $B-L$,
as well as each of the flavored charges of the $\nG$
fermion generations,
$\Delta_\alpha = B/\nG - L_\alpha$.
It is these conserved global charges that fix the baryon number
that remains
until today~\cite{Khlebnikov:1988sr,Barni:2026fvy}.

The quantitative relation between such initial flavored
baryon-minus-lepton numbers and
the baryon number, $B$,
that survives after sphaleron freeze-out is encoded in
the \textit{sphaleron conversion}~\cite{Kuzmin:1987wn,Rubakov:1996vz}.
Since the charged-lepton Yukawa couplings,
$\yeAlpha = \{h_e, h_\mu, h_\tau\}$, are
the only source of flavor dependence in the Standard Model,
the baryon number takes the form
\begin{equation}
\label{eq:Csph:def}
  B =
  \sum_\alpha \Csph^{\alpha} \Delta_\alpha^{ }
  \equiv
  \Csph^{ } (B-L)
  + \Fsph^{ } \sum_\alpha \yeAlpha^2 \Delta_\alpha^{ }
  \,,
\end{equation}
with
$\Csph^{\alpha} =
\Csph^{ } + \yeAlpha^2 \Fsph^{ }
 + \mathcal{O}(\yeAlpha^4)$ and $B-L = \sum_\alpha \Delta_\alpha$.
Throughout, $\Csph$ denotes
the flavor-blind conversion factor and
$\Fsph$ the flavored one.
In the absence of a total $B-L$ asymmetry,
flavor asymmetries still generate 
a non-zero baryon number~\cite{March-Russell:1999hpw,Shu:2006mm,Gu:2010dg,Mukaida:2021sgv}.
Determining both factors
$\Csph$ and $\Fsph$ with high precision
is essential for a broad class of
baryogenesis~\cite{%
Kuzmin:1985mm,Shaposhnikov:1986jp,Shaposhnikov:1987tw,Cohen:1990py,Cohen:1993nk%
  } and
leptogenesis scenarios~\cite{%
  Fukugita:1986hr,%
  March-Russell:1999hpw,Abada:2006ea,AristizabalSierra:2009bh,Dev:2017trv,Mukaida:2021sgv,
  Brune:2022vzd
  };
see also~\cite{Bodeker:2020ghk,vandeVis:2025efm} for recent reviews.

The standard values for
the flavor-blind factor $\Csph$
were obtained in
the instantaneous freeze-out calculations performed
more than three decades ago,
initially limited to
the symmetric (sym) phase at temperatures above the electroweak crossover, and
the broken (bro) phase at temperatures below sphaleron freeze-out~\cite{%
  Khlebnikov:1988sr,Kuzmin:1987wn,Harvey:1990qw,Nelson:1990ir,Dolgov:1991fr,
  Dreiner:1992vm,Davidson:1994gn
  },
\begin{align}
\label{eq:Csph:original}
    \Csph^\text{sym} &= \frac{28}{79} \simeq 0.3544
    \,,&
    \Csph^\text{bro} &= \frac{12}{37} \simeq 0.3243
    \,.
\end{align}
Their analytic form 
at $\mathcal{O}(g^0)$ leading order~(LO) in the SM
was derived in~\cite{Khlebnikov:1996vj} and
later extended to partial
$\mathcal{O}(\yeAlpha^2)$ next-to-leading order~(NLO) in~\cite{Laine:1999wv}.

To account for the actual thermal history of the Universe,
one needs to go beyond
these equilibrium values.
The true sphaleron freeze-out occurs during the electroweak 
crossover, where
the Higgs expectation value 
and
the sphaleron rate evolve continuously with temperature.
The equilibrium values
$\Csph^\text{sym}$ and
$\Csph^\text{bro}$ neglect this continuous evolution and
therefore misestimate the surviving baryon asymmetry
by several percent.
At higher orders, the flavor dependence of
the conversion factor also enters.
A fully time-dependent treatment of sphaleron processes during
the electroweak crossover 
is therefore required.

The goal of this work is to provide the 
\textit{first precision analysis}
of baryon number violation in the SM at the electroweak scale.
This includes
\begin{enumerate}
  \item[(i)]
    Higher-order corrections to
    the equilibrium conversion factors
    $\Csph$, $\Fsph$ and washout coefficients,
    including
    a fully perturbative determination of
    the Higgs expectation value
    at N$^2$LO
    in the broken phase below the electroweak crossover.
  \item[(ii)]
    A self-contained derivation of the baryon-number transport equation
    from first principles using linear-response theory.
  \item[(iii)]
    A precise determination of
    the sphaleron conversion factors
    $\Csph = \CsphNumErr$ and
    $\Fsph = \FsphNumErr$
    by solving the full
    transport equation across the electroweak crossover.
  \item[(iv)]
    A discussion of the flavor dependence of
    the sphaleron conversion, including
    the baryon asymmetry in the absence of
    the initial total $B-L$ asymmetry.
\end{enumerate} 
The present work details
the computation behind the results of (i)--(iv) and
is meant to be self-contained.

The paper is outlined as follows. 
Section~\ref{sec:trans}
provides a self-contained derivation of the transport equation describing
the evolution of the baryon number and its freeze-out at
the electroweak crossover,
within the framework of linear response theory.
Section~\ref{sec:Omega:2loop} details the calculation of
the pressure within
the high-temperature three-dimensional~(3d) effective field theory~(EFT)
and
the subsequent derivation of the
equilibrium baryon number $\nBeq$ and
washout coefficient $\kappa$
that enter the transport equation.
Section~\ref{sec:sphalerons} presents
the baryon asymmetry obtained by
numerically solving the transport equation.
We discuss baryogenesis starting from
both vanishing and non-zero $B-L$ number.
As a complement to the numerical results,
a formal semi-analytical solution for the transport equation is provided
in appendix~\ref{sec:formal:solution}.
Section~\ref{sec:conclusions} presents our conclusions.

%%%%%%%%%%%%%%%%%%%%%%%%%%%%%%%%%%%%%%%%%%%%%%%%%%%%%%%%%%%%%%%%%%%%%%%%%%%%%%%%%%%%%%%%%%%%%%%%%%%%
\section{Transport equation for baryon number}
\label{sec:trans}

Let us start by deriving the transport equation describing baryon number
freeze-out across the electroweak crossover.
To this end,
we define the part of the SM Lagrangian that contains 
the Higgs field as 
\begin{align}
\label{eq:SM:Lagrangian}
\mathcal{L}_\text{Higgs} =  
    \bigl(D_{\mu}\Phi\bigr)^{\dagger}D^\mu\Phi+\nu^2\Phi^\dagger\Phi 
  - \lambda\bigl(\Phi^\dagger\Phi\bigr)^2
  - \Biggl( \yt\, \bar{t}\, \tilde\Phi^\dagger q
     + \sum_{\alpha = e, \mu, \tau}
       h_{e_\alpha} \bar{\ell}_\alpha \Phi\, e_{\alpha}
  + \text{h.c.}
  \Biggr)
  \,,
\end{align}
where
$\yt$ is the top Yukawa coupling and
$h_{e_\alpha}$ are the charged-lepton Yukawa couplings,
with $\alpha \in \{e, \mu, \tau\}$.
The fields $q$ and $\ell_\alpha$ are
the left-handed quark and lepton doublets, respectively, and
$e_\alpha$ are the right-handed charged leptons.
The SM Higgs doublet is denoted by $\Phi$,
$\nu$ is its \MSbar\ mass parameter, and
$\tilde\Phi = i \sigmaPauli^2 \Phi^*$ its conjugate, where
$\sigmaPauli^a$ are the Pauli matrices.
While we neglect light-quark Yukawa couplings, since their impact
on the final baryon asymmetry is tiny,
we include all charged-lepton Yukawa couplings due to
their flavor-dependent contributions.
The values of SM
masses and couplings are taken from
the Particle Data Group~\cite{ParticleDataGroup:2026mpi}.

At the classical level,
both baryon and lepton number are conserved in the SM.
At the quantum level, however, they are violated by
the chiral anomaly~\cite{Adler:1969gk,Bell:1969ts,tHooft:1976rip},
which manifests itself
in the corresponding current equations relevant for
the baryon-number freeze-out~\cite{McLerran:1990de},
\begin{align}
\label{eq:baryon_current}
  \partial_\mu \mathcal{J}_{\rmii{$B$},\rmii{$L$}}^\mu &= \frac{\nG}{32 \pi^2}  
    \Bigl[
      g^2 \F^a_{\mu\nu} \widetilde \F^{a\, \mu\nu}  
    - g'^2 \B_{\mu\nu} \widetilde \B^{\mu\nu} 
    \Bigr]
    = \nG \Bigl[
        \mathcal{O}_{w}
      + \mathcal{O}_\Ysub
    \Bigr]
    \,, \\[2mm]
\label{eq:consv}
  \partial_\mu \mathcal{J}_\Ysub^\mu &=
  \partial_\mu \mathcal{J}_\Asub^\mu =
  \partial_\mu \mathcal{J}_{\Delta_\alpha}^\mu = 0
  \,,
\end{align}
where
$\nG = 3 $ is the number of fermion generations.
The field strengths for the
${\rm U}(1)_\Ysub$ and
${\rm SU}(2)_\Asub$ gauge groups 
are
$\B_{\mu\nu}$ and
$\F^a _{ \mu\nu}$, respectively, and
the dual field strength is
$\widetilde \F^{\mu\nu} \equiv \epsilon^{\mu\nu\rho\sigma} \F_{\rho\sigma}/2$ with
$\epsilon^{0123} = 1$.
The current associated with
$T_3$ of ${\rm SU}(2)_\rmii{$L$}$ is denoted as $\mathcal{J}_\Asub^\mu$,
the hypercharge current is $\mathcal{J}_\Ysub^\mu$, and
the currents of the flavored $B-L$ charges follow from
\begin{equation}
 \Delta_\alpha \equiv \frac{B}{\nG} - L_\alpha
  \,,\quad
  \text{with}\quad
  \alpha = e, \mu, \tau
  \,.
\end{equation}
The currents
$\mathcal{J}_\Asub^\mu$ and
$\mathcal{J}_\Ysub^\mu$
are conserved due to the gauge invariance of the SM.
Since the individual baryon and lepton currents
in eq.~\eqref{eq:baryon_current} are not conserved,
gauge configurations with non-zero topological charge,
such as a change in Chern--Simons number $\Delta N_\rmii{CS}$,
can induce a change in $B$ and $L$.
Another topological configuration is
a hypermagnetic field carrying
non-zero hypermagnetic helicity.
As its helicity decays~\cite{Giovannini:1997eg,Kamada:2016cnb},
such fields can induce a change in $B$ and $L$
through $\mathcal{O}_\Ysub$.

At finite temperature, electroweak sphaleron transitions dynamically realize
$\Delta N_\rmii{CS} \neq 0$, thereby inducing baryon and lepton
number violation in the plasma.
Since the current equation is the same for baryons and leptons,
$B-L$ and
its flavored version $\Delta_\alpha$ are individually conserved in the SM,
while
$B+L$ is not conserved.

%%%%%%%%%%%%%%%%%%%%%%%%%%%%%%%%%%%%%%%%%%%%%%%%%%%%%%%%%%%%%%%%%%%%%%%%%%%%%%%%%%%%%%%%%%%%%%%%%%%%
\subsection{Baryon number in the grand canonical ensemble}
\label{sec:grand:canonical}

While the derivation of the baryon-number transport equation
is scattered across the literature~\cite{Khlebnikov:1988sr,Rubakov:1996vz,Laine:1999wv,Khlebnikov:1996vj},
we provide here a self-contained formulation following~\cite{Domcke:2020kcp}.
In the scenario of interest,
the derivation relies on the two assumptions that
\begin{itemize}
    \item[(i)]
      all SM interactions except weak sphalerons are in equilibrium,
    \item[(ii)]
      deviations from chemical equilibrium are characterized by
      a small $B+L$ chemical potential $\muBpL \ll T$,
      which can be treated using linear response theory.%
      \footnote{%
        General baryogenesis mechanisms may violate assumption (ii).
        For instance, in some spontaneous baryogenesis scenarios operating across the electroweak crossover,
        a non-vanishing velocity of
        an axion-like particle could drive other chemical potentials than $\muBpL$ to be
        non-zero~(see, e.g.,~\cite{Servant:2014bla,Co:2019wyp,Domcke:2020kcp}).
        This would require a more general treatment beyond the scope of this work.
      }
\end{itemize}

The transport equation can be derived from
the current equations~\eqref{eq:baryon_current} and~\eqref{eq:consv} by
taking their expectation values with respect to an appropriate state.
Under the assumptions of
(i) and (ii),
such a state can be well approximated by
the grand canonical ensemble
with the density operator and partition function
\begin{align}
  \label{eq:GC}
  \hat\rho_\rmii{GC} &=
      \frac{1}{\mathcal{Z}_\rmii{GC}}
      e^{ - \beta\left( \hat{H} - \mu_i \hat{Q}_i \right) }
    \,,& 
  \mathcal{Z}_\rmii{GC} \equiv
      \tr\, e^{ - \beta\left( \hat{H} - \mu_i \hat{Q}_i \right) }
   \,,
\end{align}
where
$\beta = 1/T$ and
summation over the repeated index $i$ is implied.
The Hamiltonian is denoted as $\hat{H} = \int_\vec{x} \mathcal{H}$
with
$\int_\vec{x} \equiv \int \mathrm{d}^d x$
where $d = 3 - 2 \epsilon$
is the spatial dimension.
The subscript runs through 
$i = \{ \Delta_\alpha, B+L \}$;
$\mu_i$ and $\hat{Q}_i$
correspond to the chemical potentials and
associated conserved charges that commute with each other and with
the Hamiltonian.
The charges are defined by
$\hat{Q}_i = \int_\vec{x}\, \mathcal{J}_i^0$ with 
$\mathcal{J}_i^0$ being the time component of the four-vector current (i.e., the charge density).
For later convenience, we also define the chemical potentials for the baryon and lepton charges as 
\begin{align}
  \muB &= \muBpL + \sum_\alpha ^{\nG} \frac{\mu_{\Delta_\alpha}}{\nG}
  \,,&
  \muLAlpha &= \muBpL - \mu_{\Delta_\alpha}
  \,,
\end{align}
which are defined such that
$
  \muB Q_\rmii{$B$}
+ \muLAlpha Q_\rmii{$L_\alpha$}
= \muBpL Q_\rmii{$B+L$}
+ \mu_{\Delta_\alpha} Q_{\Delta_\alpha}$.
The grand canonical potential 
\begin{align}
  \label{eq:grand:canonical:potential}
  \Omega  \equiv - T \ln \mathcal{Z}_\rmii{GC} 
\end{align} 
is related to the pressure through
\begin{align}
   \label{pressure} 
    p (\mu, T) =  - \frac{\Omega (\mu, T)}{\vol}
    \,,
\end{align} 
where $\vol$ is the $\mathbb{R}^3$ spatial volume.
The expectation value of the charges $Q_i$
gives rise to the corresponding number densities
\begin{equation}
  \label{eq:number:density}
  \n_i = 
  \frac{\langle Q_i \rangle_\rmii{GC}}{\vol} 
  = \frac{\partial p}{\partial \mu_i}
  \,,
\end{equation}
where
for a generic operator $\mathcal{O}$, we have defined
$\langle \mathcal{O} \rangle_\rmii{GC} \equiv \tr [\hat\rho_\rmii{GC} \mathcal{O}]$.
Since the grand canonical ensemble is translation-invariant,
$[\mathbf P,\hat{H}]=[\mathbf P,\hat{Q}_i]=0$,
the local charge-density operator
$\mathcal{J}^0_i(t,\vec{x})$
has a spatially uniform expectation value.
Thus, the volume-averaged expectation value of
the charge density depends only on time,
\begin{equation}
  \n_i = \frac{1}{\vol}\int_\vec{x}
  \bigl\langle \mathcal{J}^0_i (t,\vec{x}) \bigr\rangle_\rmii{GC} =
  \bigl\langle \mathcal{J}^0_i (t,\vec{0}) \bigr\rangle_\rmii{GC}
  \,.
\end{equation}

The partition function can be conveniently computed as a path integral.
Here, one has to pay particular attention to the zero-momentum modes
of the Higgs and the temporal components of the 
gauge fields, as these can obtain non-vanishing values.
One first integrates over fields with non-zero momentum and denotes the
result by
$\exp(-\widetilde\Omega/T)$,
where
$\widetilde\Omega$ now depends on the zero-momentum modes
of the electroweak gauge fields  
$\Abg_{0}^{a}$,
$\Bbg_{0}$, and
of the Higgs field $\Phibar$.
In a second step, the partition function is then obtained by
integrating over
these bosonic zero-momentum modes,
\begin{align} 
\label{zmi} 
     \exp ( -  \Omega/T )
   =
   \int 
     {\rm d} \Abg _ 0 ^ a \,
     {\rm d} \Bbg _ 0 \, 
     {\rm d} \Phibar   \,
     \exp \Bigl(
      -  \widetilde \Omega ( \Abg  _ 0 ^ a, \Bbg  _ 0 , \Phibar    ) /T
    \Bigr)
    \,,
\end{align} 
which is manifestly gauge-invariant.
In the infinite-volume limit,
the integral is dominated by its saddle point,
\begin{align}
   \frac { \partial \widetilde \Omega  } { \partial \Abg  _ 0 ^ a }
   =
   \frac { \partial \widetilde \Omega  } { \partial \Bbg  _ 0 }
   =
   \frac { \partial \widetilde \Omega  } { \partial \Phibar  }
   = 0
   \label{sp} 
   \,,
\end{align} 
so that
\begin{align}
  \exp (  - { \Omega   }/T )
  = 
   \exp \!
   \left ( - \widetilde{  \Omega  } (  {\rm saddle \, point })
     /T \right )
   \label{spa}
   \,.
\end{align}

The first two conditions in eq.~\eqref{sp} enforce gauge-charge neutrality.
We are free to choose
\begin{align}
  \label{phibar} 
  \Phibar  = \frac 1 { \sqrt{ 2 } } 
  \biggl( 
   \begin{array}{c} 
        0 \\ \phibar   
        \end{array} 
  \biggr) 
  \,,
\end{align} 
with $\phibar > 0$,
so that
only the third isospin direction is singled out.
The temporal zero modes
$\Bbg_0$ and
$\Abgt_0$ then act as
Lagrange multipliers for the 
weak hypercharge and the third component of the weak isospin,
and can be identified with the
corresponding
chemical potentials%
\footnote{%
  The chemical potentials are real and enter
  the Euclidean action through
  eq.~\eqref{eq:GC} as $- \mu_X^{ } \mathcal{J}^0_X$,
  whereas
  temporal gauge fields $X_0 = \{B_0, A_0\}$ couple to
  such a charge density with
  $i g_X \overline{X}^{ }_0 \mathcal{J}^0_X$.
  Therefore, only combinations of
  $g_X \overline{X}_0 + i \mu_X$ appear~\cite{Gynther:2003za}, and
  a real chemical potential for a gauge charge is nothing but a
  purely imaginary temporal zero mode~\cite{Bodeker:2015zda};
  cf.\ eq.~\eqref{eq:muY:muA:4d}.
}
\begin{align}
\label{eq:muY:muA:4d}
  g_{1}\Bbg^{ }_0
  &= i \muY
  \,, &
  g_{2}\Abgt_0
  &= i \muA
   \,.
\end{align}
Stationarity with respect to these variables therefore imposes vanishing
hypercharge and weak isospin densities in equilibrium.
The third condition in eq.~\eqref{sp} corresponds to minimization of the
finite-temperature Higgs effective potential.
We also 
introduce the zero-mode-dependent pressure,
\begin{align}
  \label{eq:p-func:tilde}
   p ( \mu, T, \Psibar )
   &\equiv
   - \frac{\widetilde \Omega ( \mu, T, \Psibar ) }{\vol}
   \,, &
   \Psibar
     =\{\Bbg_{0}^{ }, \Abgt_0, \phibar\}
   \,,
\end{align}
which is distinguished from
the pressure~\eqref{pressure} by its arguments.

The conservation laws~\eqref{eq:consv} provide additional constraints among
the chemical potentials.
By taking the expectation value of
eq.~\eqref{eq:consv} with respect
to the grand canonical ensemble defined in eq.~\eqref{eq:GC},
we obtain $\nG$ conserved charge densities together with
the two neutrality conditions~\eqref{sp}
for the gauge charges,
\begin{align}
  \label{eq:consv_relation}
  \n_{\Delta_\alpha} &= \frac{\partial p}{\partial \mu_{\Delta_\alpha}}
  \,,&
  0 &= \frac{\partial p}{\partial \muY} = \frac{\partial p}{\partial \muA}
  \,.
\end{align}
Thus, gauge-charge neutrality is enforced dynamically through the saddle-point
conditions, while the conserved flavored $B-L$ charges
$\Delta_\alpha$ remain free parameters.
Their values are set by the primordial dynamics, for instance, by
thermal leptogenesis~\cite{Abada:2006ea,AristizabalSierra:2009bh,Dev:2017trv,Mukaida:2021sgv}.

To resolve the full infrared behavior
of the thermodynamic function $p$ across the electroweak crossover,
an effective 3d description~\cite{Ginsparg:1980ef,Appelquist:1981vg,Braaten:1995cm,Kajantie:1995dw} combined with
non-perturbative lattice simulation~\cite{Kajantie:1995kf,DOnofrio:2015gop} is required.
However, since the sphaleron freeze-out temperature $\Tsph$ is sufficiently
far below the pseudocritical temperature $\Tc$,
perturbative methods are applicable to correctly describe $p$
in the infrared~\cite{Gould:2023ovu}.
Under the assumption~(ii),
we compute $p$ in
two perturbative steps.
First, we determine the temperature evolution of the order parameter
at vanishing chemical potentials~\cite{DOnofrio:2015gop}.
Second, we compute the effect of chemical potentials on $p$~\cite{Gynther:2003za}
while treating the order parameter as fixed.
In this approximation, the backreaction of chemical potentials on 
the order-parameter evolution is neglected,
following~\cite{Kamada:2016eeb}.

The temperature evolution of the order parameter,
$\langle \Phi^\dagger \Phi \rangle$,
can be either computed by
lattice simulations~\cite{DOnofrio:2015gop,Kamada:2016eeb} or
perturbatively
at vanishing chemical potentials.
Here, we follow a perturbative approach and compute the Higgs expectation value
$\phibarMin $ directly from the $\mu_i = 0$ effective potential as documented
in appendix~\ref{sec:Tevolution:vev}.
Subsequently,
by keeping the
Higgs zero mode
fixed,
one may compute the effect of chemical potentials
perturbatively~\cite{Laine:1999wv}.
To this end, we evaluate
the zero-mode-dependent pressure~\eqref{eq:p-func:tilde}
using eq.~\eqref{eq:muY:muA:4d}.
For the ideal gas plus the $\mathcal{O}(\yeAlpha^2)$ terms of
ref.~\cite{Laine:1999wv}, it
reads 
\begin{eqnarray}
  \label{eq:p-func}
    p (\mu, T, \phibar) &=&
    p ( 0, T, \phibar )
  + \frac{\nG}{9} \muB^{2} T^{2}
  + \frac{1}{4} \sum_{\alpha} \muLAlpha^{2} T^{2}
  + \frac{1}{3}\Bigl(\frac{\nG}{3}\muB^{ }-\sum_{\alpha} \muLAlpha^{ } \Bigr) \muY^{ } T^{2}
  \nn &&
  + \frac{1}{2}\frac{{\mDi{1}}^2}{g_{1}^2}\muY^{2}
  + \frac{1}{2}\frac{{\mDi{2}}^2}{g_{2}^2}\muA^{2}
  + \frac{1}{8} \bigl(\muY^{ } - \muA^{ }\bigr)^{2} \phibar^{2}
  \nn &&
  + \frac{1}{(4\pi)^2} \biggl[
      \muA^{ } \phibar^2 \sum_\alpha \yeAlpha^2 \muLAlpha^{ }
    - \Bigl( \phibar^2 + \frac{T^2}{2} \Bigr) \sum_\alpha \yeAlpha^2
      \bigl( - 3 \muY^{ } \muLAlpha^{ } + 2 \muLAlpha^2 \bigr)
  \biggr]
  \nn &
  \stackrel{\eqref{eq:muY:muA:4d}}{=}&
    p(\mu, T, \Psibar)
  \,.
\end{eqnarray}
Since the last bracket covers only the charged-lepton Yukawa part of
that order, eq.~\eqref{eq:p-func} is \emph{not} complete at
$\mathcal{O}(g^2\mu^2)$; we refer to this approximation as NLO$^\star$
in the following.
The complete result,
together with the $\mathcal{O}(g\mu^2)$
soft one-loop terms, is derived in
sec.~\ref{sec:Omega:2loop} and given in
sec.~\ref{sec:nb:higher:equilibrium}.
In eq.~\eqref{eq:p-func},
$\mDi{1}^2 = g_1^2 T^2(\frac16 + \frac{5\nG}{9})$ and
$\mDi{2}^2 = g_2^2 T^2(\frac56 + \frac{\nG}{3})$
are the leading-order Debye masses of the temporal components of the
${\rm U}(1)_\Ysub$ and
${\rm SU}(2)_\rmii{$L$}$ gauge fields, respectively.
The neutral component of the Higgs zero mode is denoted by
$\phibar$ as defined in eq.~\eqref{phibar}.
The signs of the terms in eq.~\eqref{eq:p-func} that are odd in $\muY$
or $\muA$ follow from the convention in eq.~\eqref{eq:muY:muA:4d},
but have no physical consequence because both chemical potentials are
eliminated by the neutrality conditions~\eqref{eq:consv_relation}.%
\footnote{%
  For a field with charges $T^3$ and $Y$,
  the Euclidean action contains
  $\mu_\text{eff}=T^3\muA+\frac{Y}{2}\muY$.
  The neutral Higgs background
  with $T^3 = -1/2$, $Y = 1$
  contributes $p \supset \frac{1}{8}(\muY-\muA)^2\phibar^2$.
  The $\muY$ sign follows from the hard-mode hypercharge density;
  the $\muA$ sign follows from the $\phibar^2$-background term.
}
Contributions from
charged-lepton Yukawa couplings
provide an additional source of the baryon asymmetry
induced by lepton flavor asymmetries.

By solving the five
constraints~\eqref{eq:consv_relation}
using eq.~\eqref{eq:p-func},
one may express
$(\muY, \muA, \mu_{\Delta_\alpha})$ as functions of
$(\muBpL$, $\n_{\Delta_\alpha}$, $\phibar/T)$.
Differentiating eq.~\eqref{eq:p-func} with respect to $\muB$,
we obtain the baryon number density as
\begin{align}
  \label{eq:q_B-def}
  \nB = \frac{\partial p}{\partial \muB} = 
  \frac{\nG}{9}\Bigl(
      2\muBpL
    + 2\sum_\alpha \frac{\mu_{\Delta_\alpha}}{\nG}
    + \muY
  \Bigr) T^2
  \,.
\end{align}
Substituting the solution of
eqs.~\eqref{eq:consv_relation} and \eqref{eq:p-func} into
eq.~\eqref{eq:q_B-def}, one may express
$\nB = \nB(\muBpL,\nal,\phibar/T)$.

%%%%%%%%%%%%%%%%%%%%%%%%%%%%%%%%%%%%%%%%%%%%%%%%%%%%%%%%%%%%%%%%%%%%%%%%%%%%%%%%%%%%%%%%%%%%%%%%%%%%
\subsection{Sphaleron conversion in equilibrium}
\label{sec:Csph:equilibrium}

Together with the sphaleron equilibrium condition
for a general number of generations~\cite{Khlebnikov:1996vj,Laine:1999wv},
\begin{align}
\label{eq:eq:cond}
  \nG\muB + \sum_{\alpha}^{\nG} \muLAlpha = 0
  \,,
\end{align}
we can obtain the equilibrium baryon number density $\nBeq$ as a function of
the flavored $B-L$ charge densities $\nal$,
the total $B-L$ charge density $\sum_\alpha \nal = \nBmL$,
and
the order parameter $x = \phibar/T$.
It can be defined by the two coefficients anticipated in
eq.~\eqref{eq:Csph:def},
\begin{align}
\label{eq:Csph:eq:split}
  \nBeq(x) &=
    \Cspheq(x)\, \nBmL^{ }
  + \Fspheq(x) \sum_\alpha \yeAlpha^2 \nal^{ }
  \,,
\end{align}
the flavor-blind $\Cspheq(x)$ and
the flavored $\Fspheq(x)$.
Both are computed to N$^2$LO in
sec.~\ref{sec:nb:higher:equilibrium}.

At LO in couplings,
we reproduce~\cite{Khlebnikov:1996vj}
\begin{align}
  \label{eq:nBeq:LO}
  \CspheqLO(x) &=
  \frac{
    4 (4\nG^2 + 12\nG + 5 + 3(2\nG + 3) x^2)}{
    44\nG^2 + 136\nG + 65 + (72\nG + 117) x^2}
  + \mathcal{O}(g^2)
  \,,\nn[1mm]
  \FspheqLO(x) &=
  \mathcal{O}(\yeAlpha^0)
  \,.
\end{align}
The standard equilibrium values for the sphaleron conversion factor are
$\Csph^\text{sym}$ in the symmetric phase and
$\Csph^\text{bro}$ in the broken phase.
They are given in eq.~\eqref{eq:Csph:original} and
obtained in the limits
\begin{align}
  \Csph^\text{sym} &= \lim_{x \to 0} \CspheqLO(x) = \frac{28}{79}
  \,,&
  \Csph^\text{bro} &= \lim_{x \to \infty} \CspheqLO(x) = \frac{12}{37}
   \,.
\end{align}

At NLO in the charged-lepton Yukawa couplings,
the equilibrium baryon number density receives additional contributions.
For $\nG = 3$ generations,
we obtain
the NLO$^\star$ result
at partial $\mathcal{O}(g^2)$
(cf.\ eq.~\eqref{eq:power:counting:couplings})
\begin{align}
  \label{eq:nBeq:NLOstar}
  \CspheqNLOstar(x) &=
      \frac{4 (77 + 27 x^2)}{869 + 333 x^2}
    - \frac{ 126082 + 328295 x^2 + 182277 x^4 + 30456 x^6 }{12 \pi^2 (869 + 333 x^2)^2 }
    \sum_\beta \yeBeta^2
  \,,
  \nn[2mm]
  \FspheqNLOstar(x) &=
      \frac{1034 + 2473 x^2 + 792 x^4}{(4\pi)^2 (869 + 333x^2)}
  + \mathcal{O}(\yeAlpha^2)
  \,.
\end{align}

At NLO$^\star$,
both Yukawa-suppressed structures in $\nBeq$
enter at $\mathcal{O}(\yeAlpha^2)$,
but they multiply different charge combinations.
While
the term
$\propto \FspheqNLOstar$
is obtained in~\cite{Laine:1999wv},
the term
$\propto \bigl( \sum_\beta \yeBeta^2 \bigr) \nBmL^{ }$
is flavor-blind and
was therefore not accounted for in~\cite{Laine:1999wv}.
At large field values, both
Yukawa structures grow
with the charged-lepton masses as
$\yeAlpha^2 x^2 \sim m_\alpha^2/T^2$.
Nevertheless, higher-order terms remain suppressed by
$\mathcal{O}(m_\alpha^4/T^4) \lesssim 10^{-6}$.
However,
a restriction of eq.~\eqref{eq:nBeq:NLOstar}
to moderate field values,
$x = \phibar/T \lesssim 3$,
which covers the freeze-out region,
arises from truncating
higher-dimensional operators in the 3d~EFT;
see sec.~\ref{sec:EFT:validity}.

Even when the total $B-L$ charge vanishes,
$\nBmL = 0$,
the individual flavored charges $\nal$ need not vanish.
Flavor-dependent effects can then generate a non-zero equilibrium
baryon number $\nBeqNLOstar$,
as seen from the second line of
eq.~\eqref{eq:nBeq:NLOstar}.
These effects are further discussed in sec.~\ref{sec:flavor:effects:B-L:zero}.

%%%%%%%%%%%%%%%%%%%%%%%%%%%%%%%%%%%%%%%%%%%%%%%%%%%%%%%%%%%%%%%%%%%%%%%%%%%%%%%%%%%%%%%%%%%%%%%%%%%%
\subsection{Sphaleron conversion across the electroweak crossover}
\label{sec:Csph:crossover}

By lifting the equilibrium condition~\eqref{eq:eq:cond} and
reusing the equilibrium distribution~\eqref{eq:nBeq:NLOstar},
one can recast eq.~\eqref{eq:q_B-def} into
\begin{equation}
\label{eq:3muB+summuL}
  \nG^2\, \muBpL^{ } T^2
  =
  \kappa(x) \bigl[ \nB^{ }  - \nBeq(x) \bigr]
  \,,
\end{equation}
where $\kappa$ will be identified as
the baryon-number washout coefficient.
At LO and for general $\nG$,
it reads~\cite{%
  Khlebnikov:1988sr,Khlebnikov:1996vj,Rubakov:1996vz,
  Burnier:2005hp}%
\footnote{%
  By approximating
  $\kappa \simeq 13\nG/4$~\cite{Bochkarev:1987wf},
  as typically done in the literature,
  one
  assumes the Yukawa interactions with the Higgs particles to be faster than
  the sphaleron transitions~\cite{Rubakov:1996vz}.
}
\begin{align}
  \label{eq:kappa:LO}
  \kaLO(x) &=
    \frac{
      3\nG \bigl[ 44\nG^2 + 136\nG + 65 + (72\nG + 117) x^2 \bigr]
    }{
      2 \bigl[ 20\nG^2 + 62\nG + 30 + (33\nG + 54) x^2 \bigr]
    }
  \,.
\end{align}
Specializing to $\nG = 3$ and
including the charged-lepton Yukawa couplings,
we find the partial
$\mathcal{O}(g^2)$ result of NLO$^\star$
\begin{align}
  \label{eq:kappa:NLOstar}
  \kaNLOstar(x) &=
    \frac{869 + 333 x^2}{88 + 34 x^2}
  + \frac{ (77 + 27 x^2) (418 + 1025 x^2 + 360 x^4)}{72 (44 + 17 x^2)^2 \pi^2} \sum_\alpha \yeAlpha^2
  + \mathcal{O}(\yeAlpha^4)
  \,,
\end{align}
where the $\mathcal{O}(\yeAlpha^2)$ term
agrees with~\cite{Laine:1999wv}.

A dynamical equation for the baryon number
can be derived by taking
the expectation value of eq.~\eqref{eq:baryon_current},
which leads to
\begin{align}
  \label{eq:transport_temp}
  \partial_t \nB = 
  \partial_t \bigl\langle \mathcal{J}_\rmii{$B$}^0(t,\vec{0})\bigr\rangle_\rmii{GC} 
  &=
  \frac{1}{\vol}\int_\vec{x} \big\langle \partial_\mu \mathcal{J}_\rmii{$B$}^\mu (t,\vec{x}) \bigl\rangle_\rmii{GC}
  =
  \frac{\nG}{\vol}\int_\vec{x} \bigl\langle \mathcal{O}_{w}^{ }(t,\vec{x}) \bigr\rangle_\rmii{GC}
  \,.
\end{align}
In the second equality,
the surface integral
$\int_\vec{x} \partial \cdot \langle \mathcal{J}_\rmii{$B$} \rangle_\rmii{GC} = 0$, and
in the last step
we assumed that there is no background of
helical hypermagnetic fields,
so that the hypercharge contribution
$\langle \mathcal{O}_\Ysub \rangle_\rmii{GC}
  \propto
  \langle \B_{\mu\nu} \widetilde \B^{\mu\nu} \rangle_\rmii{GC}$
vanishes.
Loosening this assumption reinstates it as
a source term in the evolution of $\nB$.
We will return to this case in sec.~\ref{sec:conclusions}.

By following assumption~(ii),
one may compute the linear response of the system to a small perturbation
$\mu_i / T \ll 1$ and expand the right-hand side of
eq.~\eqref{eq:transport_temp}
to linear order
in chemical potential
\begin{equation}
\label{eq:ws_kubo_formula}
  \frac{\nG}{\vol}\int_\vec{x} \bigl\langle \mathcal{O}_w(t,\vec{x}) \bigr\rangle_\rmii{GC}
  = - \nG^2\, \Gammaws \frac{\muBpL}{T}
  \,.
\end{equation}
The transport coefficient $\Gammaws$ is defined as
the weak sphaleron diffusion rate induced by
the $\mathrm{SU}(2)_\rmii{$L$}$ gauge fields.%
\footnote{%
  One can also define the linear-response coefficient of interaction processes
  to a chemical potential as $\Gamma_\mu$.
  The Chern--Simons diffusion rate differs by a factor of 2,
  $\Gammaws = 2\Gamma_{\mu}$, because the fluctuation-dissipation relation averages
  over forward and backward sphaleron transitions~\cite{McLerran:1990de,Moore:1996qs}.
}
After having expanded in $\mu_i / T$,
transport coefficients such as
$\Gammaws$ are independent of the chemical potentials, and
expectation values can instead
be taken with respect to the canonical ensemble (C), where
$\hat\rho_\rmii{C} = e^{-\beta \hat{H}}/\mathcal{Z}_\rmii{C}$.
We further express $\Gammaws$ in terms of
the spectral function~\cite{Bodeker:2014hqa}
\begin{align}
\label{eq:transport_coef}
  \Gammaws &\equiv
    \lim_{\omega \to +0}\frac{T G_\rho^\text{ws} (\omega, \bm{0})}{\omega}
  \,, \nn[1mm]
  G_\rho^\text{ws} (\omega, \vec{\pMom}) &\equiv
  \int_\mathcal{X}
  e^{i \mathcal{K}\cdot \mathcal{X}}
  G_\rho^\text{ws} (\mathcal{X})
  \,, & 
  G_\rho^\text{ws} (\mathcal{X}-\mathcal{Y}) &\equiv
  \bigl\langle \bigl[
      \mathcal{O}_w(\mathcal{X}),  
      \mathcal{O}_w(\mathcal{Y})
    \bigr] \bigr\rangle_\rmii{C}
  \,,
\end{align}
where
$\mathcal{X} \equiv (x^0,\vec{x})$ and
$\int_\mathcal{X} \equiv \int {\rm d}x^0 \int_\vec{x}$,
using the mostly-minus metric
$\mathcal{K}\cdot \mathcal{X} = \omega x^0 - \vec{\pMom}\cdot \vec{x}$.
This transport coefficient $\Gammaws$ is also known as
the Chern--Simons or sphaleron diffusion rate,
which can be computed with
non-perturbative methods~\cite{Ambjorn:1990pu,Arnold:1996dy}.
While a previous fitting formula was based on
the lattice data~\cite{DOnofrio:2014rug},
we will adopt a more recent result for the sphaleron diffusion
rate~\cite{Annala:2023jvr}
\begin{equation}
\label{fitnew} 
  \frac{\Gammaws}{T^4} =
  \begin{cases}
    (6.23 \pm 0.05) \times 10^{-7} & \text{for} \quad T \gtrsim \Tc
    \,, \\
    \exp \left[-(\betaIntercept \pm 0.9)+(\betaSlope \pm 0.01) \frac{T}{\mathrm{GeV}}\right] &\text{for} \quad T \lesssim \Tc
    \,,
    \end{cases}
\end{equation}
and henceforth employ
$\Tc = 159.6 \pm 1.5$~GeV as the pseudocritical temperature~\cite{Laine:1998jb,DOnofrio:2015gop,Laine:2015kra}, and
$\Tsph = \TsphNum \pm 0.97$~GeV as the sphaleron freeze-out temperature~\cite{Annala:2023jvr}.
A perturbative computation
of $\Gammaws$ was performed in~\cite{Burnier:2005hp},
in good agreement with the lattice results in the broken phase.
More recently, QCD corrections to the weak-isospin conductivity $\sigma$,
which enters the computation of $\Gammaws$,
were computed in~\cite{Bodeker:2025ahg}.
These increase the rate in eq.~\eqref{fitnew} by about
$\Delta\sigma/\sigma \sim 6\%$,
which we will take into account in our numerical analysis.%
\footnote{%
  Strictly speaking, the equation of motion
  used to compute the rate in eq.~\eqref{fitnew}
  is only valid at small Higgs values.
  The same applies to the computation of
  the QCD corrections in~\cite{Bodeker:2025ahg}.
}

In the broken phase for $T \lesssim \Tc$,
the uncertainties in the fitted parameters of the pure exponential function are
characterized by
the best-fit values of the exponential parameters (the intercept and slope) and
their covariance matrix as follows~\cite{Annala:2023jvr}:
\begin{align}
\label{eq:Gamma:newfit:params}
  \mu &= \begin{pmatrix}
      \hphantom{-}\betaSlope \\
      -\betaIntercept \end{pmatrix}
  \,,&
  \text{Cov} &= \begin{pmatrix}
    \hphantom{-}3.72 \times 10^{-5} & -5.54 \times 10^{-3} \\
    -5.54 \times 10^{-3} & \hphantom{-}8.26 \times 10^{-1}
  \end{pmatrix}\,.
\end{align}
The diagonal elements of the covariance matrix provide the variances,
with standard deviations of
$\sigma_{\text{slope}} \approx 0.0061$ and
$\sigma_{\text{intercept}} \approx 0.91$,
indicating the precision of the fit.

Combining the results obtained so far
[eqs.~\eqref{eq:3muB+summuL}, \eqref{eq:transport_temp}, and \eqref{eq:ws_kubo_formula}],
we arrive at the transport equation for the baryon freeze-out
\begin{align}
\label{eq:baryon:freeze:out}
    \partial_t^{ } \nB^{ }
  + 3 H \nB^{ } &=
    \SB^{ }
  - \GammaB^{ } (\phibar/T) \bigl[ \nB^{ } - \nBeq (\phibar/T) \bigr]
  \,,&
  \GammaB (\phibar/T) &\equiv \kappa (\phibar/T) \frac{\Gammaws(T)}{T^3}
  \,.
\end{align}
Here,
we replaced $\partial_t \to \partial_t + 3H$ in eq.~\eqref{eq:transport_temp}
as the covariant time derivative in an expanding background.
In the absence of helical magnetic fields
or other sources of the $B+L$ charge,
we set the source term
$\SB = 0$.
For the opposite case, $\SB \neq 0$,
see ref.~\cite{Fujita:2016igl}.
The Hubble rate follows from the Friedmann equation
$H^2 = \frac{8\pi G}{3} e(T)$.
Since the sphaleron freeze-out takes place at
$\Tsph \gg \Teq \simeq 1$~eV,
far above the temperature scale of
matter--radiation equality,
the energy budget is entirely dominated by
the relativistic degrees of freedom of the SM plasma
\begin{align}
\label{eq:Hubble:rad}
  H^2(T) &= \frac{8\pi}{3}\frac{e(T)}{\mpl^2}
  \,,&
  e(T) &= \frac{\pi^2}{30} \geff T^4
  \,,&
  \Mpl &= \frac{\mpl}{\sqrt{8\pi}}
  \,,
\end{align}
where
$\geff$ is the effective number of relativistic degrees of freedom,
for which we employ the tabulations of~\cite{eos15}, and
$\mpl = G^{-1/2} = 1.22 \times 10^{19}$~GeV and 
$\Mpl$ are the Planck mass and reduced Planck mass, respectively.
The temperature dependence of $\phibar/T$ is calculated
fully perturbatively as detailed in appendix~\ref{sec:Tevolution:vev}.%
\footnote{%
  Previous studies~\cite{Kamada:2016eeb} utilized
  the fitted $\langle \Phi^\dagger \Phi \rangle$ correlator from
  lattice simulations~\cite{DOnofrio:2015gop}.
}

In appendix~\ref{sec:formal:solution},
we solve eq.~\eqref{eq:baryon:freeze:out} semi-analytically 
for the comoving baryon number density $\NB^{ } = a^3 \nB^{ }$.
Since $\GammaB/H$ decreases exponentially across the crossover,
the washout stops within a narrow range of temperatures,
such that
the equilibrium input~\eqref{eq:nBeq:NLOstar} and~\eqref{eq:kappa:NLOstar}
enters only at the baryon-number freeze-out temperature
$\TB \simeq \TBNum$~GeV
obtained from eq.~\eqref{eq:TB},
\begin{equation}
\label{eq:NB:freezeout}
  \NB^{ }(\Tasym) \simeq \NB^{\rm eq}(\TB)
  \,,
\end{equation}
where
$\Tasym < \Tsph$ is
an asymptotically low temperature at which
the washout has ended and
$\NB^{ }$ has become constant.
All results in this work
can thus also be obtained
by reading off the equilibrium densities at
$\TB$, cf.\ eq.~\eqref{eq:Csph:eq:values:B},
since the conversion factors~\eqref{eq:Csph:def} are
ratios of number densities
and hence independent of the scale factor $a$.

%%%%%%%%%%%%%%%%%%%%%%%%%%%%%%%%%%%%%%%%%%%%%%%%%%%%%%%%%%%%%%
\section{Perturbative computation of the grand canonical partition function}
\label{sec:Omega:2loop}

In this section,
we shall compute the zero-mode-dependent pressure
$p(\mu,T,\Psibar)$ that enters
the conserved charge densities~\eqref{eq:consv_relation} and
the baryon number density~\eqref{eq:q_B-def}.
Following eqs.~\eqref{eq:grand:canonical:potential} and~\eqref{pressure},
the pressure is obtained from the grand canonical potential, which after
the saddle-point evaluation~\eqref{spa} of the zero-mode integral is
given by
$\widetilde\Omega$ at the stationary point~\eqref{sp}.
The zero-mode-dependent pressure~\eqref{eq:p-func:tilde}
splits into
\begin{align}
\label{eq:p-func:Omega}
  p (\mu, T, \phibar) &\stackrel{\eqref{eq:muY:muA:4d}}{=}
    p_0(\mu, T)
  - \Veff(\mu, T, \Psibar)
  \,,
\end{align}
where
$\muB^{ },
\muLAlpha^{ },
\muY^{ },
\muA^{ } \sim \mu$.
We used
eq.~\eqref{eq:muY:muA:4d}
to express $\Psibar = \{\Bbg_0^{ },\Abgt_0,\phibar\} \to \{\muY,\muA,\phibar\}$.
The field-independent part of the pressure, or unit operator,
is $p_0$~\cite{Braaten:1995cm}.
All number densities follow from eq.~\eqref{eq:consv_relation} as
derivatives of this pressure,
such that every term computed below
contributes directly to $\nB^{ }$ and $\kappa$.
The Higgs zero mode $\phibar$ is carried along as a variable and set to
the minimum of the effective potential,
$\phibar = \phibarMin$,
at the end; it is determined in appendix~\ref{sec:Tevolution:vev}.

At high temperature and weak coupling,
thermal fluctuations separate into
three parametrically separated momentum scales,
\begin{align}
  \label{eq:power:counting:scales}
  \pMom_\text{hard}^{ } &\sim \pi T
  \,, &
  \pMom_\text{soft}^{ } &\sim g T
  \,, &
  \pMom_\text{ultrasoft}^{ } &\sim g^2 T
  \,.
\end{align}
Hard modes with
$\pMom_\text{hard}^{ }$
consist of all non-zero bosonic and fermionic Matsubara modes,
and yield the largest contribution to the pressure.
Since fermions have no zero Matsubara modes,
their effects, including the direct dependence on chemical potentials,
enter only through the hard scale.
Soft modes with
$\pMom_\text{soft}^{ }$ are the bosonic zero Matsubara modes.
Their hard-scale-induced masses are of the same order as
their momenta and hence have to be resummed into the propagators.
Their constant parts
$\Psibar$
consist of
the gauge-field condensates
$\Xbg_0 = \{\Bbg_0^{ }, \Abgt_0\}$ that enforce
the neutrality conditions~\eqref{eq:consv_relation} and
the scalar-field background $\phibar$.
Ultrasoft contributions stem from the spatial gauge fields, which are
screened only non-perturbatively~\cite{Linde:1980ts}.
They do not contribute to the number densities~\eqref{eq:number:density}
at the considered order, and
the ultrasoft effective Lagrangian carries
no chemical potential~\cite{Bodeker:2015zda}.
For the power counting,
we do not distinguish the different gauge couplings, and
we summarize the counting of the couplings
\begin{align}
  \label{eq:power:counting:couplings}
  g_1^{ } \,,\; g_2^{ } \,,\; \gs^{ } \,,\; \yt^{ } &\sim g
  \,, &
  \lambda &\sim g^2
  \,, &
  \yeAlpha &\ll g
  \,,
\end{align}
using a generic coupling $g$.

\begin{table}
\centering
\renewcommand{\arraystretch}{1.5}
\resizebox{\textwidth}{!}{%
\begin{tabular}{|c|c|cccc|}
  \hline
  {\bf Scale} &
  {\bf Validity} &
  {\bf Dimension} &
  {\bf Lagrangian} &
  {\bf Fields} &
  {\bf Parameters} \\
  \hline
  {\sl Hard} & $|\vec{\pMom}|=\pi T$ & $d+1$ &
  $\mathcal{L}_{\text{4d}}$~\eqref{eq:SM:Lagrangian} &
  $A_{\mu},B_{\mu},C_{\mu},\Phi,\psi_i$ &
  $\mu_h^{2},\lambda,g_1,g_2,g_3,h_i,\mu,T$
  \\
  &&\multicolumn{4}{l|}{$\Big\downarrow$
  Step~1:~{\sl Integrate out hard non-zero Matsubara modes}} \\
  {\sl Soft} & $|\vec{\pMom}|=g T$ & $d$ &
  $\mathcal{L}_{\text{3d}}$~\eqref{L3d} &
  $\At_{i},\Bt_{i}%,C_{i},
   $ &
  $\mu_{h,3}^{2},\lambda^{ }_{3},
  g^{ }_{1,3},g^{ }_{2,3},$
  \\
  &&&&
  $\At_{0},\Bt_{0},%C_{0},
    \Phi_3$
  &
  $\mDi{1},\mDi{2},\mu_{1}^{ },\rho_{1}^{ },\rho_{2}^{ },\rho_{\rmii{$G$}}^{ },$
  \\
  &&&&&
  $h_{1}^{ },h_{2}^{ },h_{3}^{ },
  \kappa_{1}^{ },\kappa_{2}^{ },\kappa_{3}^{ }$
  \\
  &&\multicolumn{4}{l|}{$\Big\downarrow$
  Step~2:~{\sl Integrate out soft non-zero momentum modes}} \\
  & $|\vec{\pMom}|=0$ & $0$ &
  $\bar{\mathcal{L}}_{\text{3d}}^{ } = \Veff^\text{soft}$ &
  $\Atbgt_{0},\Btbg_{0}^{ },%\overline{C}_{0}^{ },
    \Phibar_3$
  &
  \\\hline
\end{tabular}
}
\caption[]{%
  Dimensional reduction of the $(d+1)$-dimensional Standard Model
  based on its high-temperature scale hierarchy~\cite{Farakos:1994kx,Gynther:2003za}.
  The first step integrates out the hard non-zero Matsubara modes,
  yielding a $d$-dimensional EFT with fields
  $\Psi = \Psi_\text{3d}$, whose parameters are fixed by matching
  to the couplings and UV parameters of the parent theory.
  The second step integrates out the soft non-zero momentum modes,
  yielding the soft effective potential $\Veff^\text{soft}$.
  }
\label{tab:dr:SM}
\end{table}
Due to the scale separation~\eqref{eq:power:counting:scales},
we employ high-temperature dimensional reduction~\cite{%
  Ginsparg:1980ef,Appelquist:1981vg,Braaten:1995cm,Kajantie:1995dw}.
The latter organizes the contributions of the different scales and
the required resummations in two steps,
illustrated in tab.~\ref{tab:dr:SM}.

In the first step,
the hard modes are integrated out.
This produces a field-independent term, the hard pressure
$p_\text{hard}$ computed in sec.~\ref{sec:hard:contributions}, together
with a 3d~EFT for the remaining bosonic fields,
discussed in sec.~\ref{sec:soft:contributions}.

In the second step,
the soft non-zero momentum modes
$|\vec{\pMom}| \sim \mathcal{O}(gT)$ of that theory are
integrated out,
which generates 
the effective potential for only
the zero (Matsubara and momentum) modes.%
\footnote{
   At the order we are considering,
   gluons in the 3d theory do not contribute
   to $\Veff^\text{soft}$.
}
The resulting pressure consists of
hard and soft contributions,
\begin{align}
\label{eq:pressure:hard+soft}
  p(\mu, T, \phibar) &=
    p_\text{hard}^{ } (\mu, T)
  + p_\text{soft}^{ } (\mu, T, \phibar)
  \,.
\end{align}

The final accuracy of our computation
is $\mathcal{O}(g^2\mu^2)$ relative to
the leading $\mathcal{O}(\mu^2)$ term in the pressure.
The contributions to the pressure are organized as follows:
\begin{itemize}
  \item[$p_\text{hard}$:]
    Hard contributions to the pressure
    to $\mathcal{O}(\mu^2)$ and relative $\mathcal{O}(g^2)$:
    terms of order $g^2 \mu^2 T^2$
    computed in sec.~\ref{sec:hard:contributions}.
    Hence, the Debye masses are needed at two loops:
    $\mDi{i}^2 \sim \mathcal{O}(g^4T^2)$ multiplies
    $\Xbgt_0^2 \sim \mu^2/(g^2T)$ and
    contributes $T\,\mDi{i}^2\Xbgt_0^2$ to the pressure.
  \item[$p_\text{soft}$:]
    Soft contributions to the pressure
    to $\mathcal{O}(\mu^2)$ and relative $\mathcal{O}(g)$:
    terms of order $g\mu^2T^2$.
    The effective potential of the 3d~EFT is evaluated
    at tree level in sec.~\ref{sec:soft:tree-level} and
    at one loop in sec.~\ref{sec:soft:1loop}.
\end{itemize}
The accuracies of the individual 3d~EFT matching coefficients
are collected in appendix~\ref{sec:matching:3d:mu}.
Both the top and the lepton Yukawa couplings enter
through the matching coefficients of the 3d~EFT. 
Even though the $\yeAlpha$ are quite small,
they distinguish the different lepton flavors and can lead
to a baryon asymmetry even at vanishing $B-L$.

At the hard scale,
gauge invariance is manifest 
via the $\xi$-independent matching relations
of appendix~\ref{sec:matching:3d:mu}.
At the soft scale,
gauge invariance is manifest via
the Nielsen--Fukuda--Kugo
identity~\cite{Nielsen:1975fs,Fukuda:1975di} and
the minimum condition~\eqref{sp}.

%%%%%%%%%%%%%%%%%%%%%%%%%%%%%%%%%%%%%%%%%%%%%%%%%%%%%%%%%%%%%%%%%%%%%%%%%%%%%%%%%%%%%%%%%%%%%%%%%%%%
\subsection{Hard contributions}
\label{sec:hard:contributions}

The hard pressure consists of a single field-independent term,
the so-called unit operator~\cite{Braaten:1995cm}, to which all
fermionic and bosonic fields with hard momenta contribute.
Its terms quadratic in the chemical potentials, of
$\mathcal{O}(\mu^2T^2)$, were obtained in~\cite{Bodeker:2014hqa}
together with their relative $\mathcal{O}(g^2)$ corrections,
the largest of which is the QCD one.
The contributing diagrams at
this order are shown in fig.~\ref{fig:hard:diagrams}.
\begin{figure}[t]
\begin{center}
\begin{tabular}{r@{\qquad}cc}
  $\mathcal{O}(\mu^2T^2)$: &
  $\TopoVR(\Aqu)$
  \\[6mm]
  $\mathcal{O}(g^2\mu^2T^2)$: &
  $\ToptVS(\Aqu,\Aqu,\Lglx)$ &
  $\ToptVS(\Aqu,\Aqu,\Lsrii)$
  \\[1mm]
  &
  $g_1^2, g_2^2, \gs^2$ &
  $\yt^2, \yeAlpha^2$
\end{tabular}
\end{center}
\caption{%
  Hard contributions to the pressure at $\mathcal{O}(\mu^2)$,
  eq.~\eqref{eq:L3d:fieldindep}, ordered by their parametric size.
  Only fermion loops carry a chemical potential.
  The one-loop ring gives the free terms, the sunset with a gauge boson
  the $g_1^2, g_2^2$ and $\gs^2$ corrections, and the sunset with a
  Higgs the $\yt^2$ and $\yeAlpha^2$ ones.
  Arrowed solid lines are fermions (quarks and leptons),
  wiggly lines are gauge bosons, and
  dashed lines are scalars.
}
\label{fig:hard:diagrams}
\end{figure}

Setting the quark chemical potentials to
$\mu_q = \muB/3$, the lepton chemical potentials to
$\mu_{\ell_\alpha} = \mu_{e_\alpha} = \muLAlpha $, and
$\mu_\varphi = 0$
in eq.~(4.34) of~\cite{Bodeker:2014hqa},%
\footnote{%
   We ignore field- {\em and} $\mu$-independent terms that are not
   relevant for our computation.
}
and keeping the Yukawa couplings through
$h_{u,ij}^{ } \to \yt^{ } \delta_{i3}^{ }\delta_{j3}^{ }$,
$h_{d,ij}^{ } \to 0$ and
$h_{e,ij}^{ } \to \yeAlpha^{ } \delta_{ij}^{ }$,
we obtain
\begin{align}
\label{eq:L3d:fieldindep}
   p_{\text{hard}} ( \mu, T)
   =
   \frac{T^2}{12}
      \Bigg \{
          \frac { \nG }{ 3 }
          \biggl[&
            4 - \frac{6}{ (4\pi)^ 2 }
                \left (
                    \frac { 3 }{ 2 } g_2 ^ 2
                  + \frac { 11 }{ 18 }  g_1 ^ 2
                  + \frac { 16 }{ 3 } \gs ^ 2
              \right )
          \biggr] \muB ^ 2
          - \frac 1 { 4 \pi ^ 2  }  \yt ^ 2 \muB ^ 2
    \\
    +  \biggl[&
        3 - \frac {9}{ (4\pi)^ 2 } \left ( g_2 ^ 2 + g_1 ^ 2 \right )
      \biggr]
	   \sum _ \alpha   \muLAlpha^2
   - \frac 3 { 4 \pi  ^ 2 } \sum _ \alpha  \yeAlpha ^ 2 \muLAlpha^2
   \Bigg\}
  + \mathcal{O} \bigl(T^4,\mu^4\bigr)
  \,.
  \nonumber
\end{align}
Contributions at
$\mathcal{O}(T^4)$
originate from hard three-loop vacuum diagrams.
Since they are $\mu$-independent,
they do not contribute to the number densities~\eqref{eq:number:density}
and are therefore not needed here.
For other applications,
they can readily be included from~\cite{Gynther:2005dj,Tenkanen:2022tly}
for the SM, or from
{\tt DRalgo}~\cite{Ekstedt:2022bff}
for general BSM models.

Since the
hypercharge chemical potential $\muY$ and
isospin chemical potential $\muA$
enter only through eq.~\eqref{eq:muY:muA:4d},
$p_\text{hard}(\muY,\muA,T) = 0$.
In particular, since
the temporal zero modes are fields of the soft 3d~EFT,
their background contributions
$\muY\sim g_1 \Bbg_0^{ }$ and
$\muA\sim g_2 \Atbgt_0$
enter through
$p_\text{soft}$ in
eq.~\eqref{eq:pressure:hard+soft}.%
\footnote{%
  Contributions analytic in $g^2$ enter through
  the Debye masses
  $\mDi{1}^2$ and
  $\mDi{2}^2$, cf.\ eq.~\eqref{eq:veff:tree}, and
  the term proportional to
  $m_\varphi/T$ through the mass shift~\eqref{eq:Sigma} in the one-loop
  result of sec.~\ref{sec:soft:1loop}.
  This treatment is in contrast to~\cite{Bodeker:2014hqa,Bodeker:2015zda},
  where the chemical potentials $\muY$ and $\muA$ were treated as
  external parameters and
  the corresponding contributions were computed
  in the hard sector.
}

%%%%%%%%%%%%%%%%%%%%%%%%%%%%%%%%%%%%%%%%%%%%%%%%%%%%%%%%%%%%%%%%%%%%%%%%%%%%%%%%%%%%%%%%%%%%%%%%%%%%
\subsection{Soft contributions}
\label{sec:soft:contributions}

The contribution from the soft sector to the pressure~\eqref{eq:pressure:hard+soft}
is given by
\begin{align}
\label{eq:pressure:soft}
  p_\text{soft}^{ } (\mu, T, \phibar) &\stackrel{\eqref{eq:muY:muA}}{=}
    p_{0,\text{soft}}^{ } (\mu, T)
  - T\,\Veff^\text{soft} (\mu, T, \Psibar)
  \,,
\end{align}
where
$\Veff^\text{soft}$
is the 3d effective potential of the soft sector
with mass dimension 3, and
3d background fields $\Psibar$.
There are no field-independent soft contributions,
$p_{0,\text{soft}}^{ }(\mu, T) = 0$.
Indeed, in the 3d~EFT, $\muY$ and $\muA$ are represented by
the 3d background fields
$\Btbg_0^{ }$ and
$\Atbgt_0$ and thus enter through
$\Veff^\text{soft}(\Btbg_0^{ },\Atbgt_0)$.
Its field dependence is translated into chemical-potential dependence
through eq.~\eqref{eq:muY:muA:4d}, after which $\muY$ and $\muA$ are
fixed by the neutrality conditions~\eqref{eq:consv_relation}.
The computation of
$p_\text{soft}(\mu, T, 0)$ in the symmetric phase
was carried out at two-loop order
in~\cite{Bodeker:2015zda}.

The field-dependent terms of
the 3d effective Lagrangian are listed
in~\cite{Kajantie:1995dw,Gynther:2003za,Bodeker:2015zda},
and read
\begin{align}
  \label{L3d}
  \mathcal{L}_{\rm  3d } ( \Psi, \nabla \Psi, \mu  )
    &=
      \frac{1}{4} \Bt_{ij} \Bt_{ij}
    + \frac{1}{4} \F^ a_{ij} \F ^ a_{ij}
    +  ( D_i  \Phi ) ^{\dagger}( D_i \Phi )
    + m_{3}^{2}\Phi^{\dagger}\Phi
    + \lambda_{3}\bigl(\Phi^{\dagger}\Phi\bigr)^{2}
    \nn &
    + \frac{1}{2}(\partial _ i \Bt_{0})^{2}
    + \frac{1}{2}( D _ i \At^ a _{0})^{2}
    + \frac{1}{2}\mDi{1}^{2}\Bt_{0}^{2}
    + \frac{1}{2}\mDi{2}^{2}\At _ 0 ^ a \At_0^a
    \nn[1mm] &
    + h_{1} \Phi^{\dagger}\Phi \Bt_{0}^{2}
    + h_{2} \Phi^{\dagger}\Phi \At_{0}^{a} \At_{0}^{a}
    + h_{3} \Bt_{0}^{ } \Phi ^ \dagger \At_{0}^{a} \sigmaPauli ^ a  \Phi
    \nn[1mm] &
    + \mu_{1}^{ } \Bt_0^{ }
    + \rho_{1}^{ } \Bt_{0}^{ } \Phi  ^\dagger \Phi
    + \rho_{2}^{ } \At_{0}^{a} \Phi  ^\dagger \sigmaPauli  ^ a  \Phi
    + \rho_{\rmii{$G$}}^{ } \Bt_{0}^{ } \At_{0}^{a} \At_{0}^{a}
    \nn[1mm] &
    + \kappa_{1} \Bt_{0}^{4}
    + \kappa_{2} \At_{0}^{a}\At_{0}^{a}\At_{0}^{b}\At_{0}^{b}
    + \kappa_{3} \At_{0}^{a}\At_{0}^{a}\Bt_{0}^{2}
   \,.
\end{align}
The fields $\Psi \equiv \Psi_\text{3d}$ are understood to be
three-dimensional, and we dropped the subscript ``3d'' for brevity.
Their mass dimension is
$[\Psi_\text{3d}] = 1/2$ and
$\Psi_\text{4d} = \Psi_\text{3d} T^{1/2}$.
The corresponding covariant derivatives are
\begin{align}
\label{eq:covariant:derivatives}
  D_i^{ } \Phi
  &=
  \Bigl(
      \partial_i^{ }
    + i g_{1,3}^{ } \frac{Y}{2}\Bt_i^{ }
    + i g_{2,3}^{ } \frac{\sigmaPauli^a}{2} \At_i^a
  \Bigr)\Phi
  \,, &
  D_i^{ } \At_0^a
  &=
      \partial_i^{ } \At_0^a
    - g_{2,3}^{ } \epsilon^{abc} \At_i^b \At_0^c
  \,,
\end{align}
with $Y = 1$ for the Higgs doublet.

While the operator structure of eq.~\eqref{L3d}
is determined by the internal gauge symmetries and
the symmetries of the space-time manifold at finite temperature
using, e.g.,
Hilbert-series methods~\cite{Chakrabortty:2026swu},
the effective couplings of the EFT
are determined through matching calculations from the full 4d theory.
At vanishing chemical potential,
the effective couplings of the 3d~EFT
are known to two-loop order~\cite{Kajantie:1995dw,Croon:2020cgk,Ekstedt:2022bff}.
At finite chemical potential,
the reduction was carried out for
QCD~\cite{Hart:2000ha} and for the Standard Model~\cite{Gynther:2003za}
to $\mathcal{O}(g^4)$ in the counting $g_1^2 \sim g_2^3$.
The $\mathcal{O}(g_1^3)$ terms omitted by that counting are supplied in
appendix~\ref{sec:matching:3d:mu}, and the hard-mode susceptibilities at
$\mathcal{O}(g^2)$ in~\cite{Bodeker:2014hqa,Bodeker:2015zda}.
The corresponding effective couplings
$g_{1,3},g_{2,3}, \lambda_3, \mu_1, \ldots$
are collected in appendix~\ref{sec:matching:3d:mu} and
scale as
\begin{align}
  \mDi{i}^2 \,,\; h_i^{ }\phibarT^2 &\sim g^2 T^2
  \,,&
  \mu_1^{ } &\sim g\mu T^{3/2}
  \,, &
  \rho_i^{ } &\sim g \yeAlpha^2 \mu T^{1/2}
  \,, &
  m_3^2\bigl|_\mu^{ } &\sim \yt^2 \mu^2
  \,.
\end{align}
The temporal-vector quartic couplings
$\kappa_{1},\kappa_{2},\kappa_{3} \sim g^4 T$ are listed in~\cite{Croon:2020cgk}.
Alternatively, the matching can be organized with the finite-temperature
heat kernel, whose coefficients depend on the Polyakov
loop~\cite{Megias:2002vr,Megias:2003ui,Moral-Gamez:2011wcb,Bandyopadhyay:2026nrv,Biermann:2026qju}.

Since the zero-momentum modes acquire expectation values,
we split every 3d field into its constant part
$\Psibar$ and the soft modes of non-zero momentum $\Psi$,
\begin{align}
  \label{decomp}
   \Psi \to \Psi  + \Psibar
   \,.
\end{align}
After integrating over $\Psi$ in the path integral,
the effective potential for the zero modes $\Psibar$ is obtained,
\begin{align}
  \label{sint}
   \exp \left ( - \widetilde{ \Omega  }(\Psibar)  /T \right )
   =
   \int { \cal D}\Psi
   \exp \Bigl (
        - \int_\vec{x} \mathcal{L} _ \text{3d}(\Psi, \Psibar)
    \Bigr )
   \,,
\end{align}
which then enters eq.~\eqref{zmi}
and gives rise
to the final effective potential $\Omega_\text{3d}(\Psibar)$.
For an illustration of this procedure, see tab.~\ref{tab:dr:SM}.

As in the 4d case in eq.~\eqref{phibar},
the background of the Higgs doublet is
$\Phibar = \frac{1}{\sqrt{2}} (0, \phibarT)$
with $\phibarT^2 = \phibar^2/T$.
This places $\phibarT$ in the lower, electrically neutral doublet
component as in ref.~\cite{Gynther:2003za}, so that
$\Phibar^\dagger \Phibar = \phibarT^2/2$ and
$\Phibar^\dagger \sigmaPauli^a \Phibar = -\delta^{a3} \phibarT^2/2$.
Of the three isospin directions,
only the third one is singled out.%
\footnote{%
  Every operator with an odd number of $\sigmaPauli^3$ insertions therefore
  enters $\Veff$ with a relative minus sign, which is the origin of the
  negative $h_{3}^{ }$ term and of the relative sign between the
  $\rho_{1}^{ }$ and $\rho_{2}^{ }$ terms in eq.~\eqref{eq:veff:tree}.
  The opposite orientation, $\Atbgt_0 \to -\Atbgt_0$, merely flips the
  sign convention for $\muA$.
}
We use $\phibarT$ inside the 3d theory, and $\phibar$ or
$x = \phibar/T$ for 4d quantities such as
the pressure.

For the power counting,
we have to account for the parametric sizes of both
the soft space-dependent fields $\Psi$, and
the zero-momentum modes $\Psibar$
as given in the decomposition~\eqref{decomp}.
All three backgrounds are fixed by the stationarity conditions~\eqref{sp}.
The Higgs background
is fixed by minimizing the effective potential,
which we do at $\mu_i^{ } = 0$ in
appendix~\ref{sec:Tevolution:vev} such that 
$\phibar = \phibarMin$.
The latter is temperature-dependent, and
near $\Tsph$ we count%
\footnote{%
  The high-temperature expansion is
  valid until $x \sim 3$;
  cf.\ sec.~\ref{sec:EFT:validity}.
  For $x > 3$,
  the effect of higher-dimensional operators in
  the 3d~EFT Lagrangian~\eqref{L3d} becomes important.
  Around $\Tsph$, $x \sim 1$ and the expansion is valid.
}
\begin{align}
  \label{phibarsize}
   \phibar  \sim T
  \,, &&
   \phibarT^2 \sim T
  \,,
\end{align}
consistent with~\cite{Khlebnikov:1996vj},
such that
$x = \phibar/T \sim 1$ and
vector masses
$g_{2,3}^{ }\phibarT \sim gT$
are soft.
The zero modes of the temporal gauge fields
$\Xbgt_{0}^{ } = \{\Btbg_{0}^{ }, \Atbgt_0\}$ are fixed by
hypercharge and isospin neutrality.
The linear term $\mu_1^{ }\Bt_0^{ }$ of eq.~\eqref{L3d} exists only
because the chemical potentials break CP and CPT and
induce a non-zero expectation value for $\Btbg_0$.
In the broken phase,
$\phibarT \neq 0$, the mixed operator $h_3^{ }$
couples
$\Atbgt_0$ to
$\Btbg_0^{ }$ in $\Veff^\text{soft}$,
and $\Atbgt_0$ also acquires an expectation value,
while the remaining components
$\At_0^{1,2}$ decouple from $\Btbg_0^{ }$.

In terms of the 3d fields,
the identification in
eq.~\eqref{eq:muY:muA:4d}
is adapted with the normalization
\begin{align}
\label{eq:muY:muA}
  g_{1,3}^{ }\Btbg_0^{ } &=
    i \muY^{ }
  \,, &
  g_{2,3}^{ }\Atbgt_0 &=
    i \muA^{ }
  \,,
\end{align}
corresponding to
$\Atbgt_0 \sim \Btbg_0^{ } \sim \mu/(g\sqrt{T})$ with
$g_{i,3}^2 = g_i^2 T$ at leading order.
With this choice, the charge susceptibilities
$\chi_i^{ } =
\partial^2 p/\partial\mu_i^2 =
T\mDi{i}^2/g_{i,3}^2$ for the
hypercharge ($\chi_Y^{ }$) and
isospin ($\chi_{T_3}^{ }$)
are physical.
Since $\muY$ and $\muA$
merely multiply the neutrality conditions~\eqref{sp},
redefining them displaces the stationary point, but
the pressure there, as well as
$\nBeq$ and $\kappa$, are independent
of the choice of normalization in eq.~\eqref{eq:muY:muA}.%
\footnote{%
  Equivalently, one may work with $\Btbg_0$ and $\Atbgt_0$ throughout
  and never introduce $\muY$ and $\muA$.
}
The temporal-vector zero modes
couple to the soft Higgs field through
the temporal-vector couplings
$h_1^{ }, h_2^{ }, h_3^{ }$~\cite{Gynther:2003za}.

The power counting of soft fields $\Psi$
is determined by the typical size of their fluctuations,
which can be estimated by the propagator.
For the soft Higgs field $\Phi$,
the propagator amounts to
\begin{align}
    \bigl\langle \Phi^\dagger \Phi \bigr\rangle 
    \sim 
    \int _ { \vec{\pMom} \sim m } \frac {1} { \vec{\pMom} ^ 2 + m ^ 2 }
    \sim
    m
  \,,
\end{align}
with
$\int_\vec{\pMom} = \int \frac{{\rm d}^d \vec{\pMom}}{(2\pi)^d}$,
so that
$\Phi \sim \sqrt{m}$ where
$m$ is some scalar mass.
Counting $m \sim g T$ with a generic gauge coupling $g$,
we obtain
$\Phi \sim \sqrt{g T}$.
Similarly, one obtains the same estimate for
the soft gauge fields $\At_0$ and $\Bt_0$.

To summarize,
the parametric sizes used in the following are
\begin{align}
  \label{eq:power:counting:fields}
  \phibar &\sim T
  \,, &
  \phibarT^{ } &\sim \sqrt{T}
  \,, &
  \Psi^{ } &\sim \sqrt{gT}
  \,, &
  \Btbg_0^{ } \,,\; \Atbgt_0 &\sim \frac{\mu}{g\sqrt{T}}
  \,.
\end{align}
Which order of a matching coefficient is required depends on whether it
multiplies a zero mode or a soft field:
\begin{itemize}
  \item[$\Psibar$]
    \emph{Zero modes.}
    Since already tree-level terms
    $T \mu_1^{ }\Btbg_0^{ } \sim T \mDi{i}^2\Btbg_0^2 \sim \mu^2T^2$
    contribute to the leading pressure,
    their coefficients are needed to relative
    $\mathcal{O}(g^2)$ accuracy.
    The Debye masses up to
    $\mathcal{O}(g^4)$~\cite{Gynther:2005av,Gynther:2005dj,Schicho:2021gca} and
    $\mathcal{O}(g^2\mu^2)$~\cite{Gynther:2003za}.
    The tadpole $\mu_1^{ }$
    up to $\mathcal{O}(g^3\mu)$~\cite{Gynther:2003za}.
    The couplings $h_1^{ },h_2^{ },h_3^{ }$ of
    eq.~\eqref{eq:mu:remaining:couplings}
    up to $\mathcal{O}(g^4T^2)$
    corrections~\eqref{eq:h1:mu0}--\eqref{eq:h3:mu0}~\cite{Croon:2020cgk}.
    The mass parameter $m_3^2$ of eq.~\eqref{eq:m3_mu}
    up to $\mathcal{O}(g^2\mu^2)$.
  \item[$\Psi$]
    \emph{Soft fields.}
    Since every loop generated is suppressed by one power of $g$,
    leading-order values suffice:
    $m_3^2 \sim \mDi{i}^2 \sim h_i^{ }\phibarT^2 \sim g^2T^2$.
\end{itemize}

%%%%%%%%%%%%%%%%%%%%%%%%%%%%%%%%%%%%%%%%%%%%%%%%%%%%%%%%%%%%%%%%%%%%%%%%%%%%%%%%%%%%%%%%%%%%%%%%%%%%
\subsubsection{%
  Soft tree-level contributions to the 
  effective potential}
\label{sec:soft:tree-level}

Starting from the 3d~EFT Lagrangian~\eqref{L3d},
we first determine the tree-level effective potential for
the zero modes
$\Psibar = \{\Btbg_0^{ },\Atbgt_0,\phibarT\}$.
Therein,
the power counting in $\mu$
follows from the corresponding power in
$\Xbgt_{0}^{ } = \{\Btbg_0^{ }, \Atbgt_0\}$,
which each contribute one power of $\mu$ by
eq.~\eqref{eq:muY:muA}.
Terms without $\Xbgt_{0}^{ }$ 
keep their $\mu$-dependent couplings, with
$m_3^2|_\mu^{ }\phibarT^2$ of $\mathcal{O}(g^2\mu^2T^2)$.
Terms {\em linear} in the zero modes come with the CP-odd couplings
$\mu_1^{ },\rho_1^{ },\rho_2^{ }$, which are themselves of
$\mathcal{O}(\mu)$.
Terms {\em quadratic} in them require only the $\mu$-independent
couplings, the $\mu$-dependent part of $m_{\rmii{D$i$}}^2$ contributing
at $\mathcal{O}(\mu^4)$.
All three classes therefore enter at $\mathcal{O}(\mu^2)$, whereas
terms cubic or quartic in the zero modes, such as
$\rho_{\rmii{$G$}}^{ }\Btbg_0^{ }(\Atbgt_0)^2$ or
$\kappa_i^{ }$ terms, are of
$\mathcal{O}(\mu^3)$ and
$\mathcal{O}(\mu^4)$ and are of higher order.

The three-dimensional tree-level effective potential for the zero modes,
similar to~\cite{Laine:1999wv},
then amounts to
\begin{align}
\label{eq:veff:tree}
  \Veff^{\text{tree}}(\mu,T,\Psibar) &=
    \frac{1}{2} m_{3}^{2}\phibarT^2
  + \frac{1}{4}\lambda_3^{ }\phibarT^4
  \nn &
  + \frac{1}{2}\Bigl[
        \mDi{2}^{2}
      + h_{2}^{ }\phibarT^2
    \Bigr]\bigl(\Atbgt_0\bigr)^2
  + \frac{1}{2}\Bigl[
        \mDi{1}^{2}
      + h_{1}^{ }\phibarT^2
    \Bigr]\Btbg_0^2
  \nn &
  - \frac{h_{3}^{ }}{2}\Atbgt_0\Btbg_0^{ }\phibarT^2
  + \mu_{1}^{ }\Btbg_0^{ }
  + \frac{1}{2}\Bigl(
        \rho_{1}^{ }\Btbg_0^{ }
      - \rho_{2}^{ }\Atbgt_0
    \Bigr)\phibarT^2
  \,.
\end{align}
In the number densities~\eqref{eq:number:density},
only the $\mu$-dependent part of
$m_3^2|_\mu^{ } \sim \mathcal{O}(\mu^2)$ contributes,
whereas $\lambda_3^{ }|_\mu^{ } = 0$ at this order.%
\footnote{
  The $\mu$-dependent part of $\lambda_3^{ }$ is of
  $\mathcal{O}(g^4\zeta_3^{ }\mu^2/(4\pi)^4)$
  similar to 
  eqs.~\eqref{eq:veff:dim5} and~\eqref{eq:veff:dim6}.
}
Their $\mu$-independent parts enter only through
$\widetilde{m}_{\varphi}^{ }$ and $\widetilde{m}_{\rmii{$G$}}^{ }$ in
the one-loop term of sec.~\ref{sec:soft:1loop}.

Replacing the zero modes for chemical potentials
via eq.~\eqref{eq:muY:muA},
$(\Btbg_0^{ },\Atbgt_0) \to (\muY,\muA)$,
and adding
the field-independent hard contribution~\eqref{eq:L3d:fieldindep},
eq.~\eqref{eq:veff:tree} reproduces eq.~\eqref{eq:p-func}
and~\cite{Laine:1999wv}.
Beyond~\cite{Laine:1999wv},
eq.~\eqref{eq:veff:tree} yields two further terms $\propto \phibar^2$,
one $\propto\yt^2\muB^{ }$ from $\rho_{1}^{ },\rho_{2}^{ }$ and
one from $m_3^2|_\mu^{ }$.

%%%%%%%%%%%%%%%%%%%%%%%%%%%%%%%%%%%%%%%%%%%%%%%%%%%%%%%%%%%%%%%%%%%%%%%%%%%%%%%%%%%%%%%%%%%%%%%%%%%%
\subsubsection{%
  Soft one-loop contributions to
  the effective potential
  }
\label{sec:soft:1loop}

Fluctuations in the soft fields $\Psi$ around
the zero modes $\Psibar$ give rise to
soft higher-loop contributions to
the zero-mode effective potential.
By defining the mass matrix $[M^2]_{ij}$ of the soft fields $\Psi$
in the background of
the zero modes $\Psibar$,
the one-loop contribution is given by
\begin{align}
\label{eq:veff:1l}
  \Veff^{\text{1loop}} &=
      \frac{1}{2}\int_{\vec{\pMom}}
      \tr\ln(\pMom^2 + M^2)
    = \sum_i n_i^{ } J_{3}(m_i)
    \,,
\end{align}
where $m_i^2$ are the eigenvalues of $M^2$, 
$n_i^{ }$ the corresponding multiplicities, and
$J_3$ is the 3d logarithmic integral~\eqref{eq:J3}.

In the presence of the zero modes $\Psibar$,
the mass matrix $M^2$ is not block diagonal.
Since only terms up to $\mathcal{O}(\mu^2)$ are needed,
we perturbatively expand eq.~\eqref{eq:veff:1l}
in the background fields
$\Xbgt_{0}^{ } = \{\Btbg_0^{ },\Atbgt_0\}$.
To this end,
we split the mass matrix via
$M = \widetilde{M} + \Delta$ into
a diagonal part $\widetilde{M}^2$ with
$\Xbgt_0^{ } = 0$
and eigenvalues $\widetilde{m}_i^2$ and
a background-dependent part
$[\Delta]_{ij}^{ } = \Sigma_{i}^{ } \delta_{ij}^{ } + \sigma_{ij}^{ }$.
The latter, in turn, consists of
diagonal entries $\Sigma_{i}^{ }$ and
off-diagonal entries $\sigma_{ij}^{ }$.

The expanded one-loop potential,
\begin{align}
\label{eq:trlog}
  \Veff^{\text{1loop}} &=
      \Veff^{\text{1loop}}\big|_{\Xbgt_0^{ } = 0}
    + \frac{1}{2}\int_{\vec{\pMom}} \tr\bigl[ G \Delta \bigr]
    - \frac{1}{4}\int_{\vec{\pMom}} \tr\bigl[ G \Delta G \Delta \bigr]
    + \ldots
  \,,& 
  G &= \frac{1}{\pMom^2 + \widetilde{M}^2}
  \,,
\end{align}
is then a series of
$n$ one-loop propagators $G$ with
$n$ insertions of $\Delta$.
The first term still carries the full $\phibarT$.
Its only chemical-potential dependence comes from
$\widetilde{m}_i^2\big|_\mu^{ } \sim \mathcal{O}(g^2\mu^2)$, which
first enters at
$p_\text{soft} \sim \mathcal{O}(g^3\mu^2T^2)$.
This is beyond the accuracy of eq.~\eqref{eq:pressure:hard+soft}
and does not affect the number densities~\eqref{eq:number:density}. 

The diagonal terms are
$\Sigma_{i}^{ } \sim \mathcal{O}(\mu^2)$
(cf.\ eqs.~\eqref{eq:Sigma} and~\eqref{eq:m:Ai}),
the off-diagonal ones $\sigma_{ij}^{ } \sim \mathcal{O}(\mu)$.
Since $G$ is diagonal in the $\widetilde{M}^2$-basis,
the trace with a single insertion reduces to
$\tr[G\Delta] = \sum_i G_i^{ }\Sigma_{i}^{ }$.
It therefore contributes at $\mathcal{O}(\mu^2)$, whereas the
off-diagonal entries first contribute through two insertions,
$\tr[G\Delta G\Delta] =
 \sum_{ij} G_i^{ }G_j^{ }[\Delta]_{ij}^{ }[\Delta]_{ji}^{ }$.
Its diagonal part
$\Sigma_i^2 \sim \mathcal{O}(\mu^4)$ is negligible, while
its off-diagonal part
$\sigma_{ij}^{ }\sigma_{ji}^{ } \sim
\mathcal{O}(\mu^2)$ is kept.
In Landau gauge,
neither massless spatial photons nor ghosts
contribute.
Gluons contribute only at higher orders
in our power counting.

In the scalar sector,
the masses of the soft Higgs field for fluctuations
in the direction of and orthogonal to $\phibar$,
evaluated at $\Xbgt_{0}^{ } = 0$,
are
\begin{align}
  \label{eq:m:tilde:scalar}
  \widetilde{m}_{\varphi}^2 &=
      m_{3}^2
    + 3\lambda_3^{ }\phibarT^2
  \,, &
  \widetilde{m}_{\rmii{$G$}}^2 &=
      m_{3}^2
    + \lambda_3^{ }\phibarT^2
  \,.
\end{align}
In the temporal sector, the diagonal entries
$[\widetilde{M}^2]_{\rmii{$\At_0\At_0$}}$ and
$[\widetilde{M}^2]_{\rmii{$\Bt_0\Bt_0$}}$
of the $\At_0^3$--$\Bt_0$ block are
\begin{align}
\label{eq:m:tilde:temporal}
  \widetilde{m}_{\rmii{$\At_{0}$}}^2 &=
      \mDi{2}^2
    + h_{2}^{ }\phibarT^2
  \,, &
  \widetilde{m}_{\rmii{$\Bt_{0}$}}^2 &=
      \mDi{1}^2
    + h_{1}^{ }\phibarT^2
  \,,
\end{align}
where the matching relations for
$m_{3}^2$, $\lambda_3^{ }$, $\mDi{i}^2$ and $h_i^{ }$ are collected in
appendix~\ref{sec:matching:3d:mu}.
At potential order $\mathcal{O}(g^2\mu^2)$,
only the $\mu$-independent parts
$m_{3}^2|_{\mu=0}$ and
$\mDi{i}^2|_{\mu=0}$
for the field-independent masses
are used here. 

The temporal-vector mass matrix is not diagonal
since the operator $h_{3}^{ }$ of eq.~\eqref{L3d} mixes
$\Bt_0$ with the third isospin direction and
$\bigl[\widetilde{M}^2\bigr]_{\rmii{$\At_0^3\Bt_0^{ }$}}^{ } =
  - \frac{1}{2}h_{3}^{ }\phibarT^2$.
The $\At_0^{1,2}$ remain degenerate at $\widetilde{m}_{\rmii{$\At_0$}}$ and
their mixing with the charged Goldstone modes is
$\propto \Btbg_0^{ }$ and enters through $\Delta$
in eq.~\eqref{hl}.
After diagonalizing the $(\At_0,\Bt_0)$ block,
the neutral temporal mass eigenvalues are
\begin{align}
\label{eq:mZ0:mA0}
  \widetilde{m}_{\rmii{$Z_0$}}^2
  \,,\;
  \widetilde{m}_{\rmii{$\gamma_0$}}^2
  &=
  \frac{1}{2}
  \biggl[
      \widetilde{m}_{\rmii{$\At_{0}$}}^2
    + \widetilde{m}_{\rmii{$\Bt_{0}$}}^2
    \pm \sqrt{
        \bigl(
            \widetilde{m}_{\rmii{$\At_{0}$}}^2
          - \widetilde{m}_{\rmii{$\Bt_{0}$}}^2
        \bigr)^2
      + \bigl( h_{3}^{ }\phibarT^2 \bigr)^2
    }
  \biggr]
  \,,
\end{align}
with the upper (lower) sign for the
$Z_0$- ($\gamma_0$-)like eigenvalue.
The mixing angle, $\thetaE$, of
the static temporal components
is purely thermal.
It rotates the soft $(\At_0^3,\Bt_0)$ into $(Z_0,\gamma_0)$ via
$Z_0 = \cos\thetaE\,\At_0^3 - \sin\thetaE\,\Bt_0$ as in the vacuum,
and reads
\begin{align}
\label{eq:sin2theta}
  \sin^2\!\thetaE
  &=
  \frac{
      \widetilde{m}_{\rmii{$Z_0$}}^2
    - \widetilde{m}_{\rmii{$\At_{0}$}}^2
  }{
      \widetilde{m}_{\rmii{$Z_0$}}^2
    - \widetilde{m}_{\rmii{$\gamma_0$}}^2
  }
  \,, &
  \tan 2\thetaE
  &=
  \frac{
    h_{3}^{ }\phibarT^2
  }{
      ( \mDi{2}^2 - \mDi{1}^2 )
    + ( h_{2}^{ } - h_{1}^{ } )\phibarT^2
  }
  \,.
\end{align}
All masses and couplings are
$T$-dependent
effective parameters of the 3d EFT,
such that $\thetaE$ is
both $\phibar$- and $T$-dependent compared to its
zero-temperature value $\thetaw$~\cite{%
  Ghiglieri:2016xye,Ghiglieri:2018wbs,Croon:2020cgk}.
Since the Debye masses
$\mDi{1}^2, \mDi{2}^2 \sim g^2 T^2$
are parametrically of the same order as
the Higgs-induced entries $h_i^{ }\phibarT^2 \sim g^2 T^2$
for $\phibar \sim T$, cf.\ eq.~\eqref{eq:power:counting:fields},
they cannot be omitted in general.
Only deep in the broken phase $\phibar \gg gT$,
where the Higgs contributions dominate,
does the mixing angle approach the vacuum-like form
$\tan 2\thetaE \to h_{3}^{ }/( h_{2}^{ } - h_{1}^{ })$,
which at leading order reproduces
$\sin^2\!\thetaE \to g_1^2/( g_1^2 + g_2^2)$
in terms of the thermally corrected couplings.

To evaluate the expanded potential~\eqref{eq:trlog},
we need to define the $\Xbgt_{0}^{ }$-dependent
mass shifts $\Sigma_i$.
The spatial vector bosons
are affected by the zero modes through the
non-abelian coupling of $\Atbgt_0$ to $\At_i^{1,2}$,
which lowers their mass similarly to an isospin chemical potential,
\begin{align}
\label{eq:m:Ai}
  m^2_\rmii{$\bm{\At}^{1}$} =
  m^2_\rmii{$\bm{\At}^{2}$}
  &=
  \frac{g_{2,3}^2}{4}
  \Bigl[ \phibarT^2 + 4 \bigl( \Atbgt_0 \bigr)^2 \Bigr]
  \,, &
  m^2_\rmii{$\bm{Z}$}
  &=
  \frac{g_{1,3}^2 + g_{2,3}^2}{4}\, \phibarT^2
  \,,
\end{align}
where
$\bm{Z}$ denotes the neutral spatial mass eigenstate.
The spatial photon stays massless
$m^2_\rmii{$\bm{\gamma}$} = 0$
and does not contribute to $\Veff^\text{soft}$.
In the scalar sector,
the neutral fluctuations $(\chi,G^0)$ and
the two charged components $G^{1,2}$ are shifted by
\begin{align}
\label{eq:m:phi}
  m_{\chi}^2 &=
      \widetilde{m}_{\varphi}^2
    + \Sigma_{\rmii{$0$}}
  \,, &
  m_{\rmii{$G^0$}}^2 &=
      \widetilde{m}_{\rmii{$G$}}^2
    + \Sigma_{\rmii{$0$}}
  \,, &
  m_{\rmii{$G^\pm$}}^2 &=
      \widetilde{m}_{\rmii{$G$}}^2
    + \Sigma_{\rmii{$\pm$}}
  \,,
\end{align}
where the two charged components are mass-degenerate and
\begin{align}
\label{eq:Sigma}
  \Sigma_{\rmii{$0$}}
  \,,\;
  \Sigma_{\rmii{$\pm$}}
  &=
      h_{1}^{ }\Btbg_0^2
    + h_{2}^{ }\bigl(\Atbgt_0\bigr)^2
    \mp h_{3}^{ }\Atbgt_0\Btbg_0
  \stackrel{\text{LO}}{=}
    - \frac{1}{4}\bigl( \muY \mp \muA \bigr)^2
  \,,
\end{align}
with the upper (lower) signs for the neutral (charged) components.
The CP-odd couplings
$\rho_i^{ }$ and
the quartic $\kappa_i^{ }$
induce another mass shift of
$\rho_i^{ }\Xbgt_0 \sim
\kappa_i^{ }\Xbgt_0^2 \sim
 \mathcal{O}\bigl(g^2\mu^2\bigr)$.
While of relative $\mathcal{O}(g^2)$
compared with $\Sigma_{i}^{ }$,
this shift would enter at
$p_\text{soft} \sim \mathcal{O}(g^3\mu^2T^2)$.

\begin{figure}[t]
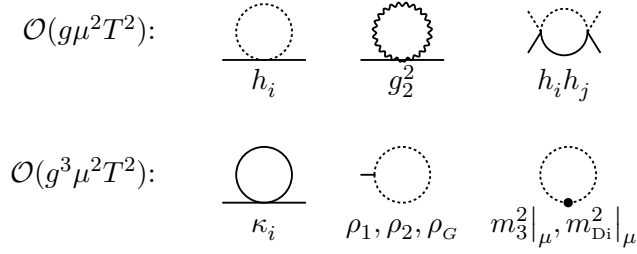

\begin{center}
\begin{tabular}{r@{\qquad}ccc}
  $\mathcal{O}(g\mu^2T^2)$: &
  $\TopoST(\Lsai,\Asrii)$ &
  $\TopoST(\Lsai,\Aglx)$ &
  $\TopoVBlr(fex(\Lsrii,\Lsai,\Lsrii,\Lsai),\Asrii,\Asai)$
  \\[1mm]
  &
  $h_i^{ }$ &
  $g_2^2$ &
  $h_i^{ } h_j^{ }$
  \\[6mm]
  $\mathcal{O}(g^3\mu^2T^2)$: &
  $\TopoST(\Lsai,\Asai)$ &
  $\TopoOT(\Lsai,\Asrii)$ &
  $\TopoVRo(\Asrii)$
  \\[1mm]
  &
  $\kappa_i^{ }$ &
  $\rho_{1}^{ },\rho_{2}^{ },\rho_{\rmii{$G$}}^{ }$ &
  $m_3^2\big|_\mu^{ }, \mDi{i}^2\big|_\mu^{ }$
\end{tabular}
\end{center}
\caption{%
  One-loop contributions to the pressure at $\mathcal{O}(\mu^2)$,
  ordered by their parametric size
  with the target accuracy $\mathcal{O}(g^2\mu^2 T^2)$.
  External solid lines are the backgrounds
  $\Xbgt_{0}^{ } = \{\Btbg_0^{ },\Atbgt_0\}$,
  external dashed lines the Higgs background $\phibarT$.
  Dashed for the soft scalars $\chi$, $G^0$ and $G^{1,2}$,
  solid for the temporal vectors
  $\At_0^{1,2}$,
  $Z_0^{ }$, and
  $\gamma_0^{ }$,
  and wiggly for the spatial vectors
  $\bm{\At}^{1,2}$,
  $\bm{Z}$.
  The blob in the last entry denotes the $\mu$-dependent mass
  insertion.
}
\label{fig:1loop:counting}
\end{figure}

The first two terms of the expanded potential~\eqref{eq:trlog} then
evaluate, with the master integrals~\eqref{eq:I3}
and~\eqref{eq:I12}, to
\begin{align}
\label{eq:rule:tadpole}
  \int_{\vec{\pMom}} \tr\bigl[ G \Delta \bigr]
  &=
  \sum_i n_i^{ }\, \Sigma_{i}^{ } \int_{\vec{\pMom}}
  \frac{1}{\pMom^2 + \widetilde{m}_i^2}
  =
  - \frac{1}{4\pi} \sum_i
    n_i^{ }\, \widetilde{m}_i^{ }\, \Sigma_{i}^{ }
  \,,
  \\[1mm]
\label{eq:rule:bubble}
  \int_{\vec{\pMom}} \tr\bigl[ G \Delta G \Delta \bigr]
  &=
  2\sum_{i<j} \bigl( \sigma_{ij}^{ } \bigr)^2
    \int_{\vec{\pMom}}
    \frac{1}{[\pMom^2 + \widetilde{m}_i^2]
             [\pMom^2 + \widetilde{m}_j^2]}
  =
  \frac{1}{4\pi} \sum_{i<j}
    \frac{\bigl( \sigma_{ij}^{ } \bigr)^2}{
      \widetilde{m}_i^{ } + \widetilde{m}_j^{ }}
  \,.
\end{align}
These contributions correspond to
the one-loop diagrams collected in
fig.~\ref{fig:1loop:counting}
together with
those that exist at one loop but are of higher order and therefore not
evaluated.
The counting is given below, and
only the first row is within the accuracy goal
$\mathcal{O}(g^2\mu^2 T^2)$.

In the broken phase,
the soft Higgs loop~\cite{Bodeker:2014hqa}
now gives the following contribution to
$
\Veff ( \Btbg_0^{ }, \Atbgt_0, \phibarT ) -
\Veff ( 0 , 0, \phibarT )$,
\begin{align}
\label{hl}
  \TopoST(\Lsai,\Asrii)
   &=
   - \frac{1}{8\pi}
   \Bigl(
        h_{2}^{ }\bigl(\Atbgt_0\bigr)^2
      + h_{1}^{ }\bigl(\Btbg_0^{ }\bigr)^2
   \Bigr)
   \Bigl(
          \widetilde{m}_{\varphi}
      + 3 \widetilde{m}_{\rmii{$G$}}
    \Bigr)
   +
   \frac{1}{8\pi}
    h_{3}^{ }\Atbgt_0\Btbg_0^{ }
    \Bigl(
        \widetilde{m}_{\varphi}
      - \widetilde{m}_{\rmii{$G$}}
    \Bigr)
  \,.
\end{align}
Solid lines denote temporal vectors and
dashed lines the soft Higgs field.
The analogous loop of temporal vectors
is of higher order,
$\vcenter{\hbox{\scalebox{0.5}{$\TopoST(\Lsai,\Asai)$}}} \sim
\mathcal{O}(g^3\mu^2T^2)$,
since
$\kappa_i^{ } \sim g^4T$.

The contribution from soft gauge-field loops
enters through
the mass shifts of the $a = 1,2$ components of the soft
$\bm{\At}^a$ given by eq.~\eqref{eq:m:Ai}.
Each component contributes
as in eq.~\eqref{eq:rule:tadpole} at
second order in
$\Atbgt_0$, {\em viz.},
\begin{align}
 \label{gfl}
  \TopoST(\Lsai,\Aglx)
  &=
   - \frac{1}{4\pi}
   g_{2,3}^{ }\phibarT
   \bigl( g_{2,3}^{ }\Atbgt_0 \bigr)^2
   \,,
\end{align}
where wiggly lines denote soft vector bosons.

Finally, a mixed contribution between
temporal vectors $\Xbgt_{0}^{ }$ and $\Phibar$ arises,%
\footnote{%
  In principle,
  the vertex of eq.~\eqref{gfl} also mixes $\At_0^{1,2}$ with
  $\bm{\At}^{1,2}$ giving rise to a diagram
  $\vcenter{\hbox{\scalebox{0.5}{$\TopoSB(\Lsai,\Asai,\Aglx)$}}} \sim
  \mathcal{O}(g\mu^2T^2)$,
  which vanishes in Landau gauge.
}
\begin{align}
\label{ml}
  \TopoVBlr(fex(\Lsrii,\Lsai,\Lsrii,\Lsai),\Asrii,\Asai)
  &=
  - \frac{\phibarT^2}{32\pi}
  \Bigl( g_{2,3}^{ }\Atbgt_0 - g_{1,3}^{ }\Btbg_0 \Bigr)^2
  \Biggl[
     \frac{
       \bigl(
           g_{2,3}^{ }\cos\thetaE
         + g_{1,3}^{ }\sin\thetaE
       \bigr)^2
     }{
         \widetilde{m}_{\rmii{$Z_0$}}
       + \widetilde{m}_{\varphi}
     }
   +
     \frac{
       \bigl(
           g_{1,3}^{ }\cos\thetaE
         - g_{2,3}^{ }\sin\thetaE
       \bigr)^2
     }{
         \widetilde{m}_{\rmii{$\gamma_0$}}
       + \widetilde{m}_{\varphi}
     }
  \Biggr]
  \nn &
  - \frac{\phibarT^2}{16\pi}
  \frac{
    \bigl( g_{1,3}^{ }g_{2,3}^{ }\Btbg_0 \bigr)^2
  }{
      \widetilde{m}_{\rmii{$\At_{0}$}}
    + \widetilde{m}_{\rmii{$G$}}
  }
  \,,
\end{align}
using the leading-order relations
$h_{1}^{ } = g_{1,3}^2/4$,
$h_{2}^{ } = g_{2,3}^2/4$,
$h_{3}^{ } = g_{1,3}^{ }g_{2,3}^{ }/2$, and
$g_{i,3}^2 \sim g_{i}^2 T$.

The first bracket in eq.~\eqref{ml}
originates from the mixing of $\chi$ with the neutral
temporal mass eigenstates $Z_0$ and $\gamma_0$, and
the last term from
the mixing of
the charged Goldstone modes
$G^{1,2}$ with
the charged temporal vectors $\At_0^{1,2}$
with mass $\widetilde{m}_{\rmii{$\At_{0}$}}$.
Only the combination $\muY - \muA$ appears in the neutral part,
consistent with the tree-level potential~\eqref{eq:veff:tree}.
In the deep broken phase,
$g_{1,3}^{ }\cos\thetaE - g_{2,3}^{ }\sin\thetaE$ vanishes and
$\gamma_0$ becomes the temporal photon.
At sphaleron freeze-out, where
$\thetaE(\Tsph) \simeq 0.18\,\thetaw$, both
the $\gamma_0$ and $Z_0$ channels of eq.~\eqref{ml} are relevant.

As a check, the symmetric limit $\phibarT \to 0$ at $\Atbgt_0 = 0$ sets
$\widetilde{m}_{\varphi} = \widetilde{m}_{\rmii{$G$}} = m_3^{ }$ and
$\Sigma_{\rmii{$0$}} = \Sigma_{\rmii{$\pm$}} = -\mu_\varphi^2$ with
$\mu_\varphi^{ } = \muY/2$, so that eqs.~\eqref{gfl} and~\eqref{ml}
vanish and the four Higgs components of eq.~\eqref{hl} collapse to
$p_\text{soft}^{ } = T(m_3^2-\mu_\varphi^2)^{3/2}/(3\pi)$,
which is the one-loop soft contribution
of ref.~\cite{Bodeker:2015zda} in closed form.

%%%%%%%%%%%%%%%%%%%%%%%%%%%%%%%%%%%%%%%%%%%%%%%%%%%%%%%%%%%%%%%%%%%%%%%%%%%%%%%%%%%%%%%%%%%%%%%%%%%%
\subsubsection{%
  On soft two-loop contributions to
  the zero-mode effective potential}
\label{sec:soft:2loop}

\begin{figure}[t!]
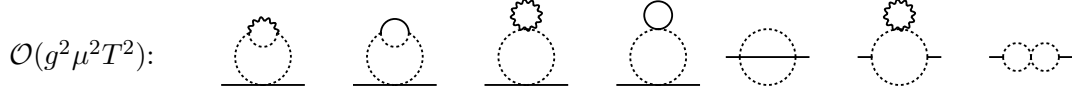

\begin{center}
\begin{tabular}{r@{\qquad}ccccccc}
  $\mathcal{O}(g^2\mu^2T^2)$: &
  $\ToptSTB(\Lsai,\Asrii,\Asrii,\Aglx,\Asrii)$ &
  $\ToptSTB(\Lsai,\Asrii,\Asrii,\Asai,\Asrii)$ &
  $\ToptSTT(\Lsai,\Asrii,\Asrii,\Aglx)$ &
  $\ToptSTT(\Lsai,\Asrii,\Asrii,\Asai)$
  $\ToptSS(\Lsai,\Asrii,\Asrii,\Lsai)$ &
  $\ToptSBT(\Lsai,\Asrii,\Asrii,\Asrii,\Aglx)$ &
  $\ToptSE(\Lsai,\Asrii,\Asrii,\Asrii,\Asrii)$ &
\end{tabular}
\end{center}
\caption{%
  Two-loop soft contributions to the pressure at
  $\mathcal{O}(g^2\mu^2T^2)$.
  Line notation is as in fig.~\ref{fig:1loop:counting}.
}
\label{fig:2loop:topologies}
\end{figure}

Following the power counting of eq.~\eqref{eq:power:counting:fields},
the soft two-loop contributions to the zero-mode effective potential
are of $\mathcal{O}(g^2\mu^2T^2)$
and are therefore, in principle, within the accuracy goal of this work.
These terms are genuinely soft and
require a two-loop computation in the 3d theory.
The procedure is similar to eq.~\eqref{eq:trlog}:
\begin{itemize}
  \item
    shift the fields of eq.~\eqref{L3d} into zero modes and fluctuations,
    $\Psi \to \Psi + \Psibar$ as in eq.~\eqref{decomp},
  \item
    generate the two-loop vacuum diagrams
    from the topologies
    \raisebox{\depth}{\scalebox{0.4}{$\ToptVE(\Asrii,\Asrii)$}}
    and
    \raisebox{\depth}{\scalebox{0.4}{$\ToptVS(\Asrii,\Asrii,\Lsrii)$}}
    using the vertices of the shifted Lagrangian,
  \item
    expand these diagrams to second order in $\Psibar$.
\end{itemize}
The background then enters both through the mass shifts $\Delta$ in the
propagators, as at one loop, and through vertices that are themselves
linear in $\Psibar$, which have no one-loop counterpart.
The resulting topologies are collected in
fig.~\ref{fig:2loop:topologies}.

For the symmetric phase, $\phibar = 0$, this computation was carried
out in~\cite{Bodeker:2015zda} and turned out to be numerically small
compared to the hard $\mathcal{O}(\gs^2)$ corrections.
While the soft sector depends only on
$\muY$ and $\muA$ via eq.~\eqref{eq:pressure:soft},
the neutrality conditions~\eqref{eq:consv_relation}
induce a numerical hierarchy between
hard- and soft-scale chemical potentials
$\muY,\muA < \muB, \muLAlpha$.
At $x \sim 1$,
we find
$(\muY/\muBmL)^2 \approx 0.03$,
which in turn suppresses the soft two-loop contribution to
$\Veff^\text{soft}$ in comparison to
the hard $\mathcal{O}(g^2\mu^2)$ terms.
We therefore refrain from including
the soft two-loop contributions in the numerical evaluation of
sec.~\ref{sec:sphalerons}.

%%%%%%%%%%%%%%%%%%%%%%%%%%%%%%%%%%%%%%%%%%%%%%%%%%%%%%%%%%%%%%%%%%%%%%%%%%%%%%%%%%%%%%%%%%%%%%%%%%%%
\subsection{Equilibrium higher-order corrections}
\label{sec:nb:higher:equilibrium}

The equilibrium baryon number density $\nBeq$,
the two conversion factors $\Cspheq$ and $\Fspheq$, and
the washout coefficient $\kappa$ (cf.\ eq.~\eqref{eq:3muB+summuL})
follow from the zero-mode potential of sec.~\ref{sec:Omega:2loop}
by imposing the conservation laws~\eqref{eq:consv_relation}
together with the equilibrium condition~\eqref{eq:eq:cond}.
These conditions can be solved order by order in
the couplings using the power counting of
eq.~\eqref{eq:power:counting:couplings}.
The different contributions
to the pressure~\eqref{eq:pressure:hard+soft}
are of the orders: 
\begin{itemize}
  \item[LO:]
    \emph{Leading order,}
    $\mathcal{O}(g^0)$.
    \\
    The free-gas terms of eq.~\eqref{eq:p-func},
    the leading-order Debye
    masses~\eqref{eq:mD1:mu0} and~\eqref{eq:mD2:mu0}, and
    the Higgs background term
    $\frac{1}{8}(\muY-\muA)^2\phibar^2$.
  \item[NLO$^\star$:]
    \emph{Charged-lepton Yukawa couplings},
    $\mathcal{O}(\yeAlpha^2\mu^2)$.
    \\
    Constituting the previous state of the art,
    $\nBeqNLOstar$ and
    $\kaNLOstar$ of
    eqs.~\eqref{eq:nBeq:NLOstar}
    and~\eqref{eq:kappa:NLOstar} are obtained via 
    eq.~\eqref{eq:p-func}
    of ref.~\cite{Laine:1999wv}.
  \item[NLO:]
    \emph{Hard one- and two-loop matching},
    $\mathcal{O}(g^2\mu^2)$.
    \\
    Soft tree-level potential
    $\Veff^\text{tree}$~\eqref{eq:veff:tree}
    with the complete set of matching
    coefficients~\eqref{k}--\eqref{eq:mD2:mu}.
    Beyond NLO$^\star$, this order supplies
    $\yt^2$ and $\gs^2$, the two largest
    SM couplings at $T\sim\Tsph$, together with the
    remaining $g_1^2$, $g_2^2$ and $\lambda$ terms.
  \item[N$^2$LO:]
    \emph{Soft one-loop potential},
    $\mathcal{O}(g\mu^2)$.
    \\
    Soft one-loop contributions
    $\Veff^\text{1loop} \supset \eqref{hl}, \eqref{gfl}, \eqref{ml}$.
    While parametrically larger than NLO,
    this order is
    numerically suppressed compared to NLO~\cite{Bodeker:2015zda}.
\end{itemize}

\begin{figure}[t]
\centering
\includegraphics[width=0.5\textwidth]{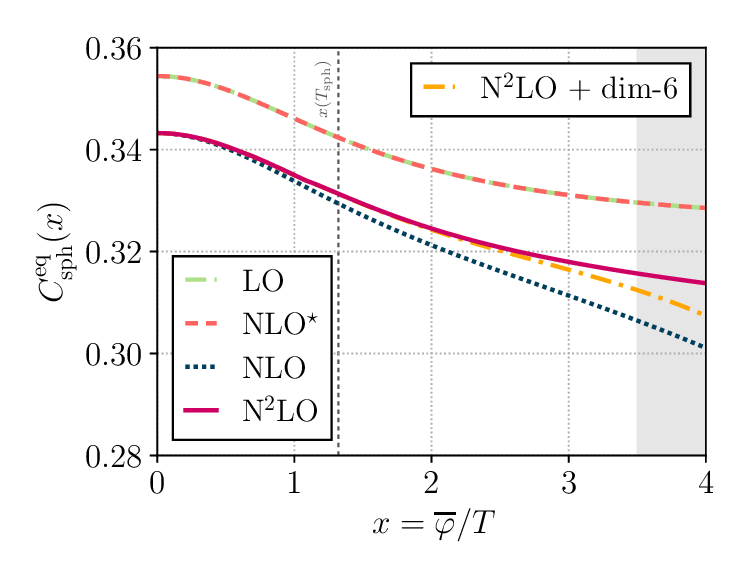}%
\includegraphics[width=0.5\textwidth]{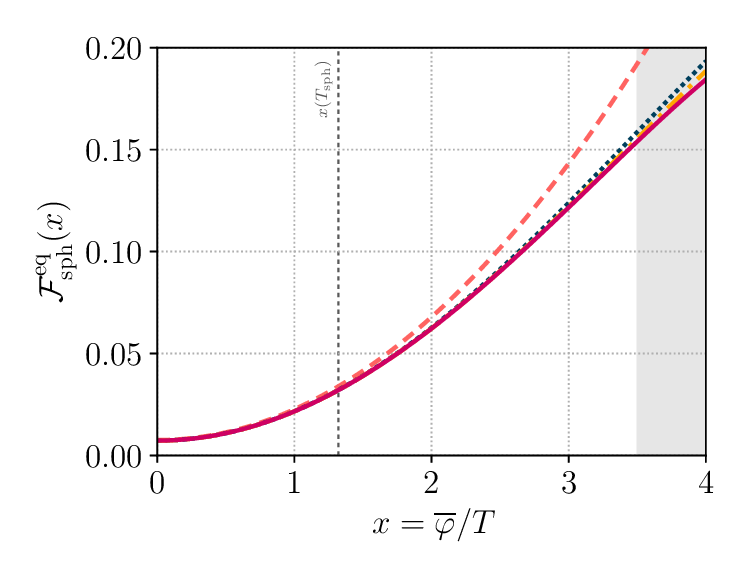}%
\caption{%
  The equilibrium sphaleron conversion factors
  $\Cspheq$~(left) and
  $\Fspheq$~(right)
  as functions of $x = \phibar/T$ at
  $\Tsph\simeq \TsphNum$~GeV.
  The orders LO, NLO$^\star$, NLO and N$^2$LO
  are the ones itemized in sec.~\ref{sec:nb:higher:equilibrium};
  the dimension-six (dim-6) curve adds
  the operators~\eqref{eq:veff:dim6} on top of N$^2$LO.
  The dotted line marks the field value $x(\Tsph)$ at sphaleron freeze-out,
  eq.~\eqref{eq:vT}, and
  the gray band the region where the dimension-six shift of
  $\Cspheq$
  exceeds $1\%$, cf.\ sec.~\ref{sec:EFT:validity}.
  The 4d reference RG scale is fixed at $\LamDRef = \Tbar$
  of eq.~\eqref{eq:RGscale:X}.
  }
\label{fig:Csph:Fsph:x}
\end{figure}

\begin{figure}[t]
\centering
\includegraphics[width=0.5\textwidth]{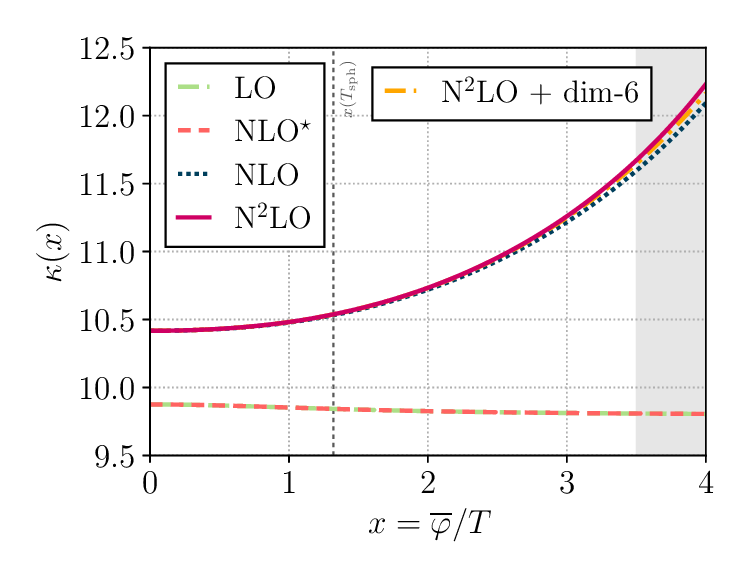}
\caption{%
  Washout coefficient $\kappa$
  as a function of $x = \phibar/T$
  at
  $\Tsph\simeq \TsphNum$~GeV.
  The figure specifications are the same as in fig.~\ref{fig:Csph:Fsph:x}.
  }
\label{fig:kappa:x}
\end{figure}

The full one-loop result is then
numerically computed from
the pressure
$p = p^\rmii{LO}+ \dots + p^\rmii{N$^2$LO}$.
We then further expand
$\mathcal{E} =
    \mathcal{E}^\rmii{LO}
  + \dots
  + \mathcal{E}^\rmii{N$^2$LO}$
for
$\mathcal{E} = \{\nBeq, \Cspheq, \Fspheq, \kappa\}$,
while also expanding in small charged-lepton Yukawa couplings
up to $\mathcal{O}(\yeAlpha^2)$
as in eq.~\eqref{eq:Csph:def}.
We show the impact of different contributions for
$\Cspheq$ and $\Fspheq$ in fig.~\ref{fig:Csph:Fsph:x}
and for $\kappa$ in fig.~\ref{fig:kappa:x}.
At LO, only $\Cspheq$ is non-zero.
At NLO, both $\Cspheq$ and $\Fspheq$ receive corrections, and
the strong coupling $\gs$ dominates,
lowering $\Cspheq$ by $\simeq 4\%$
almost independently of $x = \phibar/T$,
similarly to the top Yukawa with
$\Delta\Cspheq \simeq 0.1\text{--}2\%$.
The $g_1^2$, $g_2^2$ and $\lambda$ terms yield 
$\Delta\Cspheq < 0.2\%$.
The N$^2$LO soft one-loop corrections
remain at $\mathcal{O}(1\text{\textperthousand})$ and become
relevant only at large $x$ where
the high-temperature expansion is invalidated.

By strictly expanding
the full one-loop results for
$\Cspheq$,
$\Fspheq$, and
$\kappa$
also
in powers of the SM couplings~\eqref{eq:power:counting:couplings},
we obtain the corrections up to
$\mathcal{O}(g^2)$,
for the flavor-blind coefficient of
eq.~\eqref{eq:Csph:eq:split} and
for the washout coefficient,
\begin{align}
  \label{eq:nBeq:N2LO}
  \Cspheq(x) &=
        \frac{4 (77 + 27 x^2)}{869 + 333 x^2}
      + \frac{18}{\pi}\frac{\mathcal{S}(x)}{(869 + 333 x^2)^2}
  \nn[2mm] &
    - \frac{1}{(4\pi)^2}
      \frac{1}{(869 + 333 x^2)^2}
  \biggl[
      \frac{4}{3}\bigl(126082 + 328295 x^2 + 182277 x^4 + 30456 x^6\bigr)
      \sum_\alpha \yeAlpha^2
    \nn &
    \hspace{1.5cm}
    + 9\bigl(3630 + 9823 x^2 + 9405 x^4 + 1944 x^6\bigr)\yt^2
    \nn[2mm] &
    \hspace{1.5cm}
    - 9\bigl(38599 + 27346 x^2 + 5373 x^4\bigr)g_1^2
    \nn[2mm] &
    \hspace{1.5cm}
    - 27\bigl(5445 + 3542 x^2 + 467 x^4 - 72 x^6\bigr)g_2^2
    + 432\bigl(2904 + 2365 x^2 + 441 x^4\bigr)\gs^2
    \nn[2mm] &
    \hspace{1.5cm}
    + 144\bigl(242 + 792 x^2 + 141 x^4 + 18 x^6\bigr)\lambda
    \nn[2mm] &
    \hspace{1.5cm}
    - 288\bigl(242 + 66 x^2 + 9 x^4\bigr)\frac{\nu^2}{T^2}
  \biggr]
  \,, \\[3mm]
\label{eq:kappa:N2LO}
  \kappa(x) &=
    \frac{869 + 333 x^2}{2(44 + 17 x^2)}
  + \frac{1}{8\pi}\frac{\mathcal{S}(x)}{(44 + 17 x^2)^2}
  \nn[2mm] &
  + \frac{1}{(4\pi)^2}
    \frac{1}{(44 + 17 x^2)^2}\biggl[
      \frac{2}{9}\bigl(77 + 27 x^2)(418 + 1025 x^2 + 360 x^4\bigr) \sum_\alpha \yeAlpha^2
    \nn &
    \hspace{1.5cm}
    + \frac{1}{4}\bigl(61226 + 168432 x^2 + 100935 x^4 + 17496 x^6\bigr) \yt^2
    \nn &
    \hspace{1.5cm}
    + \frac{1}{8}\bigl(243936 + 185350 x^2 + 35487 x^4\bigr) g_1^2
    \nn &
    \hspace{1.5cm}
    + \frac{1}{8}\bigl(385748 + 294690 x^2 + 56145 x^4 - 108 x^6\bigr) g_2^2
    \nn &
    \hspace{1.5cm}
    + 12\bigl(8833 + 6765 x^2 + 1305 x^4\bigr) \gs^2
    \nn &
    \hspace{1.5cm}
    - \bigl(242 + 792 x^2 + 141 x^4 + 18 x^6\bigr) \lambda
    \nn &
    \hspace{1.5cm}
    + 2\bigl(242 + 66 x^2 + 9 x^4\bigr) \frac{\nu^2}{T^2}
  \biggr]
  \,.
\end{align}
The soft one-loop potential
at N$^2$LO enters both
$\nBeq$ and $\kappa$ through
\begin{align}
\label{eq:S:1loop}
  \mathcal{S}(x) &=
    \frac{
        121\,\widetilde{m}_{\varphi}
      + 3\bigl(121 + 44 x^2 + 6 x^4\bigr)\widetilde{m}_{\rmii{$G$}}
    }{T}
  \nn[2mm] &
  + 121\, x^2
  \Biggl[
     \frac{
       \bigl(
           g_{2,3}^{ }\cos\thetaE
         + g_{1,3}^{ }\sin\thetaE
       \bigr)^2
     }{
         \widetilde{m}_{\rmii{$Z_0$}}
       + \widetilde{m}_{\varphi}
     }
   + \frac{
       \bigl(
           g_{1,3}^{ }\cos\thetaE
         - g_{2,3}^{ }\sin\thetaE
       \bigr)^2
     }{
         \widetilde{m}_{\rmii{$\gamma_0$}}
       + \widetilde{m}_{\varphi}
     }
  \Biggr]
  \nn[2mm] &
  + \frac{x^2\bigl(22 + 3 x^2\bigr)^2}{2}
    \frac{
      g_{2,3}^2
    }{
        \widetilde{m}_{\rmii{$\At_{0}$}}
      + \widetilde{m}_{\rmii{$G$}}
    }
  + 18\, x^4\, \frac{g_{2,3}^{ }\phibarT}{T}
  \,.
\end{align}
Since
$\widetilde{m}_i \sim gT$ and
$g_{i,3}^2 \sim g_i^2 T$,
each term is $\mathcal{O}(g)$, and
it is sufficient to use
leading-order matching.
The flavored $\Fspheq$ receives no correction at this order
in a strict expansion,
since a correction relative to
eq.~\eqref{eq:nBeq:NLOstar} would be
$\mathcal{O}(g^2\yeAlpha^2)$, such that
$\Fspheq = \FspheqNLOstar$.

\begin{figure}[t]
\centering
\includegraphics[width=0.5\textwidth]{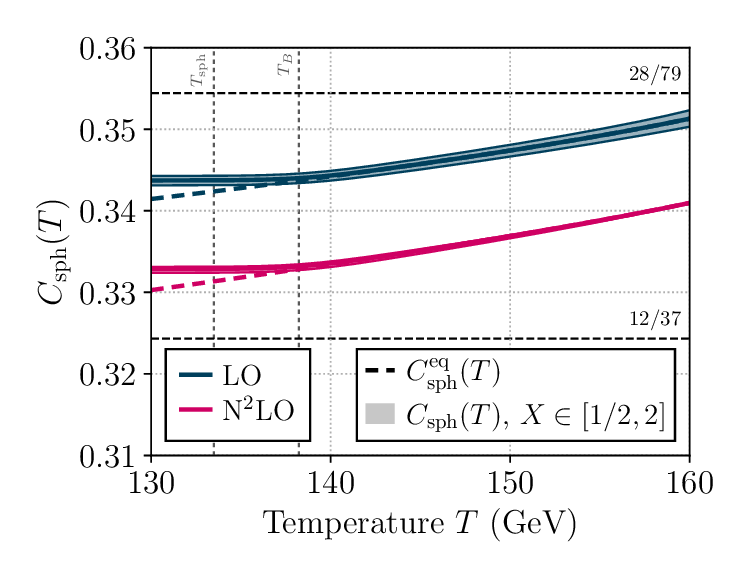}%
\includegraphics[width=0.5\textwidth]{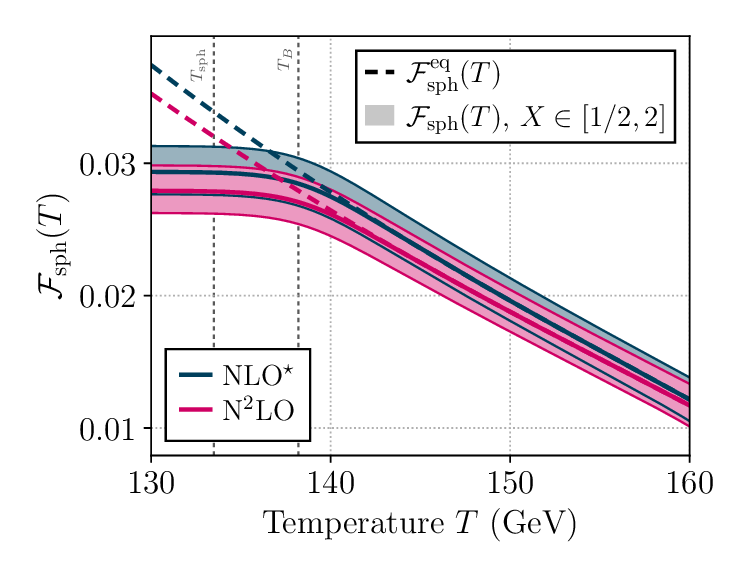}%
\caption{%
  Evolved sphaleron conversion factors
  $\Csph(T)$ (left) at LO and N$^2$LO and
  $\Fsph(T)$ (right) at NLO$^\star$ and N$^2$LO.
  Dashed lines show the equilibrium values
  $\Cspheq$ and $\Fspheq$ of sec.~\ref{sec:nb:higher:equilibrium},
  while solid lines show
  the solutions for $\Csph$ and $\Fsph$ from
  the transport equation~\eqref{eq:baryon:freeze:out}.
  The bands show the variation of the 4d scale
  $\LamDRef$~\eqref{eq:RGscale:X} over $X\in [1/2,2]$.
  The vertical lines mark the
  sphaleron and
  baryon-number freeze-out temperatures
  $\Tsph$ and $\TB$~\eqref{eq:TB};
  the horizontal lines mark
  the standard equilibrium values~\eqref{eq:Csph:original}
  in the symmetric and broken phases.
  Below the crossover,
  the solid curves remain frozen close to the equilibrium values at $\TB$,
  as given by eq.~\eqref{eq:NB:freezeout}.
}
\label{fig:nb:scaleVariation}
\end{figure}
After also using the temperature evolution of
the Higgs expectation value
from eq.~\eqref{eq:vT},
the equilibrium conversion factors are
displayed by the dashed lines in
fig.~\ref{fig:nb:scaleVariation}.
Higher-order corrections lower both factors
across the entire temperature range.
Evaluated at
the sphaleron freeze-out temperature $\Tsph$, where $x(\Tsph) \simeq 1.32$, and at
the baryon-number freeze-out temperature $\TB$, where $x(\TB) \simeq 1.20$,
they take the values%
\begin{align}
\label{eq:Csph:eq:values}
  \Cspheq(\Tsph) &= \CspheqTsphNum
  \,,&
  \Fspheq(\Tsph) &= \FspheqTsphNum
  \,,\\[2mm]
\label{eq:Csph:eq:values:B}
  \Cspheq(\TB) &= \CspheqTBNum
  \,,&
  \Fspheq(\TB) &= \FspheqTBNum
  \,,
\end{align}
at the central value $X=1$ of eq.~\eqref{eq:RGscale:X}.
The baryon-number freeze-out temperature $\TB$
is determined in appendix~\ref{sec:formal:solution}.

Throughout, the \MSbar{} parameters are evolved from
the electroweak input scale $\LamDInput = \mZ$ to
the thermal reference scale
$\LamDRef = X\Tbar$ with
$\Tbar = 4\pi e^{-\gammaE} T$ of eq.~\eqref{eq:RGscale:X},
fixed at $X = 1$ unless stated otherwise.
This setup is detailed in
appendix~\ref{sec:renorm:beta} 
together with 
the referenced SM $\beta$-functions.

%%%%%%%%%%%%%%%%%%%%%%%%%%%%%%%%%%%%%%%%%%%%%%%%%%%%%%%%%%%%%%%%%%%%%%%%%%%%%%%%%%%%%%%%%%%%%%%%%%%%
\subsection{Validity of the EFT}
\label{sec:EFT:validity}

At large field values $x = \phibar/T$,
the expressions of
$\Cspheq$ in eq.~\eqref{eq:nBeq:N2LO} and
$\kappa$ in eq.~\eqref{eq:kappa:N2LO}
scale as 
$\mathcal{O}(x^2)$,
similarly to
$\Fspheq$ in eq.~\eqref{eq:nBeq:NLOstar}
at $\mathcal{O}(\yeAlpha^2)$.
This indicates a
breakdown of the high-temperature expansion
due to neglected higher-dimensional operators.

The dimension-five and -six operators within
the 3d EFT of the SM
are listed in~\cite{Chala:2025aiz,Chakrabortty:2026swu,Bernardo:2026nyq}.
In the 3d Lagrangian~\eqref{L3d},
the following operators enter at
$\mathcal{O}(\mu^2)$,
\begin{align}
\label{eq:L3d:dim5}
  \mathcal{L}_\text{3d}^\text{dim-5} &\supset
    d_{1}^{ }\,
      \bigl(\Phi^\dagger\Phi\bigr)^2 \Bt_0^{ }
  + d_{2}^{ }\,
      \bigl(\Phi^\dagger\Phi\bigr)
      \bigl(\Phi^\dagger\sigmaPauli^a\Phi\bigr)
      \At_0^a
  \,,\\[1mm]
\label{eq:L3d:dim6}
  \mathcal{L}_\text{3d}^\text{dim-6} &\supset
    c_{1}^{ }\,
      \bigl(\Phi^\dagger\Phi\bigr)^2 \Bt_0^2
  + c_{2}^{ }\,
      \bigl(\Phi^\dagger\Phi\bigr)^2 \At_0^a \At_0^a
  \nn &
  + c_{3}^{ }\,
      \bigl(\Phi^\dagger\sigmaPauli^a\Phi\bigr)
      \bigl(\Phi^\dagger\sigmaPauli^b\Phi\bigr)
      \At_0^a \At_0^b
  + c_{4}^{ }\,
      \bigl(\Phi^\dagger\sigmaPauli^a\Phi\bigr)
      \bigl(\Phi^\dagger\Phi\bigr)
      \At_0^a \Bt_0^{ }
  \,,
\end{align}
with
$d_i^{ } \sim g^5\mu/\sqrt{T}$ and
$c_i^{ } \sim g^6$
in the 3d normalization of eq.~\eqref{L3d}.
Inserting the background gives, in the notation of
eq.~\eqref{eq:veff:tree},
\begin{align}
\label{eq:veff:dim5}
  \Veff^\text{dim-5} &=
  \frac{\phibarT^4}{4}
  \Bigl[
      d_{1}^{ }\Btbg_0^{ }
    - d_{2}^{ }\Atbgt_0
  \Bigr]
  \,,\\
\label{eq:veff:dim6}
  \Veff^\text{dim-6} &=
  \frac{\phibarT^4}{4}
  \Bigl[
      c_{1}^{ }\Btbg_0^2
    + \bigl(c_{2}^{ }+c_{3}^{ }\bigr)\bigl(\Atbgt_0\bigr)^2
    - c_{4}^{ }\Atbgt_0\Btbg_0^{ }
  \Bigr]
  \,,
\end{align}
which are of
$\mathcal{O}(g^4\mu^2)$ and
parametrically beyond eq.~\eqref{eq:veff:tree}.

These contributions
can still become relevant numerically at large field values.
Taking the dimension-six operators alone,
the induced shift of $\Cspheq$ stays below
$0.1\%$ at $x \lesssim 2$ and reaches
$1\%$ only at $x \simeq 3.5$,
which we take as the domain of validity of
eqs.~\eqref{eq:nBeq:N2LO} and~\eqref{eq:kappa:N2LO}
and indicate by the gray band in fig.~\ref{fig:Csph:Fsph:x}.
The corresponding shift of $\kappa$ stays below $1\%$
over the entire range displayed there;
see fig.~\ref{fig:kappa:x}.
A definite bound on higher-dimensional operator effects
would require
the matching of $c_{i}^{ },d_i^{ }$ at finite chemical potential.
This is left for future work, which will extend
{\tt DRalgo}~\cite{Ekstedt:2022bff} to finite-$\mu$.

%%%%%%%%%%%%%%%%%%%%%%%%%%%%%%%%%%%%%%%%%%%%%%%%%%%%%%%%%%%%%%%%%%%%%%%%%%%%%%%
\section{Baryon number from weak sphalerons}
\label{sec:sphalerons}

%%%%%%%%%%%%%%%%%%%%%%%%%%%%%%%%%%%%%%%%%%%%%%%%%%%%%%%%%%%%%%%%%%%%%%%%%%%%%%%%%%%%%%%%%%%%%%%%%%%%
\subsection{Sphaleron conversion out of equilibrium}
\label{sec:sphaleron:outOfEq}

One can now solve the transport
equation~\eqref{eq:baryon:freeze:out} for
the baryon-number freeze-out by 
using the expressions for
$\Cspheq$,
$\Fspheq$, and
$\kappa(T)$
at the different perturbative orders in the EFT.
Here, we use the updated parameterization of
the sphaleron diffusion rate~\eqref{fitnew}.

Since the transport equation~\eqref{eq:baryon:freeze:out}
is linear in $\nB$, and
only the source $\nBeq(T)$ depends on the conserved charges $\nal$,
we can directly solve
for $\Csph(T)$ and $\Fsph(T)$.
Hence,
the general solution of the transport equation~\eqref{eq:baryon:freeze:out}
can be written as
\begin{align}
\label{eq:Csph:out:split}
  \nB^{ }(T) &=
    \Csph^{ }(T)\, \nBmL
  + \Fsph^{ }(T) \sum_\alpha \yeAlpha^2 \nal^{ }
  \,,
\end{align}
which generalizes the equilibrium expression for
$\nBeq$~\eqref{eq:Csph:eq:split}.
Deviations from the two-term form are
$\mathcal{O}(\yeAlpha^4)$, inherited from
eq.~\eqref{eq:Csph:eq:split}.
The evolution equations for the two conversion factors
are
\begin{align}
\label{eq:Csph:evolution}
    \partial_t^{ } \Csph^{ }(T)
  + \GammaB^{ } (\phibar/T) \bigl[ \Csph^{ }(T) - \Cspheq(T) \bigr] &= 0
  \,,\\[2mm]
\label{eq:Fsph:evolution}
    \partial_t^{ } \Fsph^{ }(T)
  + \GammaB^{ } (\phibar/T) \bigl[ \Fsph^{ }(T) - \Fspheq(T) \bigr] &= 0
  \,,
\end{align}
which we solve numerically from
the equilibrium initial conditions
$\Csph = \Cspheq$ and
$\Fsph = \Fspheq$
imposed at $\Tc$.
The charged-lepton Yukawa couplings $\yeAlpha$ in
eqs.~\eqref{eq:Csph:eq:split} and~\eqref{eq:Csph:out:split} are
the \MSbar{} couplings at the reference scale
$\LamDRef$ of eq.~\eqref{eq:RGscale:X}.

To evaluate precision values for
$\Csph(T)$ and $\Fsph(T)$,
we first have to consider two sources of uncertainty in the computation:
\begin{itemize}
  \item
    RG-scale dependence.\\
    In the previous discussion,
    the RG scale was kept fixed at
    $\LamDRef = X \Tbar$ with
    $X = 1$.
    However, we will now use it as a proxy to estimate the importance
    of higher-order corrections in the computation and vary it between
    $X\in [1/2,2]$.
    The impact of this scale variation
    on the equilibrium baryon number density
    is shown in fig.~\ref{fig:nb:scaleVariation}.
  \item
    Uncertainty of $\Gammaws$ in eq.~\eqref{eq:Gamma:newfit:params}.\\
    We propagate the uncertainties of
    the sphaleron diffusion rate~\eqref{fitnew}
    of ref.~\cite{Annala:2023jvr}
    to the final results for
    $\Csph(T)$ and $\Fsph(T)$.
    For each of the three choices of the 4d RG scale
    $X = \{1/2,1,2\}$,
    we Monte Carlo sample $\Nsamp = 10^4$ random values of the fit parameters
    jointly from the multivariate normal distribution defined by
    $\mu$ and $\text{Cov}$ of eq.~\eqref{eq:Gamma:newfit:params}
    and then solve the transport equation~\eqref{eq:baryon:freeze:out}
    for each draw.
\end{itemize}

\begin{figure}[t]
\centering
\includegraphics[width=0.5\textwidth]{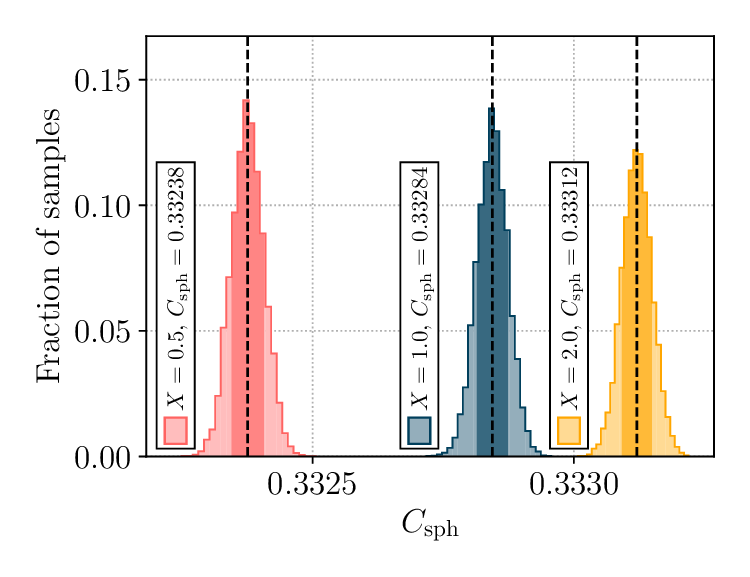}%
\includegraphics[width=0.5\textwidth]{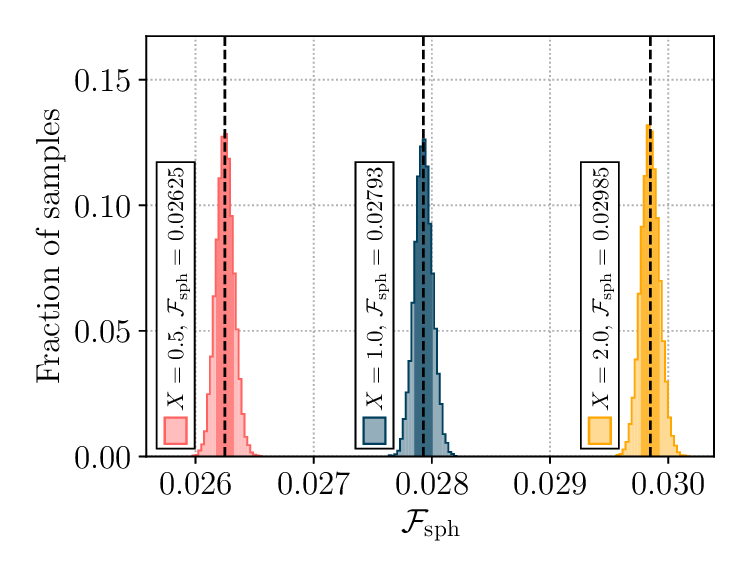}%
\caption{%
  Distributions of the evolved sphaleron conversion factors
  $\Csph$~(left) and $\Fsph$~(right)
  at $\Tasym < \Tsph$ for $X = \{1/2,1,2\}$,
  each obtained from
  $\Nsamp=10^4$ Monte Carlo draws of $\Gammaws$~\eqref{fitnew},
  sampled from its covariance matrix~\eqref{eq:Gamma:newfit:params}.
  The dark bands mark the $1\sigma$ intervals listed in
  tab.~\ref{tab:Csph:Fsph:sampling},
  showing that
  the scale variation dominates over the $\Gammaws$-uncertainty.
}
\label{fig:nb:sampling}
\end{figure}

\begin{table}[t]
\centering
\renewcommand{\arraystretch}{1.1}
\begin{tabular}{|c|cc|cc|}
  \hline
  $X$ &
  $\Csph$ &
  $\delta_{\rmii{$X$}}^{ }\Csph$ &
  $\Fsph$ &
  $\delta_{\rmii{$X$}}^{ }\Fsph$
  \\
  \hline
  \hline
  $0.5$
  & $ 0.332376 \pm 0.000031$ & $ -0.000468$
  & $ 0.026249 \pm 0.000075$ & $ -0.001680$
  \\
  $1.0$
  & $ 0.332844 \pm 0.000030$ & ---
  & $ 0.027929 \pm 0.000077$ & ---
  \\
  $2.0$
  & $ 0.333121 \pm 0.000029$ & $ +0.000277$
  & $ 0.029848 \pm 0.000079$ & $ +0.001919$
  \\
  \hline
\end{tabular}
\caption[]{%
  Evolved sphaleron conversion factors
  $\Csph$ and $\Fsph$
  at asymptotic $\Tasym < \Tsph$, computed at N$^2$LO
  for the three choices of the 4d RG scale
  $X = \{1/2,1,2\}$ of eq.~\eqref{eq:RGscale:X}.
  The entries for $\Csph$ and $\Fsph$ give the means and
  standard deviations obtained from $\Nsamp=10^4$ Monte Carlo
  samples of
  the sphaleron rate~\eqref{fitnew} shown in fig.~\ref{fig:nb:sampling}.
  The columns $\delta_{\rmii{$X$}}^{ }$ list
  the shifts relative to the $(X=1)$-scale.
  For both $\Csph$ and $\Fsph$, the RG-scale variation exceeds
  the uncertainty of $\Gammaws$ by an order
  of magnitude and dominates
  the uncertainties in eq.~\eqref{eq:Csph:out:values}.
  }
\label{tab:Csph:Fsph:sampling}
\end{table}
Combining the results from all three scale choices,
we obtain a total of
$3\Nsamp$ values of
$\Csph$ and $\Fsph$,
summarized in tab.~\ref{tab:Csph:Fsph:sampling}.
Both coefficients are determined at
an asymptotically low temperature
$\Tasym < \Tsph$,
where the transport equation~\eqref{eq:baryon:freeze:out} has
reached its asymptotic solution.
As seen in fig.~\ref{fig:nb:sampling},
the distributions for each $X$ are approximately Gaussian and
the uncertainty induced by the sphaleron rate is
subdominant compared to the scale variation.
The dominant uncertainty is thus theoretical, and
the residual $\LamDRef$-dependence enters only through
the running couplings and through
the Higgs expectation value $\phibar(T)$ of eq.~\eqref{eq:vT}.
The relative scale variation of $\Fsph$ in tab.~\ref{tab:Csph:Fsph:sampling}
exceeds that of $\Csph$ by a factor of $\sim 45$.
Since $\Fsph \sim \mathcal{O}(g^0)$,
its $\LamDRef$-dependence propagates almost entirely through $x(T)$.
Extending the computation of $\phibar(T)$ to
N$^3$LO would reduce the scale dependence of $\Fsph$,
but is beyond the scope of this work.

The evolved sphaleron conversion factors
at asymptotic $\Tasym < \Tsph$,
\begin{align}
\label{eq:Csph:out:values}
  \Csph^{ } &=
    \CsphNum \pm \CsphErrNum
  \,,&
  \Fsph &=
    \FsphNum \pm \FsphErrNum
  \,,
\end{align}
are our main results.
The errors in eq.~\eqref{eq:Csph:out:values} reflect
the theoretical uncertainty due to
the $\LamDRef$-dependence.
We report the largest values of
$\left|\delta_\rmii{$X$}\right|$ that we find for
$X = \{1/2,1,2\}$ in tab.~\ref{tab:Csph:Fsph:sampling}.
While at the sphaleron freeze-out temperature $\Tsph$,
$\Csph^{ } > \Cspheq(\Tsph)$ and
$\Fsph^{ } < \Fspheq(\Tsph)$,
the same solutions can be obtained
in equilibrium at the
baryon-number freeze-out temperature $\TB$
determined in appendix~\ref{sec:formal:solution},
yielding the values~\eqref{eq:Csph:eq:values:B}.
The equilibrium values agree with
the evolved ones within their uncertainties,
as anticipated by eq.~\eqref{eq:NB:freezeout}.

%%%%%%%%%%%%%%%%%%%%%%%%%%%%%%%%%%%%%%%%%%%%%%%%%%%%%%%%%%%%%%%%%%%%%%%%%%%%%%%%%%%%%%%%%%%%%%%%%%%%
%
\subsection[%
  Flavor effects and baryogenesis at finite \texorpdfstring{$B-L \neq 0$}{B-L != 0}]{%
  Flavor effects and baryogenesis at finite \texorpdfstring{\boldmath{$B-L\neq 0$}}{B-L != 0}%
}
\label{sec:flavor:effects}

Different initial flavor asymmetries
can affect the distributions of
the conserved charges
$n_{\Delta_\alpha}$ and consequently
the final baryon number density
after freeze-out
at $\Tasym < \Tsph$.
Similar ideas have been put forward
in~\cite{%
  Abada:2006ea,AristizabalSierra:2009bh,
  Dev:2017trv,Mukaida:2021sgv}.

\begin{figure}[t]
\centering
\includegraphics[width=0.6\textwidth]{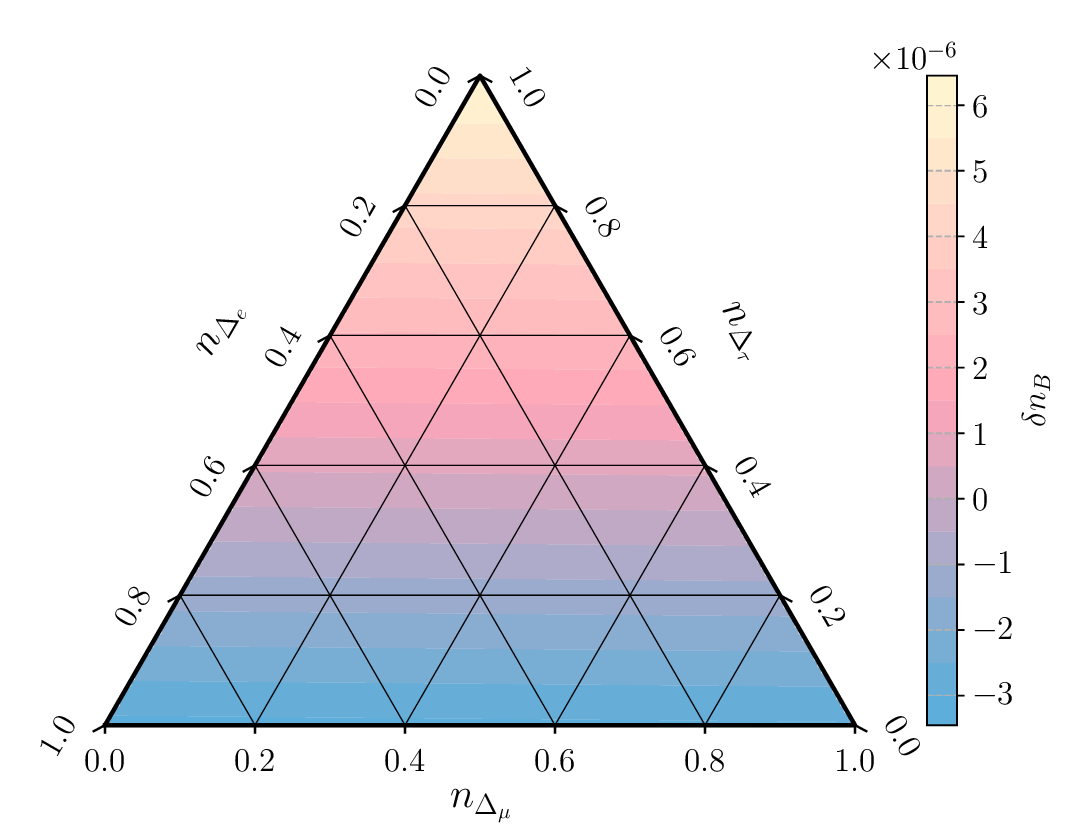}%
\caption{%
  Relative differences of
  the baryon number density
  $\delta\nB(\vec{n}_\Delta) = (\nB(\vec{n}_\Delta)-\nB^\text{ref})/\nB^\text{ref}$
  from the barycentric
  $\nB^\text{ref} = \nB^{ }(\nal=1/3)$
  at asymptotic $\Tasym < \Tsph$
  for different
  $\vec{n}_\Delta = (n_{\Delta_e}, n_{\Delta_\mu}, n_{\Delta_\tau})$
  with the constraint that
  $n_{\Delta_e} + n_{\Delta_\mu} + n_{\Delta_\tau} = \nBmL$
  is fixed at
  $\nBmL = 1$.
}
\label{fig:nb:flavorVariation}
\end{figure}
In the triangle plot in fig.~\ref{fig:nb:flavorVariation},
we show the baryon number density
after freeze-out
for different
$\vec{n}_\Delta = (n_{\Delta_e}, n_{\Delta_\mu}, n_{\Delta_\tau})$
with the constraints that
\begin{align}
  \label{eq:nBmL:constraint}
  \nBmL &= n_{\Delta_e} + n_{\Delta_\mu} + n_{\Delta_\tau} = 1
  \,,&
  \nal &> 0
  \,,
\end{align} 
where the value $\nBmL = 1$ merely defines the unit of charge;
we are not interested in physical scenarios with extremely large charges
or chemical potentials. 
The different choices
of $\vec{n}_\Delta$ lead to variations of up to
$\delta\nB(\vec{n}_\Delta) = (\nB(\vec{n}_\Delta)-\nB^\text{ref})/\nB^\text{ref}$
$\sim \mathcal{O}(10^{-6})$
in the final baryon number density
relative to the barycentric reference
$\nB^\text{ref} = \nB^{ }(\nal=1/3)$.
Such a variation is well below the error of our calculation.
Given the constraints in eq.~\eqref{eq:nBmL:constraint} and
the hierarchy of the Yukawa couplings,% 
\footnote{%
  For \MSbar{} couplings at $\LamDRef = \Tbar$, where
  the ratios are scale-independent to this accuracy.
}
\begin{align}
\label{eq:yukawa:hierarchy}
  h_e^2 : h_\mu^2 : h_\tau^2 &\simeq
  8\times10^{-8} :
  3.5\times10^{-3} :
  1
  \,,
\end{align}
the maximal variation of
$\nB$ in fig.~\ref{fig:nb:flavorVariation}
is bounded by the flavored term in
eq.~\eqref{eq:Csph:out:split} as
\begin{align}
\label{eq:nB:flavor:bound}
  \delta\nB^{ }\bigr|_\text{max} &=
    \nB^\text{max} - \nB^\text{min}
    \leq
    \Fsph \bigl( h_{\tau}^2 - h_{e}^2 \bigr) n_{\Delta}
    \simeq 3\times 10^{-6}
  \,,
\end{align}
for $n_\Delta = \nBmL$.
Flavor effects therefore only become
relevant once
the total $B-L$ is suppressed relative to
the individual flavor asymmetries,
\begin{align}
\label{eq:nBmL:threshold}
  \frac{\nBmL}{\max_\alpha \bigl| \nal^{ } \bigr|} &\lesssim
    \frac{\Fsph h_{\tau}^2}{\Csph}
  \simeq 10^{-5}
  \,.
\end{align}
We discuss
the limiting case $\nBmL = 0$
in sec.~\ref{sec:flavor:effects:B-L:zero}.

\subsection[%
  Flavor effects and baryogenesis from vanishing \texorpdfstring{$B-L = 0$}{B-L = 0}]{%
  Flavor effects and baryogenesis from vanishing \texorpdfstring{\boldmath{$B-L = 0$}}{B-L = 0}%
}
\label{sec:flavor:effects:B-L:zero}

Finally, we investigate
another scenario where
the lepton flavor asymmetries sum to
a vanishing total $B-L$ asymmetry, but
some individual flavored asymmetries remain finite,
{\em viz.},
\begin{align}
  \nBmL = n_{\Delta_e} + n_{\Delta_\mu} + n_{\Delta_\tau} &= 0
  \,,&
  \max_\alpha \bigl| \nal^{ } \bigr| &\neq 0
  \,.
\end{align}
In this case,
no net baryon or lepton asymmetry would be generated
at LO.
However, non-zero asymmetries arise at
NLO$^\star$ due to
the flavor-diagonal contributions of the form
$\yeAlpha^2 \nal$ in $\nBeq$;
see eqs.~\eqref{eq:Csph:eq:split} and~\eqref{eq:nBeq:NLOstar}.
This mechanism,
known as leptoflavorgenesis~(LFG),
was studied in~\cite{%
  Kuzmin:1987wn,Khlebnikov:1988sr,March-Russell:1999hpw,
  Laine:1999wv,Shu:2006mm,Gu:2010dg,Mukaida:2021sgv,Akita:2025zvq
  }.
Here we present a quantitative analysis.

The flavored asymmetries enter only
weighted by the squared charged-lepton Yukawa couplings,
which are strongly hierarchical
as shown in eq.~\eqref{eq:yukawa:hierarchy}.
The relevant distinction is therefore
whether the $\tau$ flavor carries an asymmetry or not, and
we consider the two corresponding benchmark scenarios:
\begin{itemize}
  \item[(i)]
  $\tau$-philic case ($\tau$).\\
  $\tau$ carries the largest asymmetry, and
  the dominant source term scales as $+\,h_\tau^2\,n_\Delta$:
  \begin{align*}
    n_{\Delta_\tau} &= n_\Delta, &
    n_{\Delta_e} = n_{\Delta_\mu} &= -\frac{1}{2}n_\Delta \;.
  \end{align*} 
  \item[(ii)]
  $\tau$-phobic case ($\msl\tau$).\\
  No asymmetry is stored in $\tau$, and
  the dominant source term scales as $+\,h_\mu^2\,n_\Delta$:
  \begin{align*}
    n_{\Delta_\tau} &= 0, &
    n_{\Delta_e} = -n_{\Delta_\mu} &= -n_\Delta \;.
  \end{align*}
\end{itemize}

Since $\nBmL = 0$,
only the flavor-diagonal contributions
remain in eq.~\eqref{eq:Csph:eq:split}, and
the baryon asymmetry can be approximated by
evaluating
$\nBeq \sim \Fspheq \sum_\alpha \yeAlpha^2 \nal^{ }$
at the sphaleron freeze-out temperature $\Tsph$,
where $\Fspheq(\Tsph) \simeq \FspheqTsphNum$.
In the $\tau$-philic case,
the $\tau$ contribution dominates, whereas
in the $\tau$-phobic case,
the $\mu$ contribution does,
\begin{align}
\label{eq:nB:LFG:estimate}
  \frac{\nB}{s} &\sim
    10^{-10}
    \biggl(\frac{h_\tau^2}{10^{-4}}\biggr)
    \biggl(\frac{n_{\Delta_\tau}/s}{3\times10^{-5}}\biggr)
  \,,&
  \frac{\nB}{s} &\sim
    10^{-10}
    \biggl(\frac{h_\mu^2}{3.5\times10^{-7}}\biggr)
    \biggl(\frac{n_{\Delta_\mu}/s}{10^{-2}}\biggr)
  \,.
\end{align}
Hence, 
the observed asymmetry requires
$n_{\Delta_\tau}/s \sim 3\times10^{-5}$ and
$n_{\Delta_\mu}/s \sim 10^{-2}$.

In both cases,
the baryon number as computed from
eq.~\eqref{eq:Csph:out:split}
reduces to its flavored term.
Leptoflavorgenesis is therefore governed by
$\Fsph$ and the Yukawa couplings,
\begin{align}
\label{eq:nBeq:LFG}
  \nB^{ }\big|_{\nBmL = 0} &=
    \Fsph \sum_\alpha \yeAlpha^2 \nal^{ }
  \,.
\end{align}
The coefficient
$\Fsph = \FsphNumErr$
of eq.~\eqref{eq:Csph:out:values}
is the corresponding solution of
eq.~\eqref{eq:Fsph:evolution} and
is independent of the flavor decomposition.

\begin{figure}[t]
\centering
\includegraphics[width=0.5\textwidth]{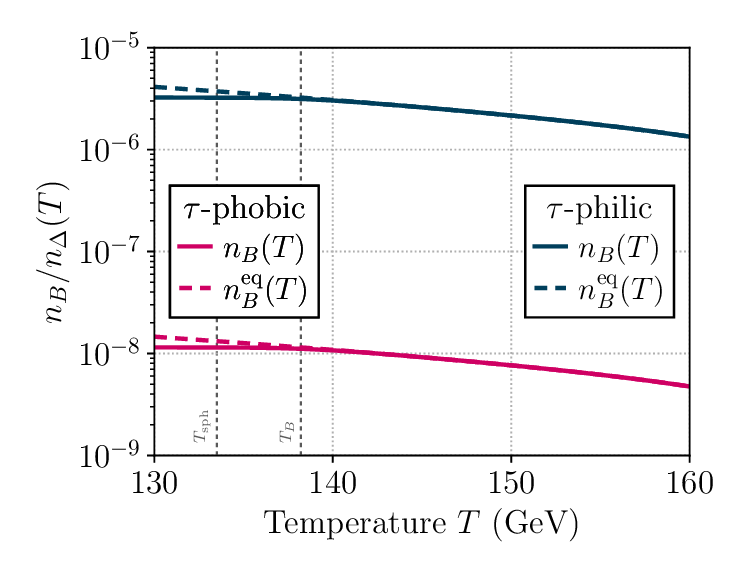}%
\caption{%
  Normalized baryon number density as a function of temperature
  for the two benchmark scenarios of leptoflavorgenesis
  with vanishing total $B-L$ asymmetry,
  the $\tau$-philic case (dark blue) and
  the $\tau$-phobic case (magenta).
}
\label{fig:nb:LFG}
\end{figure}

Using
$\Fsph$ rather than
$\Fspheq(\Tsph)$
for the two benchmarks,
\begin{align}
\label{eq:nB:LFG:values}
  \frac{\nB}{n_{\Delta_\tau}}\bigg|_{\tau} &=
    \Fsph
    \Bigl[ h_\tau^2 - \frac{1}{2}\bigl( h_\mu^2 + h_e^2 \bigr) \Bigr]
  \,,&
  \frac{\nB}{n_{\Delta_\mu}}\bigg|_{\msl{\tau}} &=
    \Fsph
    \Bigl[ h_\mu^2 - h_e^2 \Bigr]
  \,,
\end{align}
decreases the baryon asymmetry by
$\delta\Fsph = 1 - \Fsph/\Fspheq(\Tsph) \simeq 13\%$.
The corresponding evolved left-hand side of
eq.~\eqref{eq:nB:LFG:values}
is shown in fig.~\ref{fig:nb:LFG}.
Inverting eq.~\eqref{eq:nB:LFG:values} yields 
the flavor asymmetry required to reproduce
the observed baryon asymmetry
$\nB/s\big|_\text{obs} =
(\nBobsNum \pm \nBobsErrNum)\times10^{-11}$~\cite{Planck:2015fie,ParticleDataGroup:2026mpi},
\begin{align}
\label{eq:nDelta:LFG:required}
  \frac{n_{\Delta_\tau}}{s}\bigg|_{\tau} &=
    \bigl(3.1 \pm 0.2\bigr)\times10^{-5}
    \biggl(\frac{10^{-4}}{h_\tau^2}\biggr)
  \,,&
  \frac{n_{\Delta_\mu}}{s}\bigg|_{\msl{\tau}} &=
    \bigl(8.9 \pm 0.6\bigr)\times10^{-3}
    \biggl(\frac{3.5\times10^{-7}}{h_\mu^2}\biggr)
  \,.
\end{align}
Due to $\delta\Fsph$,
the estimates~\eqref{eq:nDelta:LFG:required} exceed
the ones of eq.~\eqref{eq:nB:LFG:estimate}
by $\mathcal{O}(10\%)$, and
show a $\simeq 7\%$ uncertainty
from the RG-scale variation of $\Fsph$,
in agreement with~\cite{Mukaida:2021sgv,Akita:2025zvq}.

Finally, we comment on the impact of such flavor asymmetries. 
While the asymmetry required in
the $\tau$-philic case,
$n_{\Delta_\tau}/s \sim 3\times10^{-5}$,
is too small to have significant consequences for
other cosmological phenomena, 
the larger asymmetry required in
the $\tau$-phobic case,
$n_{\Delta_\mu}/s \sim 10^{-2}$,
can lead to several interesting effects. 

First, if such an asymmetry is generated at high temperatures,
$T\gtrsim 10^2$~TeV,
it can typically trigger
the chiral plasma instability~\cite{%
  Joyce:1997uy,Akamatsu:2013pjd,Rogachevskii:2017uyc,
  Schober:2017cdw},
resulting in the generation of helical hypermagnetic fields. 
These helical hypermagnetic fields may in turn generate
an additional baryon asymmetry at
the electroweak crossover~\cite{Giovannini:1997eg,Kamada:2016cnb},
potentially leading to baryon overproduction~\cite{Kamada:2018tcs,Domcke:2022uue};
see the $\SB^{ }$ source term
in eq.~\eqref{eq:baryon:freeze:out}.
However,
the magnitude of the resulting baryon asymmetry generated through
this mechanism remains under debate~\cite{%
  Hamada:2025cwu,Hamada:2025ooy,Fukuda:2025nmc}. 
In the most optimistic scenario,
no significant additional baryon asymmetry is produced,
while primordial magnetic fields survive until today
as intergalactic magnetic fields~\cite{%
  Neronov:2010gir,Tavecchio:2010mk,Dolag:2010ni,Fermi-LAT:2018jdy,
  MAGIC:2022piy,Durrer:2013pga}.

Second,
a large electron neutrino asymmetry can affect
Big Bang nucleosynthesis (BBN)
by modifying the primordial helium abundance. 
In the case of vanishing total asymmetry,
lepton flavor asymmetries tend to be partially washed out 
by neutrino oscillations.
However, the washout is not complete, and
a finite fraction of flavor asymmetry can survive
until the BBN epoch, typically
$n_{\nu_e}^\rmii{BBN} \sim 0.1 \times n_\Delta$~\cite{%
  Dolgov:2002ab,Pastor:2008ti,Mangano:2010ei,Castorina:2012md,
  Froustey:2021azz,Froustey:2024mgf,Domcke:2025lzg}. 
The current observational constraint on the deviation of the helium abundance 
from the standard BBN prediction can be translated into
$|n_{\nu_e}^\rmii{BBN}/s| \lesssim 10^{-3}$~\cite{%
  ParticleDataGroup:2026mpi,Aver:2026dxv}.
Therefore,
$\tau$-phobic leptoflavorgenesis
is compatible with this constraint,
while still predicting a potentially observable
deviation in the primordial helium abundance.
Through
the precise requirement~\eqref{eq:nDelta:LFG:required}
and
$n_{\nu_e}^\rmii{BBN}/s \sim 0.9\times10^{-3}$,
the $\tau$-phobic benchmark 
is just below the current observational constraint.
Interestingly, recent analyses beyond
the $\Lambda$CDM framework of cosmic 
microwave background observations, such as
the Atacama Cosmology Telescope~(ACT)~\cite{%
  AtacamaCosmologyTelescope:2025nti} and
the South Pole Telescope~\cite{SPT-3G:2025bzu}, 
as well as the studies of metal-poor galaxies by
the EMPRESS survey~\cite{Yanagisawa:2025mgx}, 
suggest a relatively low helium abundance compared to
the values of~\cite{ParticleDataGroup:2026mpi} and
also another recent study 
of metal-poor galaxies~\cite{Aver:2026dxv}. 
This may provide a hint of a sizable electron neutrino asymmetry
originating from a large primordial lepton flavor asymmetry of roughly
$n_{\nu_e}^\rmii{BBN}/s \sim 10^{-3}$.

A precise identification of the parameter space of
the primordial lepton flavor asymmetry that explains
the present baryon asymmetry of the Universe
through leptoflavorgenesis
and predicts the resulting helium abundance
would give a target for future observations
of primordial helium.

%%%%%%%%%%%%%%%%%%%%%%%%%%%%%%%%%%%%%%%%%%%%%%%%%%%%%%%%%%%%%%%%%%%%%%%%%%%%%%%%%%%%%%%%%%%%%%%%%%%%
\section{Conclusions}
\label{sec:conclusions}

We computed the baryon-number freeze-out in
the Standard Model across the electroweak crossover
at a new precision level,
accounting for the fact that both
the Higgs expectation value and
the sphaleron rate vary as functions of temperature during the crossover. 

We derived the transport equation governing
the time evolution of the baryon number from
first principles using linear-response theory.
To this end,
we computed
the temperature-dependent equilibrium baryon number density and
the washout coefficient within
the dimensionally reduced 3d effective theory,
including higher-order corrections from the hard and soft momentum scales.
These corrections consist of
the one-loop corrections in the effective potential and
the contributions from the
strong,
top, and
charged-lepton Yukawa couplings.

Solving the transport equation with
updated lattice results of the sphaleron diffusion rate,
we determined the baryon number after freeze-out.
The resulting baryon number
is a linear function of the conserved charges, and
the charged-lepton Yukawa couplings are
the only source of flavor dependence in the Standard Model.
As a result,
the baryon number depends on
two constants,
the flavor-blind $\Csph$ and
the flavored $\Fsph$ of eq.~\eqref{eq:Csph:out:split}.
For these conversion factors,
we determined the most precise values to date,
\begin{align}
\label{eq:Csph:final}
  \Csph^{ } &= \CsphNum \pm \CsphErrNum
  \,,&
  \Fsph^{ } &= \FsphNum \pm \FsphErrNum
  \,,
\end{align}
at asymptotic $\Tasym < \Tsph$,
after sphaleron freeze-out has fully completed.
Deviations from
the two-term form of eq.~\eqref{eq:Csph:out:split}
are $\mathcal{O}(\yeAlpha^4)$.
We also determined
the washout-weighted baryon-number freeze-out temperature
$\TB \simeq \TBNum$~GeV,
which relates the values~\eqref{eq:Csph:final}
to their equilibrium ones,
{\em viz.},
$\Csph = \Cspheq(\TB)$ and
$\Fsph = \Fspheq(\TB)$
within uncertainties.

Our result for $\Csph$ is between
the equilibrium values in
the electroweak-symmetric and electroweak-broken phases,
$\Csph^\text{sym} = 28/79 \simeq 0.3544$
and
$\Csph^\text{bro} = 12/37 \simeq 0.3243$,
and differs from both at the percent level.
Neither limiting value therefore describes
the baryon-number freeze-out in the thermal history of the Universe,
which requires a dynamical treatment across the crossover.

The uncertainties
of the reported sphaleron coefficients are dominated by
the variation of the 4d renormalization scale~\eqref{eq:RGscale:X},
which exceeds the uncertainty propagated from
the fitted sphaleron diffusion rate~\eqref{fitnew}~\cite{Annala:2023jvr}
by more than an order of magnitude.
Further progress therefore depends on
the determination of $\phibar(T)$ and on
higher orders in the soft-scale effective potential and
the hard-to-soft matching relations,
rather than on reducing
the errors of the sphaleron rate fit.

We also examined
the impact of lepton flavor asymmetries
on the baryon number at our updated precision level.
At fixed non-zero total asymmetry,
redistributing it among same-sign flavor charges changes
the final baryon number only at $\mathcal{O}(10^{-6})$,
since flavor enters only via
the small charged-lepton Yukawa couplings.
Conversely,
at vanishing total asymmetry,
the flavored term becomes the only source of
the baryon asymmetry.
In this limit,
leptoflavorgenesis is therefore governed by
$\Fsph \ll 1$, and
the observed asymmetry requires
$n_{\Delta_\tau}/s \simeq 3.1\times10^{-5}$
in the $\tau$-philic case and
$n_{\Delta_\mu}/s \simeq 8.9\times10^{-3}$
in the $\tau$-phobic one.

In the present study,
we do not take into account
the source term $\SB$ coming from
the $\langle \B_{\mu\nu} \widetilde \B^{\mu\nu} \rangle_\rmii{GC}$ term.
It acts identically on $B$ and $L$ and
therefore sources $B+L$,
leaving the conserved $B-L$ charges, and hence
$\nBeq$ and $\kappa$, intact.
It may give a sizable contribution around the electroweak crossover
if there exists a long-range background helical hypermagnetic field
beforehand~\cite{Kamada:2016cnb}.
In this case, another systematic study solving
the transport equation~\eqref{eq:baryon:freeze:out} with
a corresponding source term $\SB$ is necessary.
For this purpose, however,
an improved estimate of $\SB$,
whose overall magnitude remains under
debate~\cite{Hamada:2025cwu,Hamada:2025ooy,Fukuda:2025nmc},
is required.
This is left for future work.

\section*{Acknowledgements}
We thank
Jacopo Ghiglieri,
Philipp Klose,
Kari Rummukainen,
Tuomas V.I.~Tenkanen,
and
Jorinde van de Vis
for useful discussions.
We thank
J.~Annala and
Kari Rummukainen
for providing us with the covariance matrix~\eqref{eq:Gamma:newfit:params}
of the fits from~\cite{Annala:2023jvr}.
PS was supported by
the Swiss National Science Foundation (SNSF) under grant
\href{https://data.snf.ch/grants/grant/215997}{\tt PZ00P2-215997}.
DB, LS, and PS were supported by the Deutsche Forschungsgemeinschaft (DFG, German Research Foundation) through
the CRC-TR 211 ``Strong-interaction matter under extreme conditions'' --
project number 315477589 --
TRR 211.
KK was supported by the National Natural Science Foundation of China (NSFC) under
grant nos.~W2532007 and~12547104, and by
JSPS KAKENHI Grant-in-Aid for Challenging Research (Exploratory) JP23K17687.
KM was supported by JSPS KAKENHI Grant nos.~JP22K14044 and JP26K07096.
KS and LS are members of the NANOGrav Collaboration.
NANOGrav is supported by NSF
Physics Frontier Center award \#2020265.
KS is an affiliate member of the Kavli Institute for the
Physics and Mathematics of the Universe at the University
of Tokyo and supported by the World Premier International
Research Center Initiative (WPI), MEXT, Japan (Kavli IPMU).
\\

\noindent
{\bf Data availability statement.}
Diagrams were generated with {\tt Axodraw}~\cite{Collins:2016aya}.
The effective potential expressions used to determine
the temperature-dependent Higgs expectation value
are publicly available via the software
{\tt DRalgo}~\cite{Ekstedt:2022bff}.

%%%%%%%%%%%%%%%%%%%%%%%%%%%%%%%%%%%%%%%%%%%%%%%%%%%%%%%%%%%%%%%%%%%%%%%%%%%%%%%%%%%%%%%%%%%%%%%%%%%%
\appendix
\renewcommand{\thesection}{\Alph{section}}
\renewcommand{\thesubsection}{\Alph{section}.\arabic{subsection}}
\renewcommand{\theequation}{\Alph{section}.\arabic{equation}}

%%%%%%%%%%%%%%%%%%%%%%%%%%%%%%%%%%%%%%%%%%%%%%%%%%%%%%%%%%%%%%%%%%%%%%%%%%%%%%%%%%%%%%%%%%%%%%%%%%%%
\section{Formal solution of the transport equation}
\label{sec:formal:solution}

In this appendix,
we construct a formal, semi-analytical solution of
the transport equation~\eqref{eq:baryon:freeze:out} and make explicit how
the equilibrium quantities derived in the main text enter the final baryon number.
Our starting point is the transport equation for
the physical baryon number density $\nB$ as a function of time $t$,
\begin{equation}
  \partial_t \nB + 3 H\nB =
      \SB
    - \Gamma_\rmii{$B$} \left(\nB^{ } - \nB^{\rm eq}\right)
  \,.
\end{equation}
The Hubble dilution term $3 H\nB$ is removed by passing to
the comoving baryon number density
$\NB = a^3 \nB$ and using
the number of $e$-folds $x = \ln a$ as the time variable,
\begin{align}
  \frac{\dd \NB}{\dd x} &= F - G \NB
  \,,&
  x &= \ln a
  \,,&
  F &= \frac{a^3 \SB}{H} + G\NB^{\rm eq}
  \,,&
  G &= \frac{\GammaB}{H} 
\,.
\end{align}
This linear first-order equation admits the formal solution
\begin{align}
    \NB\left(x\right) &= \NB\left(x_{\rm ini}\right)E\left(x,x_{\rm ini}\right) 
  + \int_{x_{\rm ini}}^x \dd y\,E\left(x,y\right) F\left(y\right)
  \,,\nn
  E\left(x,y\right) &= \exp\left[-\int_y^x \dd z\, G\left(z\right)\right]
\,,
\end{align}
where the propagator
$E\left(x,y\right)$ measures
the sphaleron washout accumulated between $e$-folds $y$ and $x$.
For a radiation-dominated Universe with constant
effective degrees of freedom for
the energy density, $\geff$, and
the entropy density, $\heff$, we have
\begin{align}
  \dd x &= H\,\dd t
  \,,&
  H &= \frac{1}{2t}
  \,,
\end{align}
so that the propagator $E$ reduces to
\begin{equation}
  E\left(t,t'\right) =
  \exp\biggl[-\int_{t'}^t \dd t''\, \GammaB\left(t''\right)\biggr] =
  \exp\Bigl[- \overline\Gamma_{\!\rmii{$B$}}\left(t,t'\right)\left(t-t'\right)\Bigr]
  \,,
\end{equation}
where
$\overline{\Gamma}_{\!\rmii{$B$}}$
is the average rate of baryon-number violation in the time interval from $t'$ to $t$,
\begin{equation}
  \overline{\Gamma}_{\!\rmii{$B$}}\left(t,t'\right) =
  \frac{1}{t-t'}\int_{t'}^t \dd t''\, \GammaB\left(t''\right)
\,.
\end{equation}
The solution can therefore equivalently be written as
\begin{align}
  \NB\left(t\right) &=
    \NB\left(t_{\rm ini}\right)\exp\left[- \overline{\Gamma}_{\!\rmii{$B$}}\left(t,t_{\rm ini}\right)\left(t-t_{\rm ini}\right)\right]
  \nn &
  + \int_{t_{\rm ini}}^t \dd t'\,\exp\left[- \overline{\Gamma}_{\!\rmii{$B$}}\left(t,t'\right)\left(t-t'\right)\right] H\left(t'\right) F\left(t'\right) 
  \,.
\end{align}

For practical purposes, we will, however, use the temperature $T$ as our time variable in the following.
For constant $\heff$, we have
\begin{equation}
  \dd x = - \frac{\dd T}{T}
\,.
\end{equation}
At $T \lesssim \Tc \simeq 159.6$~GeV,
the propagator $E$ can then be written as
\begin{align}
\label{eq:E:GoverT}
  E\left(T,T'\right) &= \exp\biggl[-\int_T^{T'}\dd T''\,\frac{G\left(T''\right)}{T''}\biggr]
  \,,&
  \frac{G\left(T\right)}{T} &= \frac{\Lambda e^{\beta T}}{T^2}
  \,.
\end{align}
The explicit form of $G/T$, and hence
the scale $\Lambda$ and the slope $\beta$,
follow from the definition $G = \GammaB/H$ by inserting
the broken-phase sphaleron rate and the Hubble rate.
Below the crossover, $T \lesssim \Tc$,
the weak-sphaleron rate is well described by
the broken-phase branch of the fit in eq.~\eqref{fitnew}~\cite{Annala:2023jvr}
with
$\widetilde{\Gamma}_\text{ws} = f_\rmii{QCD}\Gamma_\text{ws}$
incorporating the QCD prefactor
$f_\rmii{QCD}(\Tsph) \simeq \fqcdFoNum$~\cite{Bodeker:2025ahg}.

Using
$\GammaB = \nG^2\,\tilde{\kappa}\,\widetilde{\Gamma}_\text{ws}/T^3$,
where the factor $\nG^2 = 9$ ($\nG = 3$) originates from
the weak-sphaleron rate in eq.~\eqref{eq:ws_kubo_formula} and
$\tilde\kappa = \kappa/\nG^2$ is
the flavor-normalized washout coefficient of eq.~\eqref{eq:3muB+summuL},
we obtain
\begin{align}
\label{eq:GoverT:explicit}
  \frac{G(T)}{T} &=
  \frac{\GammaB}{H\,T} =
  \nG^2\,\tilde{\kappa}\,f_\rmii{QCD}\,e^{-\betaIntercept}
  \left(\frac{90}{\pi^2 \geff}\right)^{1/2}\Mpl\,
  \frac{e^{\betaSlope\,T/\textrm{GeV}}}{T^2}
  \,,
\end{align}
where we have used
the radiation-dominated Hubble rate in eq.~\eqref{eq:Hubble:rad}.
Matching eq.~\eqref{eq:GoverT:explicit} onto the parameterization in eq.~\eqref{eq:E:GoverT},
the slope $\beta$ is inherited directly from
the exponential of the sphaleron-rate fit, whereas
$\Lambda$ collects its prefactor $e^{-\betaIntercept}$,
the factor $\nG^2 = 9$, and
the Hubble normalization,
\begin{align}
  \label{eq:Lambda:beta}
  \Lambda &= \nG^2\,\tilde{\kappa}\,f_\rmii{QCD}\,e^{-\betaIntercept}
  \left(\frac{90}{\pi^2 \geff}\right)^{1/2}\,\Mpl
  \,,&
  \beta &= \frac{\betaSlope}{\textrm{GeV}}
\,.
\end{align}
The temperature dependence of the scale $\Lambda$ is very weak. 
At the temperatures of interest,
$\Tsph \sim \TsphNum$~GeV~\cite{Annala:2023jvr},
$\tilde{\kappa}(\Tsph) \simeq \kappaFo$ and
$\geff(\Tsph) \simeq \geffFo$,
\begin{align}
\label{eq:Lambda:explicit}
  \Lambda &\simeq
    1.064\times 10^{-66}
    \biggl(\frac{\tilde{\kappa}}{\kappaFo}\biggr)
    \biggl(\frac{\geffFo}{\geff}\biggr)^{1/2}
    \Mpl
    \simeq
    2.588 \times 10^{-48}\,\textrm{GeV}
  \,.
\end{align}
For temperature variations of
a few GeV around $\Tsph$,
$\kappa$ and $\geff$ can be treated as constant.

We can now explicitly compute the function $E$.
First, we note that
\begin{equation}
  \int_T^{T'}\dd T''\,\frac{e^{\beta T''}}{T''^2} =
  \biggl[\beta\,\textrm{Ei}\left(\beta T''\right)- \frac{e^{\beta T''}}{T''}\biggr]_T^{T'}
  \,,
\end{equation}
where 
$\textrm{Ei}\left(x\right) = \int_{-\infty}^x \dd y\,\frac{e^y}{y} =
  \frac{e^x}{x}\left[1 + \frac{1}{x} + \frac{2}{x^2}+\frac{6}{x^3}
  + \mathcal{O}\left(x^{-4}\right)\right]
$
is the exponential integral.
This result grows exponentially in $T$,
such that it is dominated by the upper integration boundary.
Up to exponentially small corrections,
we can therefore write
\begin{equation}
  E\left(T,T'\right) = \exp\biggl[
      -\Lambda\biggl(\beta\,\textrm{Ei}\left(\beta T'\right)- \frac{e^{\beta T'}}{T'}\biggr)
  \biggr]
  \,,
\end{equation}
which no longer depends on $T$.
Similarly, we define $E(T) = E(0,T)$.

For vanishing initial conditions and
zero hypermagnetic helicity,
we therefore obtain
\begin{equation}
\label{eq:NBT}
  \NB(T) =
    \int_T^{T_{\rm ini}} \dd T'\,\frac{G\left(T'\right)}{T'}\,E\left(T'\right) \NB^{\rm eq}\left(T'\right)
    =
    \int_T^{T_{\rm ini}} \dd T'\,P\left(T'\right) \NB^{\rm eq}\left(T'\right)
  \,,
\end{equation}
where the weight function $P$ is defined as
\begin{equation}
  P\left(T\right) = -\frac{\dd E\left(T\right)}{\dd T} = \frac{G\left(T\right)}{T}\,E\left(T\right) = \frac{\Lambda e^{\beta T}}{T^2}\,\exp\left[-\Lambda\left(\beta\,\textrm{Ei}\left(\beta T\right)- \frac{e^{\beta T}}{T}\right)\right]
\,.
\end{equation}%
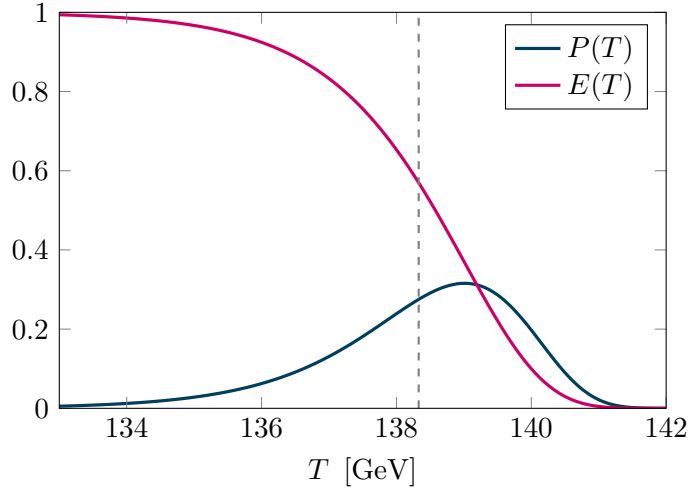
\begin{figure}[t]
\begin{center}
\begin{tikzpicture}
% ---------------------------------------------------------------------------
% Inputs: Lambda [GeV] and beta [1/GeV] from eq.~\eqref{eq:Lambda:beta},
%         and \TB [GeV] from eq.~\eqref{eq:TB}.
% The curves are evaluated in log-space to avoid the overflow of e^{beta T}
% (with beta T ~ 119) in pgfmath, using the asymptotic form
%   f(T) = Lambda e^{beta T}/(beta T^2),   ln f(T) = lnLoB + beta*T - 2 ln T,
%   E(T) = exp[-f(T)],   P(T) = beta*f(T)*exp[-f(T)],
% where the log-constant lnLoB = ln(Lambda/beta).
% For Lambda = 2.59e-48 GeV and beta = 0.86/GeV:  lnLoB = ln(2.59e-48/0.86) = -109.42 .
\pgfmathsetmacro{\betaSlope}{0.858}    % beta  [1/GeV]
\pgfmathsetmacro{\lnLoB}{-109.41998}     % ln(Lambda/beta)
\pgfmathsetmacro{\TBplot}{138.33}          % freeze-out temperature \TB [GeV], analytic eq.~(TB)
\definecolor{funcPblue}{HTML}{003f5c}
\definecolor{funcEorange}{HTML}{cf0063}
\begin{axis}[
  width=0.62\textwidth,
  height=0.44\textwidth,
  xmin=133, xmax=142,
  ymin=0, ymax=1,
  enlargelimits=false,
  axis on top,
  xlabel={$T\ \left[\textrm{GeV}\right]$},
  domain=133:142,
  xtick={134,136,138,140,142},
  samples=300,
  smooth,
  legend cell align=left,
  legend pos=north east,
]
\addplot[funcPblue, very thick] {
  \betaSlope
  * ( exp(\lnLoB + \betaSlope*x - 2*ln(x))
      * (
        + 1
        + 2/(\betaSlope*x)
        + 6/(\betaSlope*x)^2
        + 24/(\betaSlope*x)^3
      )
    )
  * exp(
    -exp(\lnLoB + \betaSlope*x - 2*ln(x))
     * (
      + 1
      + 2/(\betaSlope*x)
      + 6/(\betaSlope*x)^2
      + 24/(\betaSlope*x)^3
    )
  )
};
\addlegendentry{$P(T)$}
\addplot[funcEorange, very thick] {
  exp(
    -exp(\lnLoB + \betaSlope*x - 2*ln(x))
     * (
      + 1
      + 2/(\betaSlope*x)
      + 6/(\betaSlope*x)^2
      + 24/(\betaSlope*x)^3
    )
  )
};
\addlegendentry{$E(T)$}
\draw[gray, dashed, thick] (axis cs:\TBplot,0) -- (axis cs:\TBplot,1)
  node[pos=1, above, black, font=\footnotesize] {$\TB$};
\end{axis}
\end{tikzpicture}
\caption{%
    Weight function $P$ (dark blue) as a function of temperature for
    $\Lambda$ of eq.~\eqref{eq:Lambda:explicit} and
    $\beta$ of eq.~\eqref{eq:Lambda:beta}.
    The vertical gray dashed line marks $\TB$,
    the expectation value of $T$ computed with respect to the weight function $P(T)$.
    The exponential function $E(T)$ (magenta) represents
    a smooth version of the Heaviside function $\Theta\left(\TB-T\right)$.
  }
\label{fig:function:P}
\end{center}
\end{figure}%
As illustrated in fig.~\ref{fig:function:P},
$P$ is sharply peaked in $T$, and its integral is normalized to unity,
$\int_0^\infty \dd T\, P\left(T\right) = E\left(0\right) - E\left(\infty\right) = 1$.
We can therefore use $P$ to define the baryon-number freeze-out temperature $\TB$,
\begin{equation}
\label{eq:TB}
\TB = \int_0^\infty \dd T\, T\,P\left(T\right) \,,
\end{equation}
which corresponds to the mean temperature with respect to $P$.
Since the comoving equilibrium number density is approximately constant across
the width of the peak in $P$,
it can be moved in front of the integral, 
and
\begin{equation}
\label{eq:nbnbeq}
  \NB^{ }\left(T \ll \TB^{ } \right) = \NB^{\rm eq}\left(\TB^{ }\right)
\,,
\end{equation}
as anticipated in eq.~\eqref{eq:NB:freezeout}.

Alternatively,
the result in eq.~\eqref{eq:nbnbeq} can be motivated by going
one step back to eq.~\eqref{eq:NBT}.
Taylor-expanding $\NB^{\rm eq}\left(T'\right)$
in the integrand of eq.~\eqref{eq:NBT} around $\TB$
allows us to write
\begin{equation}
\label{eq:NBTaylor}
  \NB(T) =
    \int_T^{T_{\rm ini}} \dd T'\,P\left(T'\right)
    \biggl[
        \NB^{\rm eq}\left(\TB\right)
      + \frac{{\rm d}\NB^{\rm eq}\left(T'\right)}{{\rm d}T'}\biggr|_{\TB}\left(T'-\TB\right)
      + \mathcal{O}\left((T'-\TB)^2\right)
    \biggr]
  \,,
\end{equation}
which reproduces
eq.~\eqref{eq:nbnbeq} up to \textit{quadratic} corrections
in $T'-\TB$ in the integrand of the temperature integral,
precisely if the expansion point of the Taylor expansion, $\TB$,
is defined as in eq.~\eqref{eq:TB}.
This observation singles out the freeze-out temperature $\TB$
as the unique reference temperature
for which the {\em linear} correction
in eq.~\eqref{eq:NBTaylor} vanishes.

For
$\Lambda$ and
$\beta$ as given in eq.~\eqref{eq:Lambda:beta},
eq.~\eqref{eq:TB} yields
$\TB \simeq 138.33$~GeV,
within $0.1$~GeV of $\TB = \TBNum$~GeV obtained from
the full numerical solution,
the difference reflecting the constant-$\tilde\kappa$ and
constant-$\geff$ approximation used here.
Unfortunately, we are not able to solve the integral for $\TB$ analytically;
instead,
we need to evaluate it numerically.
Evaluated at $\TB$,
the equilibrium conversion factors read
$\Csph = \Cspheq(\TB) = \CspheqTBNum$ and
$\Fsph = \Fspheq(\TB) = \FspheqTBNum$, and
agree with eq.~\eqref{eq:Csph:out:values} within
uncertainties.

%%%%%%%%%%%%%%%%%%%%%%%%%%%%%%%%%%%%%%%%%%%%%%%%%%%%%%%%%%%%%%%%%%%%%%%%%%%%%%%%%%%%%%%%%%%%%%%%%%%%
\section{Details of dimensional reduction}

%%%%%%%%%%%%%%%%%%%%%%%%%%%%%%%%%%%%%%%%%%%%%%%%%%%%%%%%%%%%%%%%%%%%%%%%%%%%%%%%%%%%%%%%%%%%%%%%%%%%
\subsection[Renormalization and \texorpdfstring{$\beta$}{beta}-functions at zero temperature]{%
  Renormalization and \texorpdfstring{\boldmath{$\beta$}}{beta}-functions at zero temperature%
  }
\label{sec:renorm:beta}

For the SM parameters,
we encode their running with respect to
the \MSbar{} renormalization scale $\LamD$ via corresponding $\beta$-functions
from~\cite{Jenkins:2013zja,Jenkins:2013wua,Alonso:2013hga}
up to two-loop order~\cite{Buttazzo:2013uya,Croon:2020cgk}.
To this end, we
relate the \MSbar{} parameters to
their corresponding physical parameters
taken from~\cite{ParticleDataGroup:2026mpi}
at
the electroweak input scale $\LamDInput = \mZ$,
using NLO vacuum renormalization relations~\cite{Kajantie:1995dw}.
We further evolve
the \MSbar{} parameters to
the thermal reference (or matching) scale
\begin{align}
\label{eq:RGscale:X}
  \LamDRef &= X \Tbar
  \,,&
  \text{using}\quad
  \Tbar &= 4\pi e^{-\gammaE} T
  \,,
\end{align}
where $\gammaE$ is the Euler--Mascheroni constant and
$X$ parameterizes residual higher-order corrections.

The matching relations listed below furthermore involve the two
logarithms that the bosonic and fermionic Matsubara sums generate in
the high-temperature expansion~\cite{Kajantie:1995dw},
\begin{align}
\label{eq:Lb:Lf}
  \Lb &\equiv
    \ln\frac{\LamD^2}{T^2} - 2\ln 4\pi + 2\gammaE
  \,, &
  \Lf &\equiv
  \Lb + 4\ln 2
  \,,
\end{align}
both evaluated at the scale $\LamD = \LamDRef$.
Choosing $X = 1$ thus sets $\Lb = 0$, and
since the leading order~\eqref{eq:nBeq:LO} is coupling-independent,
explicit logarithms at NLO in $\nBeq$ must cancel.
The remaining RG-scale dependence
is encoded in the Higgs expectation value $\phibar(T)$ and
the running couplings.
We therefore fix the 4d RG scale at $\LamDRef = \Tbar$ and set $X = 1$,
varying it over $X \in [1/2,2]$ to estimate
the size of missing higher orders;
cf.\ sec.~\ref{sec:sphaleron:outOfEq}.

%%%%%%%%%%%%%%%%%%%%%%%%%%%%%%%%%%%%%%%%%%%%%%%%%%%%%%%%%%%%%%%%%%%%%%%%%%%%%%%%%%%%%%%%%%%%%%%%%%%%
\subsection{Matching relations at finite chemical potential}
\label{sec:matching:3d:mu}

This section lists the matching relations
of the thermal EFT employed in
sec.~\ref{sec:Omega:2loop}.
In this context,
we used results from ref.~\cite{Gynther:2003za},
supplemented by new contributions consistent with
our power counting~\eqref{eq:power:counting:couplings}.

\begin{figure}[t]
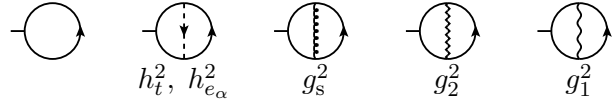

\begin{center}
\begin{tabular}{ccccc}
  \TopoOT(\Lsai,\Aqu)
  &
  \ToptOM(\Lsai,\Aqq,\Aqu,\Aqq,\Lcsi)
  &
  \ToptOM(\Lsai,\Aqq,\Aqu,\Aqq,\Lgliii)
  &
  \ToptOM(\Lsai,\Aqq,\Aqu,\Aqq,\Lglii)
  &
  \ToptOM(\Lsai,\Aqq,\Aqu,\Aqq,\Lgli)
  \\[1mm]
  & $\yt^2 ,\; \yeAlpha^2$ & $\gs^2$
  & $g_2^2$ & $g_1^2$
\end{tabular}
\end{center}
\caption[]{%
  Diagrams contributing to
  the linear-in-$\B_0$ term~\eqref{k} up to $\mathcal{O}(g^2 \mu)$
  with the corresponding relative orders below.
  Solid lines with arrows denote fermions,
  the dashed line the Higgs,
  curly lines gluons,
  zigzag lines SU(2) gauge bosons, and
  wiggly lines the hypercharge boson.
  The plain external line is $\B_0$.
  The first diagram is leading order and
  the remaining diagrams are next-to-leading order.
  The last diagram, of $\mathcal{O}(g_1^3)$,
  is omitted in ref.~\cite{Gynther:2003za}, where $g_1^2 \sim g_2^3$.
  }
\label{fig:lintrm}
\end{figure}
The coefficient of the term linear in $\B_0$
is composed of the diagrams shown in fig.~\ref{fig:lintrm} and
up to $\mathcal{O}\bigl(g^3\mu\bigr)$
reads
\begin{align}
\label{k}
  \mu_{1} = - \frac{i g_1 T^{3/2}}{3}
   \biggl( 1 - \frac{\hatPip_{\B_0}}{2} \biggr)
   \biggl\{
      & \biggl(
          1  - \frac{1}{(4\pi)^2} \biggl(
                      \frac { 15 g_1 ^ 2 } {4}
                    + \frac { 9 g_2 ^ 2 } {4}
                  \biggr)
        \biggr) \sum _ \alpha ^ {\nG} \muLAlpha
       - \frac {9} {2(4\pi)^2 }
      \sum _ \alpha ^ { \nG } \yeAlpha ^ 2 \muLAlpha^{ } 
  \nn
    - &\biggl(
      1  - \frac{1}{(4\pi)^2} \biggl(
                      \frac { 29 g_1 ^ 2 } {12}
                    + \frac { 9 g_2 ^ 2 } {4}
                    + \frac { 5\yt ^ 2 } {2}
                    + 6\CF^{ }\gs^{2}
                  \biggr)
      \biggr) \muB
  \biggr\}
   \,,
\end{align}
where
$\CF = (\CA^2 - 1)/(2\CA)$,
which for SU(3) is $\CF = 4/3$.
The term containing the lepton Yukawa couplings
is taken from eq.~(6) of ref.~\cite{Laine:1999wv}.
The top-Yukawa, weak-SU(2) and QCD corrections
can be found in eq.~(28) of ref.~\cite{Gynther:2003za}.
The bracket in eq.~\eqref{k} contains the two-loop tadpole
of~\cite{Gynther:2003za},
computed there without the $\mathcal{O}(g_1^3)$ terms.
With our power counting~\eqref{eq:power:counting:couplings},
the $\mathcal{O}(g_1^3)$ contributions
contain both the last diagram of fig.~\ref{fig:lintrm}, obtained from
$\mu_1^{ } = i g_1^{ } T^{-1/2}
\partial_{\muY} p_{\text{hard}}|_{\muY=0}$, and
the $\mathcal{O}(g_1^2)$ part of the field normalization.%
\footnote{
  \label{fn:fieldnorm}
  The field normalization
  $(\B_0^2)_\rmii{3d} = (\B_0^2)_\rmii{4d}(1+\hatPip_{\B_0})/T$
  is obtained from the
  $\mathcal{O}(g_1^2)$
  momentum-dependent part of the $\B_0$ self-energy
  $\hatPip_{\B_0} = \frac{g_1^2}{(4\pi)^2}\bigl(
    \frac13
  + \frac{20\nG}{9}(\Lf-1)
  + \frac{\Lb}{6}\bigr)
  $;
  see ref.~\cite{Croon:2020cgk}.
}

The cubic couplings between
the temporal vectors and the Higgs
up to $\mathcal{O}\bigl(g^3\mu\bigr)$
are
\begin{align}
   \rho_{1}^{ } &=  - \frac { 2 i g_1 }  { (4\pi) ^ 2 } T ^{ 1/2 }
   \biggl(
        \frac 53 \yt ^ 2  \muB
        - 3 \sum _ \alpha ^ {\nG} \yeAlpha ^ 2 \muLAlpha^{ }
   \biggr)
   \label{rhop}
  \,,\\
  \rho_{2}^{ }  &= - \frac{2 i g_2 }{ (4 \pi) ^ 2 } T ^{ 1/2 }
   \biggl( -  \yt ^ 2  \muB
  +
   \sum _ \alpha ^ {\nG} \yeAlpha ^ 2 \muLAlpha^{ }
   \biggr)
   \label{rho}
   \,,\\
  \rho_{\rmii{$G$}}^{ } &= - \frac{ 2 i g_1^{ } g_2^2 }{ (4\pi)^2 }
   T^{1/2}
   \biggl(
      \muB^{ } - \sum_\alpha^{\nG} \muLAlpha^{ }
   \biggr)
   \label{rhoG}
   \,.
\end{align}
The $\muB$-terms are obtained
from ref.~\cite{Gynther:2003za};%
\footnote{%
  \label{ifactors}
  Our
  $\A_0^{a}$ and
  $\B_0^{ }$ coincide with those
  of ref.~\cite{Gynther:2003za}, which are likewise purely imaginary;
  see eq.~\eqref{eq:muY:muA:4d}.
}
the $\muLAlpha$-terms of eqs.~\eqref{rhop} and~\eqref{rho}
from ref.~\cite{Laine:1999wv}.
The corresponding Chern--Simons terms vanish
due to the non-conservation of $B+L$ in eq.~\eqref{eq:eq:cond}.

The scalar mass parameter gets
chemical-potential corrections at one-loop order,%
\footnote{%
  The
  $\mu$-independent term $ m_{3,\mu=0}^2$
  is given in ref.~\cite{Kajantie:1995dw}.
  The second
  term on the right-hand side of eq.~\eqref{eq:m3_mu}
  can be found in ref.~\cite{Gynther:2003za}, while the 
  third one can be obtained from ref.~\cite{Laine:1999wv}.
}  
\begin{align}
\label{eq:m3_mu} 
    m_3^2 
    = m_{3,\mu=0}^2
      + \frac{1}{(4 \pi)^2}\biggl(
          \frac{4}{3} \yt ^ 2\, \muB ^ 2
        +  4 \sum _ \alpha ^ {\nG} \yeAlpha ^ 2\, \muLAlpha ^ 2
      \biggr)
    \,,
\end{align} 
with two-loop corrections of the form
$\mathcal{O}\bigl(g^4 \mu^2, \mu^4/T^2\bigr)$.
The Debye masses
up to $\mathcal{O}\bigl(g^2\mu^2\bigr)$
read~\cite{Gynther:2003za}
\begin{align}
  \label{eq:mD1:mu}
  \mDi{1}^2 &=
      m_{\rmii{D1},\mu=0}^2
    + \frac{4g_1^2}{(4\pi)^2}
      \biggl(
          \frac{11}{9}\muB^2
        + 3\sum_\alpha^{\nG} \muLAlpha^2
      \biggr)
  \,,\\
  \label{eq:mD2:mu}
  \mDi{2}^2 &=
      m_{\rmii{D2},\mu=0}^2 
    + \frac{4g_2^2}{(4\pi)^2}
      \biggl(
        \muB^2
        + \sum_\alpha^{\nG} \muLAlpha^2
      \biggr)
  \,.
\end{align}

The quartic and gauge coupling constants
get chemical-potential corrections as well;
see ref.~\cite{Gynther:2003za}.
Since these corrections enter at
$\mathcal{O}(g^4\mu^2/T^2)$,
only their $\mu$-independent contributions
are required
\begin{align}
\label{eq:mu:remaining:couplings}
% \label{eq:h1:mu}
  h_{1}^{ } &=
      h_{1,\mu=0}^{ }
  \,,&
% \label{eq:h2:mu}
  h_{2}^{ } &=
      h_{2,\mu=0}^{ }
  \,,&
% \label{eq:h3:mu}
  h_{3} &=
      h_{3,\mu=0}
  \,,&
  \lambda_{3}^{ } &=
      \lambda_{3,\mu=0}^{ }
  \,,\nn[1mm]
  g_{1,3}^{2} &=
      g_{1,3,\mu=0}^{2}
  \,,&
  g_{2,3}^{2} &=
      g_{2,3,\mu=0}^{2}
  \,.
\end{align}

%%%%%%%%%%%%%%%%%%%%%%%%%%%%%%%%%%%%%%%%%%%%%%%%%%%%%%%%%%%%%%%%%%%%%%%%%%%%%%%%%%%%%%%%%%%%%%%%%%%%
\subsection{Matching relations at finite temperature and zero chemical potential}
\label{app:matching:zeroMu}

The corresponding matching relations
at finite temperature and
zero chemical potential
can be found in~\cite{Croon:2020cgk}.
For completeness, we list them here again
up to $\mathcal{O}(g^4)$.
The 3d scalar mass and
self-coupling are given by~\cite{Kajantie:1995dw} 
\begin{align}
  m_{3,\mu=0}^2 &=
    - \nu^ 2 
    + \frac{T^2}{16} \Bigl(
        3 g_2^2
      + g_1^2
      + 4 \yt^2
      + 8 \lambda
    \Bigr)
  \,,\\[1mm]
  \lambda_{3,\mu=0} &=
    T \biggl[
      \lambda
      + \frac{1}{(4\pi)^2} \biggl(
        \Bigl(3 g_2^4 + 2 g_2^2 g_1^2 + g_1^4\Bigr)
        \frac{2-3\Lb}{16}
        + \frac{3\Lb}{2} \Bigl(
            3 g_2^2
            + g_1^2
            - 8 \lambda
          \Bigr)\lambda
        \nn &\hphantom{{}=T\biggl[\lambda+\frac{1}{(4\pi)^2}\biggl(}
        + 3\Lf \bigl(\yt^4 - 2\lambda \yt^2 \bigr)
      \biggr)
    \biggr]
  \,.
   \label{m3}
\end{align}
The two-loop term of $m_{3,\mu=0}^2$,
which can be found in ref.~\cite{Croon:2020cgk},
is not needed here.
Since it is
independent of the chemical potentials, it cancels from 
$\nBeq$ and $\kappa$, and
it enters the soft one-loop pressure of sec.~\ref{sec:soft:1loop} only through
$\widetilde{m}_{\varphi}^{ }$ and
$\widetilde{m}_{\rmii{$G$}}^{ }$,
where the leading order suffices.
The 3d gauge couplings and Debye masses
up to $\mathcal{O}(g^4)$ are
\begin{align}
  \label{eq:g13:mu0}
  g_{1,3,\mu=0}^{2} &=
    g_1^2 T\biggl[
      1
      + \frac{g_1^2}{(4\pi)^2} \Bigl(
        - \frac{1}{6} \Lb
        - \frac{20 \nG}{9} \Lf
      \Bigr)
    \biggr]
  \,,\\[1mm]
  g_{2,3,\mu=0}^2 &=
    g_2^2 T\biggl[
      1
      + \frac{g_2^2}{(4\pi)^2} \Bigl(
          \frac{43}{6} \Lb
        + \frac{2}{3}
        - \frac{4\nG}{3} \Lf
      \Bigr)
    \biggr]
  \,,\\[1mm]
  \label{eq:mD1:mu0}
  m_{\rmii{D1},\mu=0}^{2} &=
    g_1^2 T^2 \biggl[
        \frac{1}{6}
      + \frac{5\nG}{9}
    \nn &\hphantom{={}g_1^2 T^2}
      + \frac{1}{(4\pi)^2}\biggl(
          \biggl[
              \frac{5\bigl(9 - 204\nG + 160\nG^2\bigr)}{648}
            - \frac{(3 + 10\nG)}{108}\Lb
            - \frac{10\nG(3 + 10\nG)}{81}\Lf
          \biggr] g_1^2
        \nn &\hphantom{{}=g_1^2 T^2+ \frac{1}{(4\pi)^2}\biggl(}
        + \frac{3 - 4\nG}{8} g_2^2
        - \frac{22\nG}{9}\gs^2
        - \frac{11}{12}\yt^2
        + \lambda
        - \frac{2\nu^2}{T^2}
      \biggr)
    \biggr]
  \,,\\[1mm]
  \label{eq:mD2:mu0}
  m_{\rmii{D2},\mu=0}^2 &=
    g_2^2 T^2 \biggl[
        \frac{5}{6}
      + \frac{\nG}{3}
    \nn &\hphantom{={}g_2^2 T^2}
      + \frac{1}{(4\pi)^2}\biggl(
          \biggl[
              \frac{207 + 44\nG + 32\nG^2}{72}
            + \frac{43(5 + 2\nG)}{36}\Lb
            - \frac{2\nG(5 + 2\nG)}{9}\Lf
          \biggr] g_2^2
        \nn &\hphantom{{}=g_2^2 T^2+ \frac{1}{(4\pi)^2}\biggl(}
        + \frac{3 - 4\nG}{24} g_1^2
        - 2\nG\gs^2
        - \frac{1}{4}\yt^2
        + \lambda
        - \frac{2\nu^2}{T^2}
      \biggr)
    \biggr]
  \,.
\end{align}
The two-loop pieces of
eqs.~\eqref{eq:mD1:mu0} and~\eqref{eq:mD2:mu0}
are of the same relative order as
the next-to-leading terms of
$g_{i,3}^2$,
$h_{i}^{ }$ and
must be included at
$p_\text{soft} \sim \mathcal{O}(g^2 \mu^2 T^2)$
in the zero modes of eq.~\eqref{eq:veff:tree}.

The temporal-vector couplings 
with the soft Higgs are
\begin{align}
\label{eq:h1:mu0}
  h_{1,\mu=0} &=
    \frac{g_1^2 T}{4}\biggl[
        1
      + \frac{1}{(4\pi)^2} \biggl(
        \frac{3g_2^2}{2}
        - \Bigl[
            \frac{\Lb-1}{6}
          + \frac{20\nG}{9} (\Lf - 1)
        \Bigr] g_1^2
        - \frac{34}{3} \yt^2
        + 12 \lambda
      \biggr)
    \biggr]
  \,,\\[1mm]
\label{eq:h2:mu0}
  h_{2,\mu=0} &=
    \frac{g_2^2 T}{4}\biggl[
        1
      + \frac{1}{(4\pi)^2} \biggl(
        \Bigl[
            \frac{43}{6}\Lb
          + \frac{17}{2}
          - \frac{4\nG}{3} (\Lf - 1)
        \Bigr] g_2^2
        + \frac{g_1^2}{2}
        - 6 \yt^2
        + 12 \lambda
      \biggr)
    \biggr]
  \,,\\[1mm]
\label{eq:h3:mu0}
  h_{3,\mu=0} &=
    \frac{g_1 g_2 T}{2}\biggl[
        1
      + \frac{1}{(4\pi)^2} \biggl(
        \frac{g_1^2 - 3g_2^2}{3}
        + \Bigl[
            \frac{43}{12}g_2^2
          - \frac{g_1^2}{12}
        \Bigr]\Lb
      \nn &\hphantom{{}=\frac{g_1 g_2 T}{2}\biggl[1 + \frac{1}{(4\pi)^2}\biggl(}
        - \nG(\Lf - 1)\Bigl[
            \frac{2}{3} g_2^2
          + \frac{10}{9} g_1^2
        \Bigr]
        + 2 \yt^2
        + 4 \lambda
      \biggr)
    \biggr]
  \,.
\end{align}
The temporal-vector quartic couplings
$\kappa_{1},\kappa_{2},\kappa_{3}$ are
listed in ref.~\cite{Croon:2020cgk}, and
their effect goes beyond the accuracy of the computation.

%%%%%%%%%%%%%%%%%%%%%%%%%%%%%%%%%%%%%%%%%%%%%%%%%%%%%%%%%%%%%%%%%%%%%%%%%%%%%%%%%%%%%%%%%%%%%%%%%%%%
\subsection{Temperature evolution of the Higgs expectation value}
\label{sec:Tevolution:vev}

To compute the temperature dependence of the Higgs expectation value,
we work within
the high-temperature 3d~EFT of
the Standard Model.%
\footnote{%
  An alternative approach used in the previous
  literature is to fit an empirical formula for $\phibarMin(T)$ to
  lattice simulations of the electroweak crossover, as done, e.g., in
  appendix~B of~\cite{Kamada:2016cnb}, based on lattice results
  of~\cite{DOnofrio:2015gop}.
}
In this high-temperature regime,
the Higgs value is
$\phibarMin \sim \mathcal{O}(T)$, so no hierarchically
small mass scales appear and the perturbative expansion is
well-behaved.
In the parent 4d theory,
we count
$g^2 \sim \yt^2 \sim \lambda$ as
in eq.~\eqref{eq:power:counting:couplings}.
Upon dimensional reduction to the 3d~EFT, this
power counting shifts to $\lambda \sim g^3$~\cite{Arnold:1992rz}, and
the natural expansion parameter there becomes
$x = \overline\lambda_3/\overline g_3^2$.
The loop expansion parameter
is then either $g$ or $\overline g_3^2/m$, where $m$ denotes a generic
mass scale in the 3d~EFT.
The effective parameters $\overline m_3^2$
and $\overline\lambda_3$ are obtained by matching onto
the softer 3d~EFT of the Standard Model, and we take their values
from appendix
B.2.2 of ref.~\cite{Croon:2020cgk}. We assume chemical potentials to be
small and neglect them throughout.

For a given temperature,
we have to find the minimum of the effective
potential by solving 
\begin{align}
\label{min}
    \Veff'(\phibarMin)=0
  \,.
\end{align}
The expansions of the effective potential and
of the finite-temperature Higgs expectation value
are
\begin{align}
  \Veff^{ } &=
      \Veff^\rmii{LO}
    + \Veff^\rmii{NLO}
    + \Veff^\rmii{N$^2$LO}
    + \cdots
    \,,\\
  \phibarMin &=
      \phibar_0
    + \phibar_1
    + \phibar_2
    + \cdots
  \,,
\end{align}
where the latter is obtained by expanding
eq.~\eqref{min}.
The lowest-order term is determined by
$(\Veff^\text{tree})'(\phibar_0)=0$, which gives
\begin{align}
\label{eq:vT}
  \frac{\phibar_{0}^2}{T} &=
    -\frac{\overline{m}_3^2}{\overline\lambda_3}
  \;
  \qquad
  \text{for}
  \quad
  \overline{m}_3^2 < 0
  \,.
\end{align}

At leading order,
the effective potential is given by
the tree-level contribution and
the one-loop contribution from
the gauge bosons~\cite{Lofgren:2021ogg,Ekstedt:2022zro,Ekstedt:2024etx}.
This way, the LO minimum corresponds to
\begin{align}
\label{eq:vT:LO}
  \frac{\phibar_{0}^2}{T} &=
    \frac{
        2\overline{g}_{3}^3
      + (\overline{g}_{3}'^2+\overline{g}_{3}^2)^{\frac{3}{2}}
      + \sqrt{
        \bigl(2\overline{g}_{3}^3 + (\overline{g}_{3}'^2+\overline{g}_{3}^2)^{\frac{3}{2}}\bigr)^2
        -1024\pi^2\overline\lambda_3^{ }\overline{m}_3^2
      }}{32\pi\overline\lambda_3}
  \,.
\end{align}
The
first correction $\phibar_1$ and
second correction $\phibar_2$ are determined by the linear equations
\begin{align}
    \Veff^\rmii{LO}{}''(\phibar_0)\phibar_1 +
    \Veff^\rmii{NLO} {}'(\phibar_0) &= 0
    \,,\\
    \Veff^\rmii{LO}{}''(\phibar_0)\phibar_2
    + \frac12  \Veff^\rmii{LO}{}'''(\phibar_0)\phibar_1^2
    + \Veff^\rmii{NLO}{}''(\phibar_0) \phibar_1
    + \Veff^\rmii{N$^2$LO} {}' (\phibar_0)
    &= 0
    \,,
\end{align}
such that
\begin{align}
    \phibar_1 &=
    - \frac{\Veff^\rmii{NLO} {}'(\phibar_0)}{\Veff^\rmii{LO}{}''(\phibar_0)}
    \,,\\
    \phibar_2 &=
    - \frac12  \frac{\Veff^\rmii{LO}{}'''(\phibar_0) \bigl(\Veff^\rmii{NLO} {}'(\phibar_0)\bigr)^2}{\bigl(\Veff^\rmii{LO}{}''(\phibar_0)\bigr)^3}
    + \frac{\Veff^\rmii{NLO}{}''(\phibar_0) \Veff^\rmii{NLO} {}'(\phibar_0)}{\bigl(\Veff^\rmii{LO}{}''(\phibar_0)\bigr)^2}
    - \frac{\Veff^\rmii{N$^2$LO} {}' (\phibar_0)}{\Veff^\rmii{LO}{}''(\phibar_0)}
    \,.
\end{align}

In perturbation theory, the SM electroweak phase transition is
always found to be first order, with a critical temperature $\Tc$
separating the symmetric and broken phases. This is an artifact of
the perturbative expansion.
In reality, the transition is a smooth
crossover, and the perturbative prediction for
$\phibarMin(T)$ becomes
unreliable as $T$ approaches $\Tc$ from below. Sufficiently far below
$\Tc$, however, the perturbative expansion of the Higgs expectation
value agrees well with the true crossover behavior. Since we are
interested in temperatures $\Tsph < \Tc$ in this regime, we can
consistently use
the perturbative $\phibarMin(T)$ when solving for the baryon
number density in the broken phase.
Correspondingly, when integrating
the transport equation~\eqref{eq:baryon:freeze:out},
we set the
maximal integration temperature to $\Tc$ rather than extending it to
arbitrarily high temperatures,
beyond the critical endpoint at
$\mH \simeq 70$~GeV~\cite{Kajantie:1995kf},
where perturbation theory no longer
describes the actual crossover.

%%%%%%%%%%%%%%%%%%%%%%%%%%%%%%%%%%%%%%%%%%%%%%%%%%%%%%%%%%%%%%%%%%%%%%%%%%%%%%%%%%%%%%%%%%%%%%%%%%%%
\section{Master integrals at finite chemical potential}
\label{sec:master:integrals}

This appendix collects the master integrals used
in the effective potential and
the dimensional-reduction matching of
sec.~\ref{sec:Omega:2loop}.
The $d$-dimensional integral measure is
\begin{equation}
\label{eq:measure}
  \int_{\vec{\pMom}} \equiv \int
  \frac{{\rm d}^d \vec{\pMom}}{(2\pi)^d} =
  \frac{2}{(4\pi)^2\Gamma(\frac{d}{2})} \int_{0}^{\infty} {\rm d}\pMom\,\pMom^{d-1}
  \;.
\end{equation}
The $d$-dimensional
logarithmic integral is given by
\begin{eqnarray}
\label{eq:J3}
  J_{d} &\equiv& \frac{1}{2} \int_\vec{\pMom} \ln(\pMom^2 + m^2) =
  -\frac{1}{2}
  \Big( \frac{\Lambda^2_{\rmii{3d}}e^\gammaE}{4\pi} \Big)^\epsilon
  \frac{[m^2]^\frac{d}{2}}{(4\pi)^{\frac{d}{2}}} \frac{\Gamma(-\frac{d}{2})}{\Gamma(1)}
  \,,
  \nn
  J_{3}
  &\stackrel{d=3-2\epsilon}{=}& -\frac{(m^2)^{\frac{3}{2}}}{12 \pi} + \mathcal{O}(\epsilon)
  \;.
\end{eqnarray}
Additionally, in sec.~\ref{sec:soft:1loop},
we encounter the following class
of one-loop master integrals,
\begin{eqnarray}
\label{eq:I3}
  I_{s}^{d}(m) &\equiv& \int_\vec{\pMom} \frac{1}{[\pMom^2 + m^2]^s} =
  \Big( \frac{\Lambda^2_{\rmii{3d}}e^\gammaE}{4\pi} \Big)^\epsilon
  \frac{[m^2]^{\frac{d}{2}-s}}{(4\pi)^{\frac{d}{2}}}
  \frac{\Gamma(s-\frac{d}{2})}{\Gamma(s)}
  \,,
  \\
  I_{1}^{3}(m) &\stackrel{d=3-2\epsilon}{=}&
  -\frac{(m^2)^{\frac{1}{2}}}{4 \pi} + \mathcal{O}(\epsilon)
  \;.
\end{eqnarray}
Two propagators of different mass give
\begin{align}
\label{eq:I12}
  I(m_1^{ },m_2^{ })
  &\equiv
  \int_{\vec{\pMom}}
  \frac{1}{[\pMom^2 + m_1^2][\pMom^2 + m_2^2]}
  \stackrel{d=3-2\epsilon}{=}
  \frac{1}{4\pi ( m_1^{ } + m_2^{ } )}
  + \mathcal{O}(\epsilon)
  \,.
\end{align}

%%%%%%%%%%%%%%%%%%%%%%%%%%%%%%%%%%%%%%%%%%%%%%%%%%%%%%%%%%%%%%%%%%%%%%%%%%%%%%%%%%%%%%%%%%%%%%%%%%%%
\small
% \bibliographystyle{utphys}
% \bibliography{refs}

%%%%%%%%%%%%%

\end{document}